%% file: Draft-final-version.tex
\documentclass[12pt]{article} 
\usepackage{amsmath,amssymb,amsfonts}
\usepackage{setspace}
\usepackage{psfrag}
\usepackage{color}
\definecolor{darkblue}{rgb}{0.1,0.1,.7}
\usepackage[colorlinks, linkcolor=darkblue, citecolor=darkblue, urlcolor=darkblue, linktocpage]{hyperref} 
\usepackage[square, comma, compress,numbers]{natbib}
\usepackage[]{amsmath}
\usepackage[]{graphicx}
\usepackage[]{latexsym}
\usepackage{geometry}
\usepackage{amscd}
\usepackage[all,cmtip]{xy}
\usepackage{mathrsfs}
\usepackage{bbold}
\usepackage{dsfont}
\usepackage{subfigure}
\usepackage{anyfontsize}
\usepackage[margin=10pt,font=small,labelfont=bf]{caption}
\usepackage{changepage}
\usepackage{mymacros}
\numberwithin{equation}{section}

\newcommand{\bea}{\begin{eqnarray}}
\newcommand{\eea}{\end{eqnarray}}

\DeclareMathOperator{\Tr}{Tr}

\def\beq{\begin{equation}} 
\def\eeq{\end{equation}} 
\def\<{\langle}
\def\>{\rangle}

\def\nn{\nonumber} 
 
\def\cO {{\cal O}}

\begin{document}

\vspace*{-.6in} \thispagestyle{empty}
\begin{flushright}
\end{flushright}
\vspace{.2in} {\Large
\begin{center}
{\bf Charting the space of 3D CFTs\\ with a continuous global symmetry}\\
\end{center}
}
\vspace{.2in}
\begin{center}
{\bf 
Anatoly Dymarsky$^{a}$, 
Joao Penedones$^{b}$,
Emilio Trevisani$^{c}$,
Alessandro Vichi$^{b}$} 
\\
\vspace{.2in} 
$^a$ {\it  Department of Physics and Astronomy, University of Kentucky,\\ Lexington, KY 40506, USA}\\
 {\it Skolkovo Institute of Science and Technology, Skolkovo Innovation Center, \\ Moscow 143026 Russia}\\
$^c$ {\it Institute of Physics,
\'Ecole Polytechnique F\'ed\'erale de Lausanne (EPFL),\\
Rte de la Sorge, BSP 728, CH-1015 Lausanne, Switzerland}\\
$^d$ {\it Centro de Fisica do Porto, Universidade do Porto, 
Rua do Campo Alegre 687, 
\\4169 007 Porto, 
Portugal}\\
 {\it Perimeter Institute for Theoretical Physics, Waterloo, Ontario 
\\
N2L 2Y5, Canada}
\end{center}

\vspace{.2in}

\begin{abstract}
We study correlation functions of a conserved spin-1 current $J_\mu$ in three dimensional Conformal Field Theories (CFTs). 
We investigate the constraints imposed by permutation symmetry and current conservation on the form of three point functions $\langle J_\mu J_\nu \mathcal O_{\Delta,\ell}\rangle $ and the four point function $\langle J_\mu J_\nu J_\rho J_\sigma \rangle $ and identify the minimal set of independent crossing symmetry conditions. We obtain a recurrence relation for conformal blocks for generic spin-1 operators in three dimensions. In the process, we improve several technical points, facilitating the use of recurrence relations. By applying the machinery of the numerical conformal bootstrap we obtain universal bounds on the dimensions of certain light operators as well as the central charge. Highlights of our results include numerical evidence for the conformal collider bound and new constraints on the parameters of the critical $O(2)$ model. The results obtained in this work apply to any unitary, parity-preserving three dimensional CFT.
\end{abstract}

\newpage

\tableofcontents

\newpage


\section{Introduction}

The classification of all Conformal Field Theories (CFTs) is the utopian dream that drives the systematic development of  the conformal bootstrap program \cite{Belavin:1984vu,Polyakov:1974gs}. Almost ten years ago it was observed that the constraint of crossing symmetry can be recast into an infinite set of linear and quadratic equations, whose feasibility can be studied numerically \cite{Rattazzi:2008pe,Rychkov:2009ij,Caracciolo:2009bx,Poland:2011ey}. Since then,   the numerical conformal bootstrap has been successfully applied to four-point  functions of scalar operators in several spacetime dimensions \cite{Rattazzi:2010gj,Rattazzi:2010yc,Poland:2010wg,Vichi:2011ux,ElShowk:2012ht,Kos:2014bka,Chester:2014gqa,Li:2016wdp}, and to spin-$\frac{1}{2}$ operators in three dimensions \cite{Iliesiu:2015qra,Poland3DON}. This has led to spectacular results, such as the most precise determination of the three dimensional Ising model critical exponents and its spectrum  \cite{El-Showk:2014dwa,Kos:2016ysd,Simmons-Duffin:2016wlq}, a partial classification of $O(N)$-models in three dimensions \cite{Kos:2015mba}, and interesting insights on superconformal theories with and without Lagrangian formulation \cite{Beem:2013qxa,Chester:2014fya,Poland:2015mta,Lemos:2015awa,Beem:2015aoa,Bobev:2015jxa,Chester:2015lej,Lin:2015wcg,Beem:2016wfs,Lemos:2016xke,Lin:2016gcl}. In this paper, we consider the next step in complexity: correlators of vector operators. In particular we study the four-point function of a conserved current.

Any local CFT with a continuous global symmetry contains a conserved current $J_\mu$,
whose flux through the boundary of a region $B$ measures the total charge inside this region.\footnote{The existence of a conserved current follows from the Noether theorem in any Lagrangian CFT. However, we do not know of a more general (bootstrap) proof of this statement.} 
This property is encoded in the Ward identity,
\be
\int_{\partial B} dx\, n^\mu \langle J_\mu (x) \mathcal{O}_1(x_1)\dots \mathcal{O}_n(x_n)\rangle= -
\langle  \mathcal{O}_1(x_1)\dots \mathcal{O}_n(x_n)\rangle
\sum_{x_i\in B} q_i\,,
\ee
where $n^\mu$ is the unit normal to the boundary of the region $B$ and  $q_i$ are the charges of the local operators $\mathcal{O}_i$.
We shall study the four-point function  of $J_\mu$
which is an observable that exists in any CFT with a continuous global symmetry. 
This will allow us to constrain the spectrum of operators that appear in the Operator Product Expansion (OPE) of two currents. 
In three spacetime dimensions, these neutral operators can be classified by their  scaling dimension $\Delta$, $SO(3)$ spin $\ell$ and parity.\footnote{We shall restrict our analysis to parity invariant CFTs. It would be interesting to relax this condition since there are many examples of parity breaking 3D CFTs involving   Chern-Simons gauge fields.}  
More precisely, we will study the conformal block decomposition
\begin{align}
\langle J_{\mu_1}(x_1) \dots J_{\mu_4}(x_4)\rangle
&={\sum_{\cO} \sum_{p,q } } \lambda_{JJ\cO}^{(p)}\lambda_{JJ\cO}^{(q)}
\raisebox{1.7em}{$\xymatrix@=6pt{{J}\ar@{-}[rd]& & & &J \ar@{-}[ld]   \\  
& *+[o][F]{\mbox{\tiny p}}  \ar@{-}[rr]^{\displaystyle \cO } & & *+[o][F]{\mbox{\tiny q}}  &  \\
J \ar@{-}[ru]& & & &J \ar@{-}[lu]}$} \ ,  
\end{align}
where $\lambda_{JJ\cO}^{(p)}$ are the coefficients of the operator $\cO$ in the OPE of two currents. The index $p,q$ run over a finite range, which depends on the spin and parity of the operator $\cO$. The symbol
$\raisebox{1em}{$\xymatrix@=2pt{ \ar@{-}[rd]& & & &  \ar@{-}[ld]   \\  
& *+[o][F]{\mbox{\tiny  }}  \ar@{-}[rr]^{\displaystyle  } & & *+[o][F]{\mbox{\tiny  }}  &  \\
  \ar@{-}[ru]& & & &  \ar@{-}[lu]}$}$ stands for the conformal blocks that are labeled by $p$, $q$ and the quantum numbers  $\Delta$, $\ell$ and  parity of $\cO$. This is described in detail in section \ref{sec:setup}.
Following the usual bootstrap strategy, we then impose crossing symmetry of this four-point function. However, due to current conservation, not all crossing equations are linearly independent. In section \ref{sec:setup}, we explain how to select a minimal set of independent crossing equations to be imposed numerically.
With these ingredients and assuming unitarity, we applied the usual bootstrap semi-definite programming method (SDPB) to constrain the spectrum of neutral operators and some OPE coefficients $\lambda_{JJ\cO}^{(p)}$.

\begin{figure}[h]
 \begin{center}
 \includegraphics[width=0.6\textwidth]{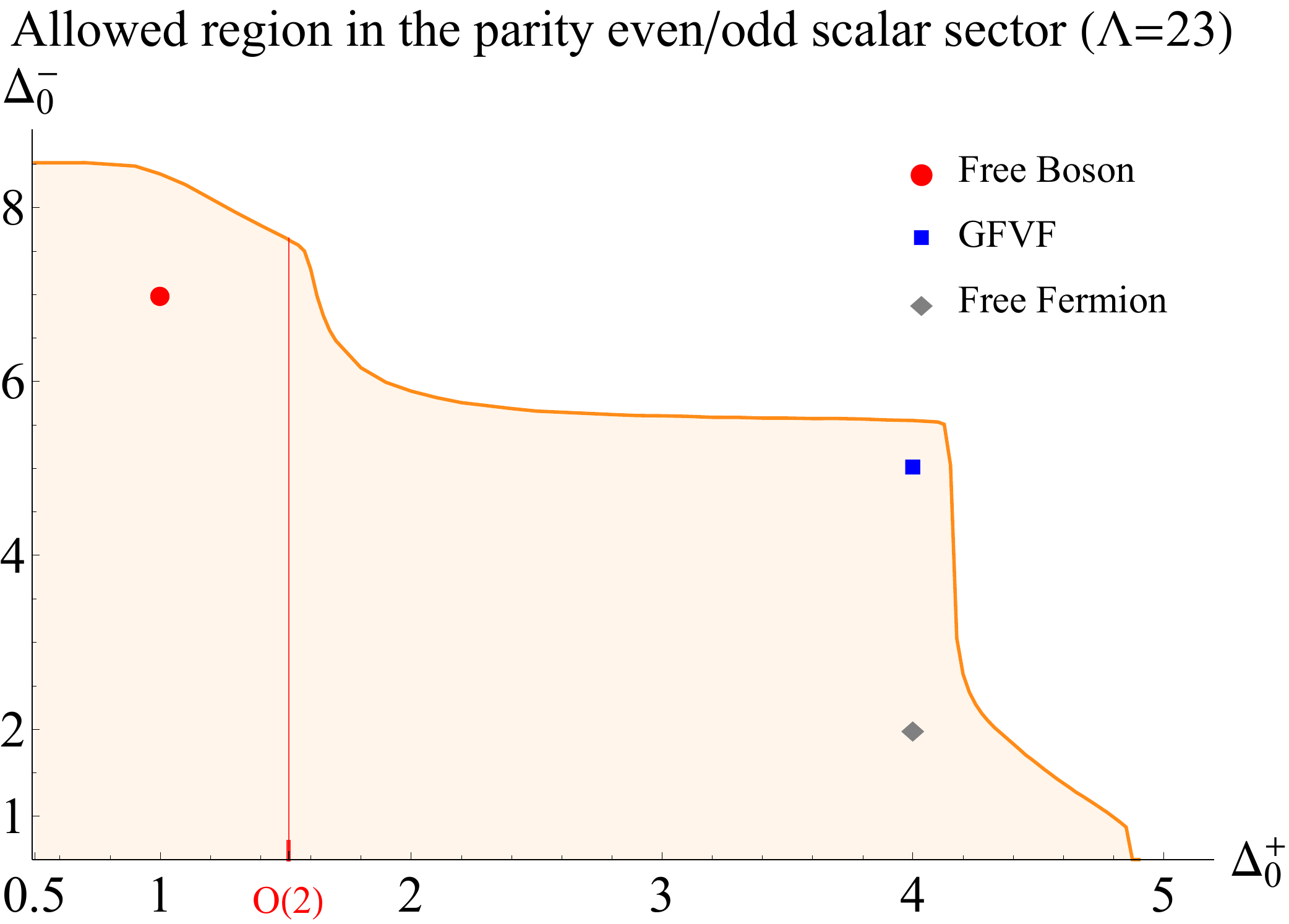}
\end{center}
\caption{Exclusion plot in the plane $(\Delta^+_{0},\Delta^-_{0})$ corresponding to 
the lightest parity even and parity odd scalars appearing in the OPE of two equal conserved currents. The shaded region is allowed.
\label{Fig:Even0Odd0}
}
\end{figure}

In figure \ref{Fig:Even0Odd0}, we show our result for the excluded region of the  plane $(\Delta_{0}^+,\Delta_{0}^-)$, where $\Delta_{\ell}^\pm$ denotes the scaling dimension of  
the lightest parity even/odd neutral spin-$\ell$ operator. This curve was calculated using up to $\Lambda=23$ derivatives of the crossing equations at the crossing symmetric point (451 components). The parameter $\Lambda$ is defined in eq.~\eqref{eq:functionaldef}.
In this plot, we represented several known theories to verify that they all fall inside the allowed region. On one hand, the theories of a free Dirac fermion and of a free complex scalar field lie well within the allowed region. On the other hand, the critical $O(2)$-model and the generalized free theory (GFVF) of a current seem to play an important role in determining the boundary of the allowed region.
Our results suggest that these theories sit at kinks of the optimal boundary corresponding to $\Lambda=\infty$.

The stress-energy tensor appears in the OPE of two currents,
\be
 J_\mu(x)J_\nu(0) = C_J\frac{\delta_{\mu\nu}-2\hat{x}_\mu\hat{x}_\nu}{x^4}   +
 \frac{3C_J}{32\pi|x|}\left[ t_{\mu\nu}^{\alpha\beta}(\hat{x})+12\gamma\, 
\tilde{t}_{\mu\nu}^{\alpha\beta}(\hat{x})\right] T_{\alpha\beta}(0) +\dots
\label{JJOPET}
\ee	
where $\hat{x}^\mu=\frac{x^\mu}{|x|}$ and the dots represent the contributions from all other operators besides the identity and the stress tensor $T_{\alpha\beta}$. 
There are two independent tensor structures
\footnote{Their explicit form is:
\begin{align}
\nonumber
t_{\mu\nu}^{\alpha\beta}(\hat{x})&=6 \hat{x}_{(\mu}\delta_{\nu)}^{\alpha}\hat{x}^\beta
+ 2\delta^\alpha_\mu\delta^\beta_\nu+3\hat{x}_\mu\hat{x}_\nu\hat{x}^\alpha\hat{x}^\beta-5\delta_{\mu\nu}\hat{x}^\alpha\hat{x}^\beta \ ,
\\
\nonumber
\tilde{t}_{\mu\nu}^{\alpha\beta}(\hat{x})&=2\hat{x}_{(\mu}\delta_{\nu)}^{\alpha}\hat{x}^\beta
- 2\delta^\alpha_\mu\delta^\beta_\nu-3\hat{x}_\mu\hat{x}_\nu\hat{x}^\alpha\hat{x}^\beta-3 \delta_{\mu\nu}\hat{x}^\alpha\hat{x}^\beta \ .
\end{align}
}
compatible with conservation and permutation symmetry. The conformal Ward identities relate the overall coefficient to   the OPE coefficient of the identity operator ($C_J$) but the relative coefficient  $\gamma$ is an independent  parameter that characterises the CFT. In particular, it controls the high frequency/low temperature behaviour of the conductivity \cite{Katz:2014rla}.
In the holographic context, $\gamma$ corresponds to a higher derivative coupling between two photons and a graviton  in the bulk. In particular, $\gamma$ vanishes for Einstein-Maxwell theory.
The conformal collider analysis of \cite{arXiv:0803.1467}  gives rise to the bounds $-1\le 12\gamma \le 1$ (see also \cite{Buchel:2009sk,arXiv:1010.0443, arXiv:1210.5247}). This bound was recently proven using only unitarity and convergence of the OPE expansion 
\cite{Hofman:2016awc} (also see \cite{Faulkner:2016mzt} for an alternative approach).
The bound is saturated by free complex bosons ($12\gamma=-1 $) and free fermions ($12\gamma=1 $).

\begin{figure}[h]
 \begin{center}
 \includegraphics[width=0.6\textwidth]{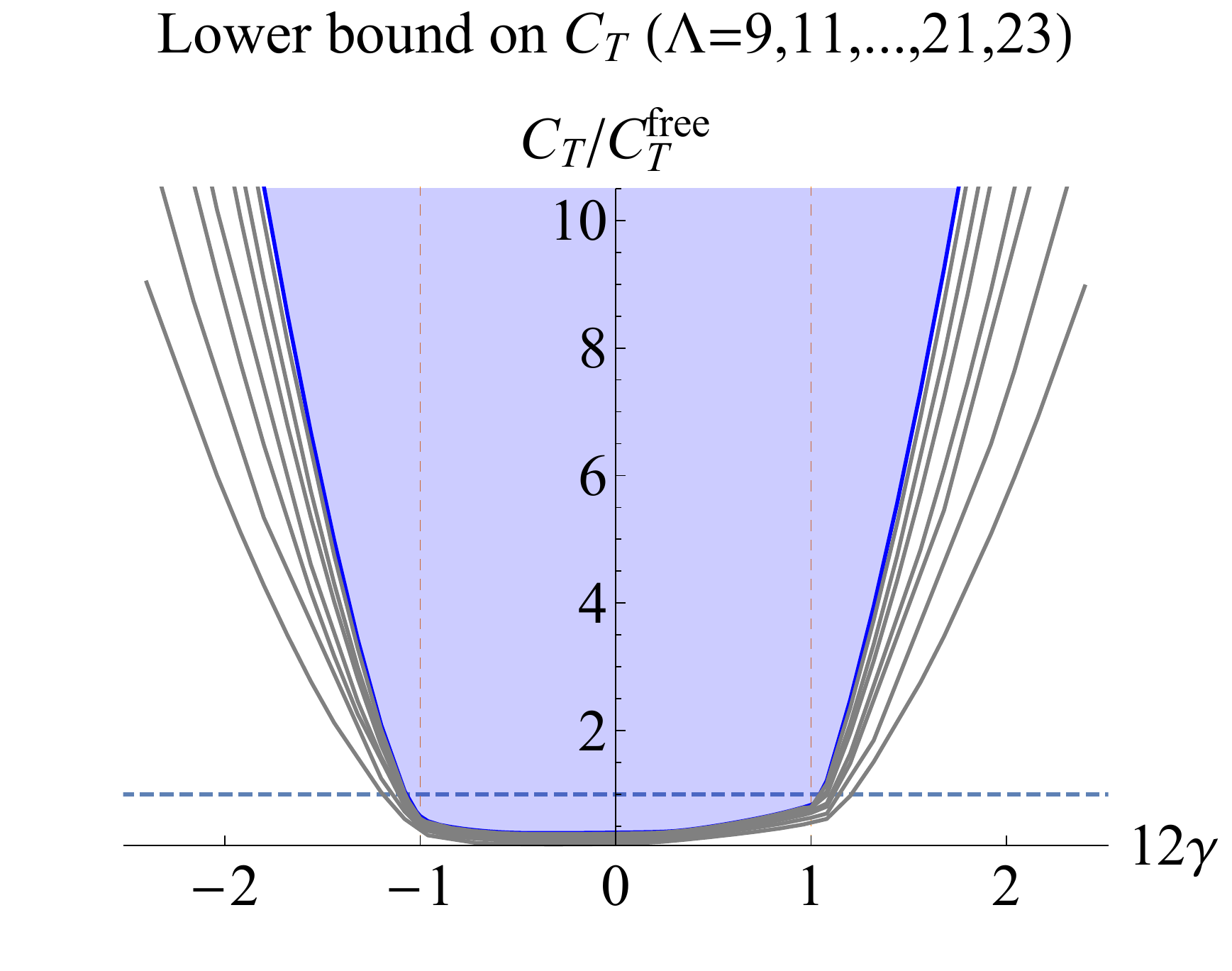}
\end{center}
\caption{Lower bound on the central charge normalized to the one of a free complex scalar as a function of $\gamma$. The vertical dashed lines corresponds to the conformal collider bound $-1\leq 12\gamma\leq 1$.
We impose that the first spin-2 operator after $T_{\mu\nu}$ has dimension larger than 3.5 (see section \ref{sec:results} for explanation).
\label{Fig:CTvsGAMMA}}
\end{figure}

In figure \ref{Fig:CTvsGAMMA}, we plot the minimal value of the central charge $C_T$  as a function of $\gamma$ and for several  values of $\Lambda$ (number of derivatives of the crossing equations imposed).
 In dashed lines we plot the conformal collider bounds and the value of the central charge $C_T$ of the minimal theories that saturate them: a free complex scalar and a free Dirac fermion.
It is encouraging to notice that the lower bound on $C_T$ grows very rapidly outside the region $-1\le 12\gamma \le 1$. We suspect it diverges when $\Lambda \to \infty$. On the other hand, for  $-1< 12\gamma < 1$ the lower bound on $C_T$ seems to be converging to a finite curve as we increase $\Lambda$. 

Figures \ref{Fig:Even0Odd0} and \ref{Fig:CTvsGAMMA} are just an appetiser for the results presented in section \ref{sec:results}.
To facilitate the interpretation of our results we listed in appendix \ref{sec:reminder} some known 3D CFTs with a continuous global symmetry.
 In section \ref{sec:setup}, we summarize the steps involved in setting up the numerical conformal bootstrap approach to the four point function of a conserved current, leaving many details to  appendices  \ref{sec:ind3pt}, \ref{sec:4ptbasis}, \ref{ap:ftilde}, \ref{app:ConformalBlocks} and \ref{sec:numdetails}.
Finally, we conclude in section \ref{sec:discussion} with a discussion of future work.
\section{Setup}
\label{sec:setup}

In this section we define our notation for three and four point correlation functions of spin 1 currents.  We will often work in general spacetime dimension $d$ and specialize to $d=3$ at the end. Through this section we will work in embedding space, see \cite{SpinningCC} for a detailed review.

In the embedding formalism each operator $\mathcal O_{\Delta,\ell}$ is associated to a field $\Phi_{\Delta,\ell}(P,Z)$, polynomial in the $(d+2)$-dimensional polarization vector $Z$, such that
\beq
\Phi(\lambda P; \alpha Z + \beta P ) = \lambda^{-\Delta} \alpha^{\ell} \Phi(P;Z)\,.
\label{eq:EmbField}
\eeq
We fix the normalization of the operators such that:
\beq
\langle\Phi_{\Delta,\ell}(P_1,Z_1)\Phi_{\Delta,\ell}(P_2,Z_2) \rangle = \frac{H_{12}^\ell}{(P_{12})^\Delta} \, ,
\label{eq:EmbField}
\eeq
where $P_{ij}\equiv -2 P_i \cdot P_j$.
The quantity $H_{12}$ entering the above equation, together with $V_{i,jk}$ are the building blocks needed to construct higher point correlation functions. They are defined as:
\begin{align}
H_{ij} &\equiv \frac{(Z_i \cdot Z_j )( P_i\cdot P_j) - (Z_i \cdot P_j )( Z_j\cdot P_i)}{(P_i \cdot P_j )}\,, \nonumber
\\
V_{i,jk} &\equiv \frac{(Z_i \cdot P_j )(P_i\cdot P_k )- (Z_i \cdot P_k )( P_i\cdot P_j)}{\sqrt{-2 (P_i \cdot P_j)(P_i \cdot P_k)(P_j \cdot P_k)}}\,.
\label{eq:HV} 
\end{align}

\subsection{Three point functions $\langle J J \mathcal O_{\Delta,\ell}^\pm\rangle$}
\label{sec:3pt}
In order to decompose the four-point function $\langle JJJJ\rangle$ in conformal blocks, we need to understand the structure of the OPE of two currents $J\times J$.
This is equivalent to classifying all the conformal invariant three-point functions 
$\langle J J \mathcal O_{\Delta,\ell}^\pm\rangle$. 
Since we are assuming the CFT is parity preserving, the three-point functions 
$\langle J J \mathcal O_{\Delta,\ell}^+\rangle$ will not involve the $\epsilon$-tensor while the three-point functions 
$\langle J J \mathcal O_{\Delta,\ell}^-\rangle$ will do.

Let us start by writing the most general form of the three-point function between two equal vector operators (of dimension $d-1$) and a parity even operator,
\begin{align}
&\langle J(P_1,Z_1) J(P_2,Z_2) \mathcal O_{\Delta,\ell}^+(P_3,Z_3)\rangle \
(P_{12})^{d-1-\frac{\Delta}{2}} 
(P_{13} )^\frac{\Delta}{2}
(P_{23})^\frac{\Delta}{2}=
\nonumber
  \\&= \left\{
  \begin{array}{ll}
\lambda_{JJ\mathcal O^+}^{(1)} V_1 V_2 V^\ell_3 +\lambda_{JJ\mathcal O^+}^{(2)} H_{12} V^\ell_3 & \text{if } \ell = 0\\
\\[-0.8em]
  \lambda_{JJ\mathcal O^+}^{(3)} ( H_{13}V_2   - H_{23}V_1 ) V^{\ell-1}_3  & \text{if } \ell \geq 1,\text{odd} \\
  \\[-0.8em]
  \lambda_{JJ\mathcal O^+}^{(1)}  V_1 V_2 V^\ell_3 
  +\lambda_{JJ\mathcal O^+}^{(2)}  H_{12}V^\ell_3 +&\\
   \qquad \qquad
  +\lambda_{JJ\mathcal O^+}^{(3)}  (H_{13}V_2  + H_{23}V_1 )V^{\ell-1}_3
  +\lambda_{JJ\mathcal O^+}^{(4)}   H_{13}H_{23}V^{\ell-2}_3   & \text{if } \ell \geq 2,\text{even} \\
 \end{array}
 \right.
  \label{eq:3pstructuresBose}
\end{align}
where $\lambda_{JJ\mathcal O^+}^{(i)}$ are undetermined constants and we used the notation
\be
V_1=V_{1,23}\ ,\qquad
V_2=V_{2,31}\ , \qquad
V_3=V_{3,12}\ .
\ee
In this expression, we only imposed conformal and permutation symmetry.  
To impose conservation of $J(P_1,Z_1)$, it is enough to demand that the embedding space differential operator $ \frac{\partial}{\partial P_1 } \cdot \frac{\partial}{\partial Z_1 }$ annihilates the three-point function.
This further reduces the number of independent constants.
In the case of a scalar operator, conservation implies $\lambda_{JJ\mathcal O^+}^{(2)} = \frac{\Delta -d+1}{\Delta} \lambda_{JJ\mathcal O^+}^{(1)}$ leaving only one free constant.
In the case of odd spin, conservation implies $\lambda_{JJ\mathcal O^+}^{(3)}=0$, which means that a parity even, spin odd operator cannot appear in the OPE $J\times J$.
Finally, in the case of even spin $\ell\ge 2$ we find 
\begin{align*}
\lambda_{JJ\mathcal O^+}^{(4)} &= \frac{ (d-\Delta -2) (d-\Delta
   -1)\lambda_{JJ\mathcal O^+}^{(1)}+ \left(\Delta  (d-\Delta
   -2)+\ell (3 d-2 \Delta
   -4)+\ell^2\right) \lambda_{JJ\mathcal O^+}^{(2)}}{(2 d-\Delta +\ell-4)
   (2 d-\Delta +\ell-2)}\,, \\
 \lambda_{JJ\mathcal O^+}^{(3)} &= 
 \frac{  (d-\Delta -1)\lambda_{JJ\mathcal O^+}^{(1)}+  (\Delta
   +\ell)\lambda_{JJ\mathcal O^+}^{(2)}}{2 d-\Delta
   +\ell-2} \,,
\end{align*}
which reduces the number of independent structures down to 2.

Let us now turn our attention to the three point function of two conserved currents and one parity odd operator $\mathcal O^-_{\Delta,\ell}$. In $d\leq4$ one can use the $\epsilon$-tensor to make parity odd conformally invariant three point functions. Indeed, any  parity odd structure can be obtained by multiplying parity even structures  by  $\epsilon$-tensors.
In $d=3$,  there are three parity odd building blocks:
\beq
\epsilon_{ij}\equiv   \epsilon(Z_i,Z_j,P_1,P_2,P_3)\qquad(i,j=1,2,3)\,.
\eeq
Conformal invariance and permutation symmetry restricts the tensor structures to be:\footnote{As explained in appendix \ref{sec:ind3pt}, the structure $\epsilon_{12}V_3^l$ is not independent of the ones we used for $\ell\ge 2$. Structures involving an $\epsilon$ tensor contracted with three polarization vectors can also be expressed in terms of $\epsilon_{ij}$.}
\begin{align}
&\langle J(P_1,Z_1) J(P_2,Z_2) \mathcal O_{\Delta,\ell}^-(P_3,Z_3)\rangle \
(P_{12})^{d-\frac{\Delta+1}{2}} 
(P_{13} )^\frac{\Delta+1}{2}
(P_{23})^\frac{\Delta+1}{2}=
\nonumber \\&= \left\{\begin{array}{ll}
\lambda_{JJ\mathcal O^-}^{(1)} \epsilon_{12} V^\ell_3 & \text{if } \ell = 0\\
\\[-0.8em]
  \lambda_{JJ\mathcal O^-}^{(2)}( \epsilon_{13} V_{2} +\epsilon_{23} V_{1} )   V_3^{\ell-1} & \text{if } \ell =1  \\
  \\[-0.8em]
\lambda_{JJ\mathcal O^-}^{(2)}( \epsilon_{13} V_{2} -\epsilon_{23} V_{1} )V_3^{\ell-1} +
\lambda_{JJ\mathcal O^-}^{(3)}( \epsilon_{13} H_{23} -\epsilon_{23} H_{13} )V_3^{\ell-2} \qquad
    & \text{if } \ell \geq 2,\text{even} \\
    \\[-0.8em]
\lambda_{JJ\mathcal O^-}^{(2)}( \epsilon_{13} V_{2} +\epsilon_{23} V_{1} )V_3^{\ell-1} +
\lambda_{JJ\mathcal O^-}^{(3)}( \epsilon_{13} H_{23} +\epsilon_{23} H_{13} )V_3^{\ell-2} \qquad
    & \text{if } \ell \geq 3,\text{odd} \\
 \end{array}\right.   \label{eq:3pstructuresOdd}
\end{align}
where $\lambda_{JJ\mathcal O^-}^{(i)}$ are undetermined constants. Current conservation then fixes
\be
\lambda_{JJ\mathcal O^-}^{(3)}=\frac{    \Delta-2-(-1)^\ell }{\Delta   -\ell-3}\lambda_{JJ\mathcal O^-}^{(2)}\,,
\ee
for $\ell \ge 2$. In the case $\ell=0$, conservation is automatic and $\lambda_{JJ\mathcal O^-}^{(1)}$ is a free parameter. In the case $\ell=1$, conservation implies that $\lambda_{JJ\mathcal O^-}^{(2)}=0$. In other words, no spin 1 operator that can appear in the OPE of two equal currents.

In summary, the number of independent constants in the three-point function 
$\langle J J \mathcal O_{\Delta,\ell}^\pm\rangle $ is given in the following table:
\begin{center}
  \begin{tabular}{| c | c | c |  c |}
\cline{3-4}
      \multicolumn{2}{c|}{ }  & \multicolumn{2}{|c|}{Parity} \\ \hline
    Spin& Dimension & $\ \ \ +\ \ \ $ & $\ \ \ -\ \ \ $ \\ \hline
 $ \ell=0$ &    $\Delta\ge 1/2  $ & {\color{blue}1} & {\color{blue}1} \\ \hline
  $ \ell=1$ &   $\Delta\ge2$ & {\color{blue}0} & {\color{blue}0} \\ \hline 
   $ \ell\ge2\text{ even}$ &          $\Delta\ge \ell+1$ & {\color{blue}2} & {\color{blue}1} \\ \hline
$ \ell\ge3\text{ odd}$ &  $\Delta\ge \ell+1$ &  {\color{blue}0} & {\color{blue}1} \\ \hline
  \end{tabular}
\end{center}

Finally, let us comment on the special cases when $\mathcal{O}$ saturates the unitarity bound.
For $\ell \ge 1$ this happens 
 when $\mathcal{O}$ is a conserved current with $\Delta=\ell+d-2$.
  It is easy to check that the three-point function (\ref{eq:3pstructuresBose}) of $\mathcal{O}^+$ are automatically conserved at $P_3$ if we set  $\Delta=\ell+d-2$.
On the other hand, conservation of $\mathcal{O}^-(P_3)$ implies that the three-point functions (\ref{eq:3pstructuresOdd}) vanish for $\ell>2$. For $\ell=2$ conservation follows from $\Delta=3$. This is consistent with the fact that it is possible to couple the currents to the stress tensor with a parity odd three-point function in theories that violate parity.
For $\ell=0$, one should impose $\partial^2\mathcal {O}=0$ when $\Delta=\frac{1}{2}$. This implies that the three-point function $\langle JJ\mathcal{O}\rangle$ must vanish for both $\pm$ parity.

\subsubsection{Special case: $\langle J J T_{\mu\nu}\rangle$}
\label{sec:JJT}

Let us study in detail the three point function of two identical conserved currents and the energy momentum tensor. As discussed in the previous section there are only two independent structures. 
The three-point function is given by \eqref{eq:3pstructuresBose} with $\ell=2$, $\Delta=d$ and
\begin{align}
\lambda_{JJ T}^{(1)}&= \left(2-3d-4  d^3 \gamma \right)C\,,\qquad
 &\lambda_{JJ T}^{(2)}= (1-2 d -4 d^2 \gamma )C\,, 
 \\
   \lambda_{JJ T}^{(3)}&=-2 d(1+4 \gamma ) C \,,\qquad
   &\lambda_{JJ T}^{(4)}=2 
   \left(\frac{1}{d-2}-4 \gamma d\right)C\,,
   \label{alphaiJJT}
\end{align}
where 
\be
C=\frac{d(d-2)}{2(d-1)^2S_d}C_J\,,
\ee
is related via the conformal Ward identity to the current two-point function
\be
\langle J(P_1,Z_1)J(P_2,Z_2)\rangle=C_J
\frac{H_{12}}{(P_{12})^{d-1}}\ .
\ee 
The symbol $S_d=2\pi^{\frac{d}{2}}/\Gamma(\frac{d}{2})$ is the volume of a $(d-1)$-dimensional sphere and $\gamma$ is an independent parameter that appears in the OPE \eqref{JJOPET}. The parameter $\gamma$  controls the anisotropy of the energy correlator of a state created by the current \cite{arXiv:0803.1467, arXiv:1010.0443, arXiv:1210.5247}.
Positivity of this energy correlator implies the bounds
\beq
\label{eq:HMbound}
-\frac{1}{4d}\le \gamma \le \frac{1}{4d(d-2)}\ ,
\eeq
which are saturated by free scalars and free fermions, respectively.
This bound was recently proven relying only on unitarity and OPE convergence \cite{Hofman:2016awc}.
The parameter $\gamma$ also has a nice physical meaning from the perspective of the dual AdS description. The current three-point function can be computed from the bulk action
\beq
S_{AdS}=C_J \int d^{d+1}x\left[
-\frac{1}{4}F_{\mu\nu}F^{\mu\nu}
+\gamma L^2 W^{\mu\nu\tau\rho}F_{\mu\nu}F_{\tau\rho}
\right]
\eeq
where $L$ is the AdS radius, $W$ is the Weyl tensor and $F$ is the field strength of the bulk gauge field dual to the current. In this form, it is clear that $\gamma$ does not contribute to the   two-point function of the current in the vacuum.

In the conformal bootstrap analysis of the four-point function $\langle JJJJ\rangle$ we normalize all operators to have unit two-point function. Recall that the stress tensor has a natural normalization due to the Ward identities,
\be
\langle T(P_1,Z_1)T(P_2,Z_2)\rangle=C_T
\frac{H_{12}^2}{(P_{12})^{d}}\ .
\ee 
That means that we should multiply the OPE coefficients \eqref{alphaiJJT} by 
\beq
\frac{1}{C_J \sqrt{C_T}}\ .
\label{eq:alphaiJJTrescaling}
\eeq
This shows that $C_J$ is not accessible in the bootstrap analysis of $\langle JJJJ\rangle$. On the other hand, $C_T$ does affect the OPE coefficients of normalized operators. For comparison, we recall the values of $C_T$ for free theories \cite{hep-th/9307010}. Each real scalar field contributes
\beq
C_T=\frac{1}{S_d^2} \frac{d}{d-1}\ .
\eeq
Each Dirac field contributes
\beq
C_T=\frac{1}{S_d^2} \frac{d}{2} 2^{\left[\frac{d}{2}\right]}\ ,
\eeq
where $2^{\left[\frac{d}{2}\right]}$ is the dimension of the Dirac $\gamma$-matrices in $d$ spacetime dimensions.
Notice that in $d=3$, a complex scalar contributes the same as a Dirac fermion
\beq
\label{eq:ctfree}
C^{ \rm free}_T=\frac{3}{S_3^2}=\frac{3}{16\pi^2}  \ .
\eeq
This is the minimal matter content of free theories with a $U(1)$ global symmetry.

\subsection{Four point function $\langle J J J J\rangle$}
\label{sec:4pt}

The general structure of the four point function is\footnote{The factor of $v^{1-d}$ is convenient  to make the crossing equations simpler.}
\bea
\langle J(P_1,Z_1) \dots J(P_4,Z_4)\rangle = 
\frac{v^{1-d}  }
  {(P_{12}\, P_{34})^{d-1}}
  \sum_s  f_s(u,v) \,Q_s (\{ P_i;Z_i\})\,,
  \label{eq:4pJJJJ}
\eea
where 
\be
u=\frac{P_{12} P_{34}}{P_{13} P_{24}}\,,\qquad \qquad 
v= \frac{P_{14} P_{23}}{P_{13} P_{24}}\,,
\ee
are the usual conformal invariant cross ratios and $Q_s$ encode tensor structures in the embedding formalism. 
In Table~\ref{4pstruct} we list all the parity even structures $Q_s$ contributing to the four point function. In general dimension they are 43. 
As explained below, when $d=3$ they reduce to  41. 
In addition, since we are considering equal conserved currents, there are two permutations which leave unchanged the conformal invariants $u,v$: $1234\to2143$ and $1234\to3412$. 
The action of these permutations simply sends one structure into another. The final effect is to reduce the number of independent functions $f_s(u,v)$ that appear in \eqref{eq:4pJJJJ} to 19 (17 for $d=3$). The transformation properties of each tensor structure, together with a list of the independent ones is reported in Table~\ref{4pstruct}.

\subsubsection{Crossing Symmetry}

The crossing symmetry $1234\to2134$ sends the cross ratios $(u,v)$ into $\left(\frac{u}{v},\frac{1}{v}\right)$. As usual in the conformal bootstrap analysis,  
this crossing symmetry follows automatically from the conformal block expansion in the $(12)(34)$ channel associated to the three-point functions studied in section \ref{sec:3pt}.

On the other hand, the crossing symmetry $1234\to3214$ is not satisfied by the conformal blocks in the $(12)(34)$ channel and gives rise to non-trivial constraints on the operator spectrum and OPE coefficients. The crossing symmetry $1234\to3214$ leads to
\footnote{These equations are derived for $\sqrt{u} +\sqrt{v} \neq 1$ where all the 43 tensor structures are linearly independent. 
By continuity, the equations also hold for any $u$ and $v$. This is indeed the case for the free theory examples discussed in appendix \ref{sec:reminder}.}
\beq
f_s(u,v)=\sum_{s' } \CC_{ss'} \ f_{s'}(v,u)\,,
\eeq
where the matrix $\CC$ is a permutation, which can be decomposed as follows
\beq
\CC \equiv  
\Mat^{-1} 
\
\Parity
\
\Mat
\label{eq:Pmatrix}
\eeq
where $\Parity$ is a diagonal matrix with diagonal entries equal to $\pm 1$. 
This leads to a simpler form of the crossing equations.
Introducing new functions $\tilde{f}_s(u,v)=\sum_{s' }\Mat_{ss'}f_{s'}(u,v)$  (see 
 appendix \ref{ap:ftilde} for the precise definitions),
the crossing equations simplify to
\begin{align}
\tilde{f}_s(u,v)&=- \tilde{f}_s(v,u)\ , 
&s&=1,2,\dots,7 \,\text{and} \, s=19\ ,\nonumber\\
\tilde{f}_s(u,v)&= \tilde{f}_s(v,u)\ , &s&=8,9,\dots,18\ .
\label{eq:crossingftilde}
\end{align}
In other words, we have 8 odd and 11 even functions under the crossing symmetry $u \leftrightarrow v$.
We will see that the functions $\tilde{f}_{18}$ and  $\tilde{f}_{19}$ will disappear in 3 dimensions, hence the choice to put them at the end of the list.\

\subsubsection{Conservation}
\label{sec:conservation}

In the numerical conformal bootstrap approach
one writes the four point function as a sum of conformal blocks
and imposes (a truncated version) of the 19 crossing equations \eqref{eq:crossingftilde}. Fortunately, we can use  conservation of the external currents to reduce this large number of crossing equations.
Imposing conservation directly on the four point function produces a set of differential constraints that the functions $\tilde{f}_s(u,v)$ must satisfy. The four point function of three vectors  and one scalar operator contains 14 independent tensor structures (in any dimension). As a consequence, each conservation condition will produce 14 first order differential equations of the form
\begin{align}
&\sum_{s=1}^{19}
\left[[K(u,v)]_{is} \tilde{f}_{s}(u,v) + [K^{u}(u,v)]_{is} \partial_u \tilde{f}_{s}(u,v)  + [K^{v}(u,v)]_{is} \partial_v \tilde{f}_{s}(u,v)\right] = 0\,,
 \label{eq:cons}
\end{align}
where $ i=1,...,14 $.
The first important observation is that the conformal block decomposition\footnote{See for instance \eqref{eq:CBexpansion} in the next section.}  automatically satisfies these equations.\footnote{In fact, we used this to cross check the computation of the conformal blocks.}
The second observation is that the equations \eqref{eq:cons} are crossing symmetric. 
In other words, applying the crossing symmetry $u \leftrightarrow v$ to \eqref{eq:cons} and using \eqref{eq:crossingftilde} we obtain an equivalent set of differential equations.
This means that if we use these differential equations to determine the functions $\tilde{f}_{s}$ evolving from a crossing symmetric "initial condition", then crossing symmetry is guaranteed everywhere. Therefore, if we start from a conformal block decomposition, it is sufficient to impose crossing symmetry on a minimal set of data about the functions $\tilde{f}_s$ that determines these functions everywhere via the differential equations (\ref{eq:cons}).

\begin{figure}[t!]
\graphicspath{{Fig/}}
\def\svgwidth{6 cm}
\centering
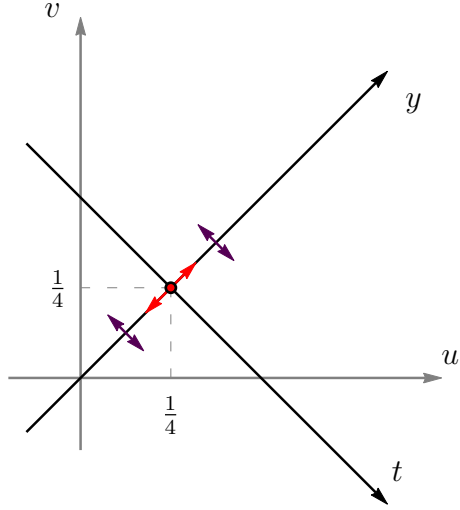 
\caption{
\label{fig:Conservation} Change of coordinates from the usual cross ratio $u$ and $v$ to the new ones $y=u+v-1/2$ and $t=u-v$. The conservation equation can be used to evolve 12 functions (6 even and 6 odd under the action of $\Parity$) from the line $t=0$ to the full plane. We can further evolve 2 functions (1 even and 1 odd) from the point $(t,y)=(0,0)$, to the line $t=0$.}
\end{figure}

To make this idea more precise it is convenient to introduce new coordinates
\beq
t=u-v\ ,\qquad\qquad
y=u+v-\frac{1}{2}\ ,
\eeq
which are represented in figure \ref{fig:Conservation}.
We will think of the  $t$ as time and $y$ as space.  Crossing symmetry \eqref{eq:crossingftilde} means that   8 functions $\tilde{f}$ are odd  under time-reversal, while the remaining 11 functions are even. The conservation equations (\ref{eq:cons})
become the following first order time evolution equations
\begin{align}
&\sum_{s=1}^{19}
\left[[K]_{is}  \tilde{f}_{s} + [K^{y}]_{is} \partial_y \tilde{f}_{s}   + [K^{t}]_{is}   \partial_t \tilde{f}_{s}  \right] = 0,\qquad
\qquad i=1,\dots,14\ ,
\label{eq:conservation}
\end{align}
where $[K^t]=[K^u]-[K^v]$ and $[K^y]=[K^u]+[K^v]$ are $14 \times 19$ matrices.

One can check that the matrix $[K^t]$ has rank 12.\footnote{In fact, this is true for a generic choice of time coordinate around the point $u=v=1/4$. The exception being the coordinate $y$. In this special case, the rank of $[K^y]$ is 10.} 
That means that we can evolve 12 functions $\tilde{f}_s$ starting from an initial time slice, which we choose to be $t=0$. 
Since the functions $\tilde{f}_s$ are either even or odd under $t\rightarrow -t$, crossing symmetric boundary conditions are obtained by simply imposing the odd ones to vanish on the line $t=0$, while the even ones are left unconstrained. 
One can explicitly check that the (7 dimensional) Kernel of $[K^t]$ decomposes in two orthogonal subspaces (of dimension 5 and 2) associated to the eigenvalues $\pm 1$ of the crossing symmetry matrix  $\Parity$ defined in \eqref{eq:Pmatrix}.
This means that we can   evolve $8-2=6$ odd functions and $11-5=6$ even functions.  One  possible choice is  $\tilde{f}_1,..., \tilde{f}_6$ and $\tilde{f}_8, \tilde{f}_9,\tilde{f}_{10},\tilde{f}_{11},\tilde{f}_{12},\tilde{f}_{14}$.   Hence, by using 12 out of the 14 conservation equations we reduced to the set of crossing symmetry conditions:
\begin{align}
\tilde{f}_s(u,u) &= 0 &  \text{for } s=&1,2,3,4,5,6 \nonumber \\
\tilde{f}_s(u,v) &= - \tilde{f}_s(v,u) &  \text{for } s=&7,19 \\
\tilde{f}_s(u,v) &=   \tilde{f}_s(v,u) &  \text{for } s=&13,15,16,17,18  \nonumber
\end{align}
Note that the boundary condition on the line doesn't constrain the even functions: any initial condition $\tilde{f}_s$, $s=8, 9, 10, 11, 12, 14$ will be automatically evolved into a crossing symmetric function.

In fact, this is still not the minimal set of data where we can impose crossing symmetry. We will use the two remaining conservation equations to reduce further the set of crossing symmetry equations. 
The remaining conservation equations are not evolution equations. They are two constraint equations on the initial data at $t=0$.  One can check that at $t=0$, the first constraint equation only involves odd functions and the second only involves even functions.
More precisely, the first constraint equation can be written as
\be
\partial_y\tilde{f}_3(0,y) =  
\sum_{s\neq 3} A_s(y) \partial_y\tilde{f}_s(0,y)+
\sum_{s} B_s(y)  \tilde{f}_s(0,y) \,,
\ee
where the sum runs over the odd functions ($s=1,2,\dots,7$ and $s=19$) and the coefficients $A_s(y)$ and $B_s(y)$ are regular at the crossing symmetric point $y=0$.
This means that it is sufficient to impose $\tilde{f}_3(t=0,y=0)=0$ because this equation will ensure that $\tilde{f}_3(t=0,y)=0$ for any $y$. Since the second constraint equation only involves even functions, which are unconstrained at the initial surface $t=0$ it is not useful to further reduce the crossing symmetry constraints.
 
In the end the minimal set of crossing symmetry conditions is:
\begin{align}
&\tilde{f}_3(1/4,1/4) = 0 && \\ 
&\tilde{f}_s(u,u) = 0  \qquad &&\text{for } s=1,2,4,5,6 \\
&\tilde{f}_s(u,v) = - \tilde{f}_s(v,u) \qquad &&\text{for } s=7,19 \\
&\tilde{f}_s(u,v) =   \tilde{f}_s(v,u) \qquad  &&\text{for } s=13,15,16,17,18
\end{align}
where we went back to the original coordinates $u$ and $v$.  In agreement with \cite{Dymarsky:2013wla} in general $d$ there are 7 equations in the ``bulk"; additionally there are five constraints on the line and one at a crossing-symmetric point. 
We remark that our analysis of the conservation equations is valid only in a local neighbourhood of the crossing symmetric point $u=v=1/4$. However, this is sufficient for the numerical bootstrap algorithm where we only consider a finite number of derivatives of the crossing equations at  $u=v=1/4$.

\subsubsection{Three dimensions}

In three dimensions not all 43 tensor structures of the four point function are linearly independent. The easiest way to see this is to consider the embedding space tensor
\be
W_{ij}^{A_1\dots A_6}= Z_i^{[A_1} Z_j^{A_2} P_1^{A_3}P_2^{A_4}P_3^{A_5}P_4^{A_6]}\,,
\ee
which vanishes identically in $\mathbb{R}^{d+1,1}$ for $d=3$.
On the other hand, for any $d$, the contraction $W_{(12)(34)}=\eta_{A_1B_1}\dots \eta_{A_6B_6}W_{12}^{A_1\dots A_6}W_{34}^{B_1\dots B_6}$ can be written as a linear combination of the 43 tensor structures $Q_s$ that form a basis for four-point functions of vector primary operators. Therefore, in $d=3$ this gives rise to a linear relation between the 43 tensor structures $Q_s$. Using the 3 invariants  $W_{(12)(34)}$, $W_{(13)(24)}$ and $W_{(14)(23)}$ we obtain 2 independent relations between the structures $Q_s$ in $d=3$.\footnote{One can check the identity $W_{(12)(34)}+W_{(13)(24)}-W_{(14)(23)}=0$ for any $d$. } These constraints can be found in appendix \ref{sec:numdetails}. We use these to express the structures $Q_{31}$ and $Q_{40}$ in terms of the other $Q_s$. According to the definitions in appendix 
\ref{ap:ftilde}, this corresponds to the functions $\tilde{f}_{18}$ and  $\tilde{f}_{19}$.
 The entire argument about the conservation equations proceeds in the same way just dropping these two functions.

In the end the minimal set of crossing symmetry conditions in $d=3$  is as follows. It includes five equations in the ``bulk'' \cite{Dymarsky:2013wla}, five constraints on a line, and one at a point: 
\begin{align}
\label{eq:crossing3D}
\tilde{f}_7(u,v) = - \tilde{f}_7(v,u) & \\
\tilde{f}_s(u,v) =   \tilde{f}_s(v,u) &\qquad  \text{for } s=13,15,16,17 \\
\tilde{f}_s(u,u) = 0 & \qquad \text{for } s=1,2,4,5,6 \\
\tilde{f}_3(1/4,1/4) = 0 & 
\end{align}

\subsection{Conformal blocks} 
\label{sec:CBs}
\begin{figure}[t!]
\graphicspath{{Fig/}}
\def\svgwidth{8.5 cm}
\centering
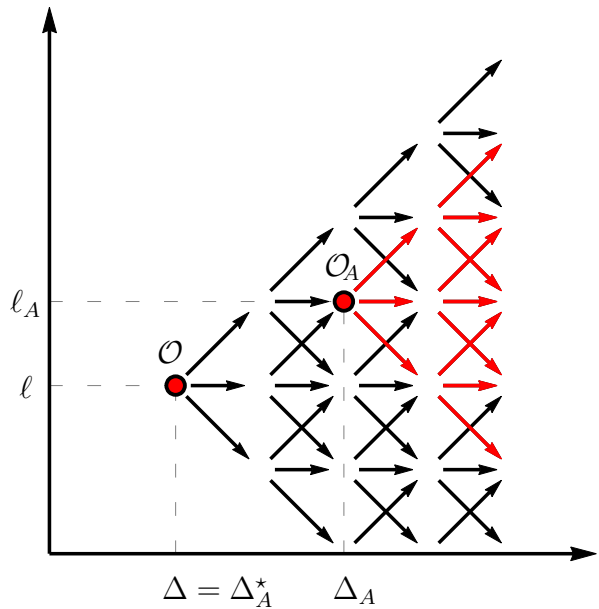 
\caption{ \label{fig:subrep} 
The picture represents the conformal multiplet $\Hcal_\Ocal$ of an operator $\Ocal$ with dimension $\D$ and spin $\ell$. The arrows represent the descendant operators and the horizontal and vertical axis are labeled respectively by the conformal dimensions and the spin.
When the primary operator $\Ocal$ has dimension $\D=\D_A^\star$, its descendant $\Ocal_A$ (with dimension $\D_A=\D^\star_A+n_A$ and spin $\ell_A$) becomes primary. The state  $\Ocal_A$ and all its descendants form a conformal multiplet of null states $\Hcal_{\Ocal_A}$.
}
\end{figure}

In this work we computed the conformal blocks (CB) for four external currents using the recurrence relation of \cite{Kos:2013tga, arXiv:1509.00428, Costa:2016xah}.
The existence of such recurrence relation comes from the study of the analytic properties of the CBs as functions of the conformal dimension $\D$ of the exchanged operator $\Ocal$. To see this, it is convenient to rewrite the CBs in radial quantization as follows
\be
\label{CB_radial_quantization}
\sum_{p,q} c_{12{\Ocal}}^{(p)} c_{34{\Ocal}}^{(q)}G_{\Ocal}^{(p,q)}=\sum_{ | \alpha \rangle \in \Hcal_{\Ocal}} \frac{\langle0| \Ocal_1 \Ocal_2 | \alpha \rangle \langle \alpha | \Ocal_3 \Ocal_4 |0 \rangle}{ \langle \alpha| \alpha \rangle} \ ,
\ee
where $\Hcal_{\Ocal}$ is the conformal multiplet associated to the primary operator $\Ocal$.
Tuning the conformal dimension $\D$ to some special values $\D_A^\star$, it happens that  one of the descendant $| O_A \rangle$ (with dimension $\D_A=\D_A^\star+n_A$ and spin $\ell_A$) becomes primary. Namely $K^\m | O_A \rangle=0$ where $K^\m$ is the generator of special conformal transformations.
When this happens, $| O_A \rangle$ becomes null, and so do all its descendants. Thus the representation $\Hcal_{\Ocal}$ becomes reducible and it contains an irreducible sub representation $\Hcal_{\Ocal_A}$ of null states as shown schematically in figure \ref{fig:subrep}. 

From formula (\ref{CB_radial_quantization}) it is clear that the conformal blocks $G_{\Ocal}^{(p,q)}$ have poles\footnote{In \cite{arXiv:1509.00428} it was shown that there can exist only simple  poles in odd dimensions. In even dimensions higher order poles can appear. However  the CBs for even dimensions can be obtained by analytic continuation from the odd dimensional case.} at $\D=\D_A^\star$  because of the contribution of all the null states in $\Hcal_{\Ocal_A}$. All these contributions together form a conformal block associated to the exchange of $\Ocal_A$.
Accordingly, the residue at the pole $\D=\D^\star_A$  is proportional to the conformal block $G_{\Ocal_A}$,
\be
G_{\Ocal }^{(p,q)} = \sum_{p',q'} \frac{(R_{A})_{p q p' q'}}{\D-\D^\star_A } \; G_{\Ocal_A}^{(p',q')} + O\left((\D-\D^\star_A)^0 \right) \ ,
\ee
where the $ (R_{A})_{p q p' q'}$ are coefficients which depend on the representation of the operator $\Ocal$.

The previous discussion explains the pole structure of the conformal blocks. 
To complete the recurrence relation we also need to obtain the asymptotic of conformal blocks when $\D\rightarrow \infty$.
To this end
it is convenient to write the conformal blocks in the basis of four-point function tensor structures as we did in (\ref{eq:4pJJJJ}),
\beq
\label{def:CBsExp}
G_{\Ocal }^{(p,q)}(\{ P_i;Z_i\}) = 
\frac{\left(\frac{P_{24}}{P_{14}} \right)^\frac{\D_{12}}{2}
\left(\frac{P_{14}}{P_{13}} \right)^\frac{\D_{34}}{2}}{(P_{12})^{\frac{\D_1+\D_2}{2}} 
(P_{34})^\frac{\D_3+\D_4}{2}} 
  \sum_s  g^{(p,q)}_{\Ocal ,s}(r,\eta) \,Q_s (\{ P_i;Z_i\})\, .
\eeq
Here $r$ and $\eta\equiv \cos \theta$ are the radial coordinates of \cite{Hogervorst:2013sma}, defined by
\be \label{radcoord}
re^{i\theta}=\frac{z}{(1+\sqrt{1-z})^2} \ , \qquad re^{-i\theta}=\frac{\bar z}{(1+\sqrt{1-\bar z})^2} \ ,
\ee
where $u= z \bar z$ and $v=(1-z)(1-\bar z)$.
The conformal blocks are not regular at $\D \rightarrow \infty$ because of the essential singularity $g_\Ocal(r,\eta) \propto (4 r)^\D$, however we can factor it out and define a new function $h_\Ocal$ which is well behaved
\be
h^{(p,q)}_{\Ocal,s}(r,\eta)  \equiv (4 r)^{-\D} g^{(p,q)}_{\Ocal,s }(r,\eta) \underset{\D \rightarrow \infty}{\longrightarrow} h^{(p,q)}_{\infty \Ocal,s}(r,\eta) \ .
\ee

So far the discussion was schematic and valid for any conformal block. We now want to give more details for the case of four external vectors in three dimensions. 
We shall construct the conformal blocks for generic external vector operators and only at the end we will specialize to the particular case of equal conserved currents.
The goal is to find the conformal blocks
\begin{align}
&h^{(p,q)}_{\D \ell +,s}(r,\eta) \qquad p,q=1\dots 5 \ , \\
&h^{(\pp, \qq)}_{\D \ell -,s}(r,\eta)  \qquad \pp,\qq=1\dots 4 \ ,
\end{align}
where\footnote{The actual independent structures are $41$, but we find it more convenient to work in the $43-$dimensional space and project out the final result into the $41-$dimensional space.} $s=1,\dots 43$. 
We obtain a set of recurrence relations for the conformal blocks which are diagonal in the label $s$ but which couple the labels $p,q$, 
\begin{align}
h^{(p,q)}_{\D \ell +,s}(r,\eta)&=h^{(p,q)}_{\infty l +,s}(r,\eta)+\sum_{p\rq{},q\rq{}=1}^5\sum_{A \in \Acal_+} \frac{(R_{+A})_{p q,p\rq{} q\rq{}} \; (4 \, r)^{n_A}}{\D-\D^\star_A} h^{(p\rq{},q\rq{})}_{\D_A \ell_A +,s}(r,\eta) \nonumber \\
& \ \ \ \ \ \ \ \ \ \ \ \ \ \ \ \ \ \ \ +\sum_{\pp\rq{},\qq\rq{}=1}^4 \sum_{A \in \Acal_-} \frac{(R_{+A})_{p q,\pp\rq{} \qq\rq{}} \; (4 \, r)^{n_A}}{\D-\D^\star_A} h^{(\pp\rq{},\qq\rq{})}_{\D_A \ell_A -,s}(r,\eta) \label{RecRelSpin+} \ ,
\\
h^{(\pp, \qq)}_{\D \ell -,s}(r,\eta)&=h^{(\pp, \qq)}_{\infty l -,s}(r,\eta)+\sum_{\pp\rq{},\qq\rq{}=1}^4\sum_{A \in \Acal_+} \frac{(R_{-A})_{\pp \qq,\pp\rq{} \qq\rq{}} \; (4 \, r)^{n_A}}{\D-\D^\star_A} h^{(\pp\rq{},\qq\rq{})}_{\D_A \ell_A -,s}(r,\eta) \nonumber \\
& \ \ \ \ \ \ \ \ \ \ \ \ \ \ \ \ \ \ \ +\sum_{p\rq{},q\rq{}=1}^5 \sum_{A \in \Acal_-} \frac{(R_{-A})_{\pp \qq,p\rq{} q\rq{}} \; (4 \, r)^{n_A}}{\D-\D^\star_A} h^{(p\rq{},q\rq{})}_{\D_A \ell_A +,s}(r,\eta)
 \ . \label{RecRelSpin-}
\end{align}
Here the label $A$ stands for $(T,n)$ where $T$ is one of the four types $\I, \II, \III, \IV$ and $n$ is an integer belonging to the set  $\Scal_T$ which can be finite or infinite depending on $T$. In particular we have 
\be
\sum_{A\in \Acal_+} \equiv \sum_{T=\I, \II, \III} \quad  \sum_{n \in \Scal_T} \ , \qquad 
\sum_{A\in \Acal_-} \equiv \sum_{T=\IV} \quad  \sum_{n \in \Scal_T} \ .
\ee
We present here a table which specifies the labels $A$,  $\Scal_T$, $\D^\star_A$,  $n_A$, $\ell_A$.
\be \label{AllTheLabels}
\begin{array}{|c | cccc|}
\hline
\phantom{\Big(}		A	\phantom{\Big(}	&\Scal_T						&\D^\star_A 					&n_A 										&\ell_{A}	 \\ 
\hline
\phantom{\Big(}  \I,  n \phantom{\Big(}& [1,\infty)				& 1-\ell-n 			 		& n 	  	   		&\quad \ell + n \quad \\ 
\phantom{\Big(} \II,  n \phantom{\Big(}& [1,\ell]  	&      \quad   2+\ell-n \quad & \quad n \quad  			&\quad \ell - n \quad  \\
\phantom{\Big(} \III,        n \phantom{\Big(}&[1,\infty)				& \frac{3}{2}-n  						& 2n       			 	 &\ell    \\ 
\phantom{\Big(} \IV,        n \phantom{\Big(}& [1,\min(2,\ell)] 				& 2-n  					& 2n -1      			 	 &\ell     \\ 
\hline
\end{array}
\ee
Further details about the table \ref{AllTheLabels} can be found in the appendix \ref{app:ConformalBlocks}.
The conformal blocks at large dimension $h_\infty$ are computed exactly by solving the Casimir equation at leading order in the large $\D$ expansion as explained in  appendix \ref{largedelta}. The coefficients $R$ can be conveniently written in terms of three contributions
\be
\label{RA=MAQAMA}
(R_{\pm A})_{p q p' q'}= (M_{\pm A}^{(L)})_{p p'} \  Q_A \ (M_{\pm A}^{(R)})_{q q'}   \ ,
\ee
where the coefficient $Q$ and the matrix $M$  arise because of the different normalization of the two and three point functions involving the primary descendants $\Ocal_A$. Schematically, 
\begin{align}
\langle  \Ocal_A | \Ocal_A \rangle ^{-1} &= \frac{Q_A}{ \D-\D^{\star}_A} \; \langle  \Ocal | \Ocal \rangle ^{-1} + O \left( (\D-\D^{\star}_A)^0 \right) \ , 
\\
\langle \Ocal_1 \Ocal_2 | \Ocal_A \rangle^{(p)} &= \sum_{p'} \; (M_{A}^{(L)})_{p p'} \;  \langle \Ocal_1 \Ocal_2 | \Ocal \rangle^{(p')}  \ .
\end{align}
In appendix \ref{RA} we further detail how to obtain the coefficients $R$.

Notice that with formulas (\ref{RecRelSpin+}-\ref{RecRelSpin-}) one can obtain all the blocks correspondent to four generic external vector operators. In this work however we only need the blocks for conserved and equal currents. To obtain them we contract the labels $p,q$ of the blocks with some matrices $m_\pm$ which come from the conservation of the $3$-point function of $\langle J J \Ocal^\pm \rangle$ explained in section \ref{sec:3pt}. We further contract the index $s$ with a matrix $\Mat$ which simplifies the crossing equation of $4$ equal vector operators as explained in section \ref{sec:4pt},
\be
\label{def:CB_conserved_equal}
 \tilde g^{(\tilde p,\tilde q)}_{\D \ell \pm, \tilde s}(r,\eta) \equiv  \, v^{1-d} \, \sum_s \Mat_{\tilde s s} \sum_{p q} (m_\pm)_{\tilde p p} \ (m_\pm)_{\tilde q q} \  g^{(p,q)}_{\D \ell \pm,s}(r,\eta) \ .
\ee
Here the matrix $m_+$ is $2 \times 5$ while the matrix $m_-$ is $1 \times 4$,
therefore $\tilde p,\tilde q=1,2$ for the parity even case and $\tilde p,\tilde q=1$ for the parity odd case.   In appendix \ref{app:conservedequal} we give the precise form of such matrices.
The matrix $\Mat$ is $19 \times 43$ and it is defined  in appendix \ref{ap:ftilde}.
It is worth to stress that since the equations  (\ref{RecRelSpin+}-\ref{RecRelSpin-}) are diagonal in $s$, it is possible to compute only some  structures, without having to compute the others.
In the following sections we will drop the tilde symbol above the labels $p,q,s$.

Using the OPE channel $(12)(34)$, one obtains the following conformal block expansion\footnote{In appendix \ref{app:ConformalBlocks}, we compute the conformal blocks in a three-point function basis which is different from the one of section \ref{sec:3pt}, therefore the coefficients $\tilde{\lambda}_{JJ\mathcal{O}}^{(p)}$ are just a linear transformation of the coefficients $\lambda_{JJ\mathcal{O}}^{(p)}$.}
\beq
 \tilde f_s(u,v) =
\sum_{\mathcal{O}} \sum_{p,q} \tilde{\lambda}_{JJ\mathcal{O}}^{(p)}
\tilde{\lambda}_{JJ\mathcal{O}}^{(q)} \,\tilde g_{\mathcal{O},s}^{(p,q)}(u,v)
\label{eq:CBexpansion} \ ,
\eeq
where the functions $\tilde f$ were defined in section \ref{sec:conservation}. %
For further details we refer the reader to  appendix \ref{app:ConformalBlocks}.

\subsection{Bootstrap equations}
\label{sec:crossingeqs}

Plugging the conformal blocks decomposition \eqref{eq:CBexpansion} into the three dimensional crossing equations \eqref{eq:crossing3D} we explicitly obtain 11 conditions which can be nicely written in vector notation as
\bea
\label{eq:crossing}
0 = \sum_{\substack{\mathcal O^+\\ \ell\geq2,even}} \left(\begin{array}{ccc} \tilde{\lambda}_{JJ \mathcal O^+}^{(1)} & \tilde{\lambda}_{JJ \mathcal O^+}^{(2)} \end{array} \right) \Vp \left( \begin{array}{c} \tilde{\lambda}_{JJ \mathcal O^+}^{(1)} \\ \tilde{\lambda}_{JJ \mathcal O^+}^{(2)} \end{array} \right) 
+\sum_{\mathcal O^+,\ell=0} \left(\tilde{\lambda}_{JJ \mathcal O^+}^{(1)}\right)^2 \Vo+\sum_{\mathcal O^-,\ell} \left(\tilde{\lambda}_{JJ \mathcal O^-}\right)^2 \Vm \ .\nn\\
\eea
Here  $\tilde{\lambda}_{JJ\mathcal O^+}^{(i)}, i=1,2$  are the OPE coefficients defined in Sec.~\ref{app:ConformalBlocks}, while $\tilde{\lambda}_{JJ \mathcal O^-}=\tilde{\lambda}_{JJ \mathcal O^-}^{(1)}$ for scalars and  $\tilde{\lambda}_{JJ \mathcal O^-}=\tilde{\lambda}{JJ \mathcal O^-}^{(2)}$ for higher $\ell$. 
In particular, for the stress energy tensor we have:
\begin{align}
\tilde{\lambda}_{JJT}^{(1)}=\sqrt{\frac{3C_T^\text{free}}{16C_T}}(1-12\gamma)\qquad\quad
\tilde{\lambda}_{JJT}^{(2)}=\sqrt{\frac{3C_T^\text{free}}{16C_T}}(5-12\gamma)
\label{eq:lambdaofgamma}
\end{align}
Finally,  $\Vo, \Vm$ are   $11$-dimensional vectors and $\Vp$ is a 11-vector of $2 \times 2$ matrices. \\
Introducing the (anti)symmetric combination of conformal blocks defined in \eqref{def:CB_conserved_equal}, 
\bea
\tilde{F}^{(p,q)}_{\Delta \ell \pm,s}(u,v) &=& \tilde{g}^{(p,q)}_{\Delta \ell \pm,s}(u,v)-\tilde{g}^{(p,q)}_{\Delta \ell \pm,s}(v,u)\,,\nn\\
\tilde{H}^{(p,q)}_{\Delta \ell \pm,s}(u,v) &=& \tilde{g}^{(p,q)}_{\Delta \ell \pm,s}(u,v)+\tilde{g}^{(p,q)}_{\Delta \ell \pm,s}(v,u)\,.
\eea 
we have
\be
\Vo \equiv \left(\!\!\begin{array}{c}  
                                            \Hp{3}(\frac14,\frac14)  \\
                                              \Hp{1}(u,u) \\ 
                                              \Hp{2}(u,u) \\
                                              \Hp{4}(u,u) \\
                                              \Hp{5}(u,u) \\
                                              \Hp{6}(u,u) \\
                                              \Hp{7}(u,v) \\
                                              \Fp{13}(u,v) \\
                                              \Fp{15}(u,v) \\
                                              \Fp{16}(u,v) \\
                                              \Fp{17}(u,v) \\
                                              \end{array}\!\! \right),
\ 
\Vm \equiv \left( \!\!
                                              \begin{array}{c}  
                                              \Hm{3}(\frac14,\frac14)  \\
                                              \Hm{1}(u,u) \\ 
                                              \Hm{2}(u,u) \\
                                              \Hm{4}(u,u) \\
                                              \Hm{5}(u,u) \\
                                              \Hm{6}(u,u) \\
                                              \Hm{7}(u,v) \\
                                              \Fm{13}(u,v) \\
                                              \Fm{15}(u,v) \\
                                              \Fm{16}(u,v) \\
                                              \Fm{17}(u,v) \\
                                              \end{array} 
                                              \!\!
                                              \right),
\ 
\Vp \equiv \left( \!\! \begin{array}{c}   
                                             \Symm[\HH{3}(\frac14,\frac14)]  \\
                                               \Symm[\HH{1}(u,u)] \\ 
                                               \Symm[\HH{2}(u,u)] \\
                                                \Symm[\HH{4}(u,u)]\\
                                                \Symm[\HH{5}(u,u)] \\
                                                \Symm[\HH{6}(u,u)] \\
                                                \Symm[\HH{7}(u,v)] \\
                                                \Symm[\FF{13}(u,v)] \\
                                                \Symm[\FF{15}(u,v)] \\
                                                \Symm[\FF{16}(u,v)] \\
                                                \Symm[\FF{17}(u,v)] \\
                                              \end{array} 
                                              \!\!
                                              \right) .
\label{eq:Vdef}
\ee 
For any $2\times2$ matrix $M$, $\Symm[M]\equiv(M+M^T)/2$ selects its symmetric part.
In the above expression we omitted the $(p,q)$ upper indices when only one conserved conformal block is allowed, namely for parity even scalars and parity odd operators.

\subsection{Setting up the Semi-Definite Programming}

The feasibility of the above set of equations can be constrained using semidefinite programming (SDP).  We refer to \cite{Kos:2014bka} for details.  To rule out a hypothetical CFT spectrum, we must find a linear functional $\alpha$ such that
\bea                              
& \vec\alpha[V_{0,+}] \geq 0,&\text{for the identity operator},\nn \\
& \vec\alpha[\Vo] \geq 0,&\text{for all scalar operators},\nn \\
&\vec\alpha[\Vp] \succeq 0,&\text{for all parity even operators in the spectrum with $\ell$ even}, \label{eq:functional} \\
&\vec\alpha[\Vm] \geq 0,&\text{for all parity odd operators in the spectrum with any $\ell\neq1$}. \nn
\eea
Here, the notation ``$\succeq 0$" means ``is positive semidefinite". Since the 11 crossing equations have a different dependence on the conformal invariants $u,v$, it's worth spelling out the explicit form of the linear functional $\alpha$ we consider in this work. Let us remind the reader the definition of the usual   coordinates $z,\bar{z}$:
\bea
u = z \bar{z} \qquad v = (1-z) (1-\bar{z}) 
\eea
Then, we define the family of linear functionals $\alpha$ acting on an 11-dimensional vector, whose entries are functions of $z,\bar{z}$
\bea
\alpha[\vec{V}] = \alpha_1 V_1\left(\frac12,\frac12\right)+\sum_{i=2}^6\sum_{m=0}^{\Lambda/2}\alpha_{m}\partial_z^{2m} V_i(z,1-z)\bigg|_{z=\frac12}+\sum_{i=7}^{11}\sum_{m+n<\Lambda}\alpha_{mn}\partial_z^m \partial_{\bar{z}}^n V_i(z,\bar{z})\bigg|_{z=\bar{z}=\frac12}\nonumber\\
\label{eq:functionaldef}
\eea
Although we didn't write it explicitly, the linear functionals are parametrized by the integer $\Lambda$, which indicates the order of derivatives considered. Notice that the action of the functional on a vector of matrices results in a matrix, while its action on a vector of scalar functions produces a number. 
The existence of such a functional for a hypothetical CFT spectrum implies the inconsistency of this spectrum with crossing symmetry. In addition to any explicit assumptions placed on the allowed values of $\Delta$, we impose that all operators must satisfy the unitarity bound
\bea
\Delta &\geq& \left\{\begin{array}{ll}
\ell + d - 2 & \ell > 0\\
\frac{d-2}{2} & \ell = 0
\end{array}
\right.,
\eea
where $d=3$ is the spacetime dimension.

The more information about the spectrum we use in \eqref{eq:functional}, the easier it is to find a functional $\vec\alpha$ that excludes the putative CFT. 
In this work we mainly focus on  assumptions about the minimal values of operator dimensions in given sectors and the value of parameter $\gamma$ defined in section \ref{sec:JJT}.

We will review the exact SDP problem to solve case by case in the next section.

\section{Results}
\label{sec:results}

In this section we present the results of our numerical investigations. In what follows $\Delta_\ell^{\pm}$ will denote the dimension of the first parity even/odd neutral spin $\ell$ operator. We will also use  $(\Delta_\ell^{\pm})'$ to denote the second operator in the same sector.

\subsection{Bounds on operator dimensions}

\begin{figure}[t]
 \begin{center}
 \includegraphics[width=0.7\textwidth]{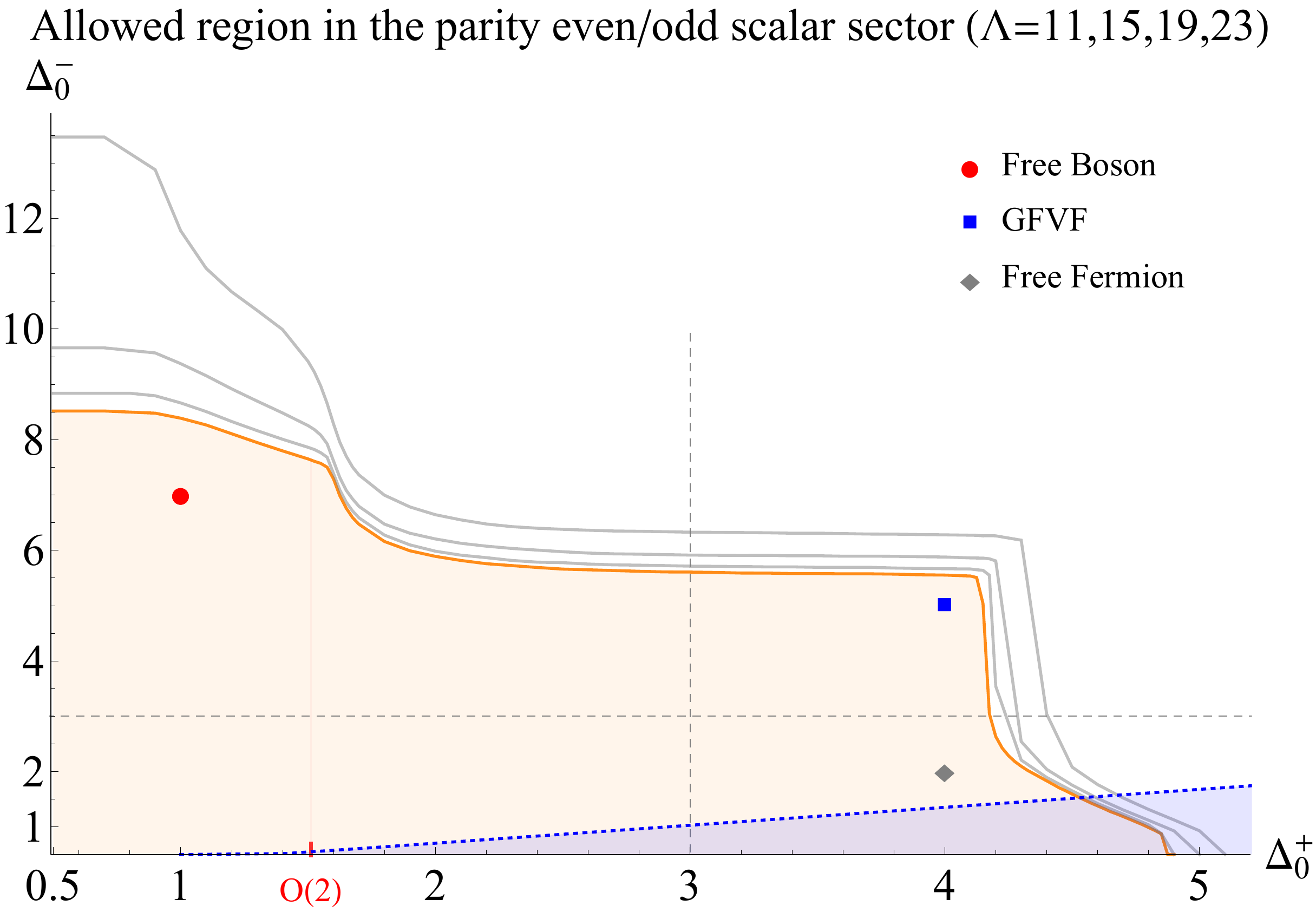}
 \end{center}
  \caption{
  Allowed region consistent with crossing symmetry assuming that all the parity-even scalars appearing in the OPE $J\times J$ have dimension larger than $\Delta_0^+$ and all parity-odd scalars have dimensions larger than $\Delta_0^-$. The orange shaded region is allowed. Marks correspond to known CFTs: free complex boson $((\Delta_0^+,\Delta_0^-)=(1,7))$,  free Dirac fermion $((\Delta_0^+,\Delta_0^-)=(4,2))$ and GFVF $((\Delta_0^+,\Delta_0^-)=(4,5))$. The red vertical line corresponds to the approximate dimension of the lightest singlet operator in the interacting $O(2)$ model: $\Delta_0^+=1.5117$. The blue shading shows the region excluded by bootstrapping the four point function of identical parity odd scalars with the dimension $\Delta_0^-$. See text for more details. The best bound has been computed at $\Lambda=23$ while gray lines correspond to $\Lambda=11,15,19$.}
\label{Fig:spin0spin0odd}
\end{figure}

We begin our journey in the space of CFTs with global symmetries by inspecting the constraints imposed by crossing symmetry on the spectrum of scalar operators. As reviewed in Sec.~\ref{sec:3pt}, the OPE $J\times J$ contains both parity even and parity odd scalars. The first issue we want to address is how large can the dimensions of these operators be. To answer  this question we solved the semi-definite problem $\eqref{eq:functional}$ with the assumption that   all scalar parity-even/odd operators have dimension larger than $\Delta_0^\pm$ correspondingly.  The allowed region is shown in figure \ref{Fig:spin0spin0odd}. The very first surprising result is that crossing symmetry is able to constrain the plane $\Delta_0^+,\Delta_0^-$ into a closed region, meaning that all CFTs with global symmetry must have parity even and parity odd scalar operators. This is completely universal: this result is only based on unitarity and associativity of the OPE. To our knowledge this is the first completely general result for 3D unitary CFT with global symmetry.\footnote{All previous results in the bootstrap literature assumed at least the presence of a scalar or fermion operator with a given fixed dimension; theories with  extended supersymmetry represent an  exception: scalars are contained in certain protected super-multiplets.}

Let us now describe the shape of figure \ref{Fig:spin0spin0odd}. If we regard the boundary of the allowed region as a function $(\Delta_0^-)^\text{max}$ of $\Delta_0^+$, then it can only be a monotonic non-increasing function.\footnote{If we can not exclude a theory with $\Delta_0^+=a$ and $\Delta_0^-=b$ then we cannot exclude theories with  $\Delta_0^+\le a$ and $\Delta_0^-\le b$.}  Hence we expect the allowed region to be shaped by existing CFTs with the largest gap in the scalar sector. 
There are three solvable models that we can place in the $\Delta_0^+,\Delta_0^-$ plane: a free massless complex scalar field $\phi$, a free massless 3d Dirac fermion $\psi$ and  a Generalized Free Vector Field (GFVF). 
In the  free scalar field case, the $U(1)$ current OPE schematically reads: 
\be
\label{eq:freebosonOPE}
J_\mu \times J_\nu \sim \underbrace{\phi^\dagger\phi}_\text{parity-even} +\underbrace{\epsilon^{\alpha\beta\rho}\phi^\dagger\partial_\alpha\phi
\partial^\sigma \partial_\beta\phi^\dagger\partial_\sigma \partial_\rho\phi}_\text{parity-odd}+\dots , \quad \Delta_0^+=1,\, \Delta_0^-=7\,.
\ee

\begin{figure}[h!]
 \begin{center}
 \includegraphics[width=.95\textwidth]{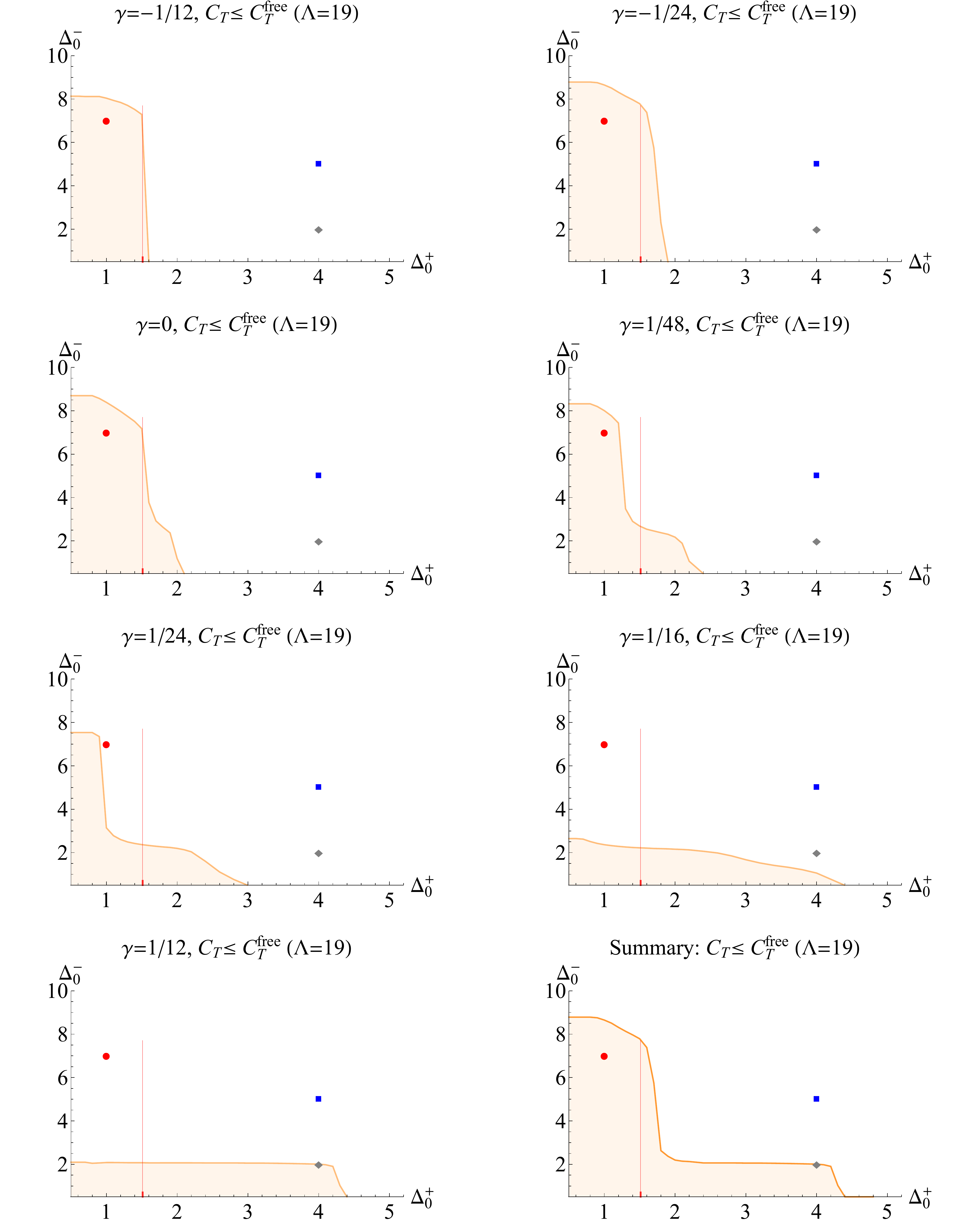}
 \end{center}
    \caption{\small{ 
    Allowed region  assuming that all parity-even scalars appearing in the OPE $J\times J$ have dimension larger than $\Delta_0^+$ and all parity-odd scalars have dimensions larger than $\Delta_0^-$.  We also impose small central charge $C_T\le C_T^{\rm free}$ and fix $\gamma$ to specific values within the range $12|\gamma|\le 1$.
  The bound has been computed at $\Lambda=19$. See figure \ref{Fig:spin0spin0odd} for marks legend.}
}
\label{Fig:spin0evenspin0oddCTsmall}
\end{figure}

\begin{figure}[h!]
 \begin{center}
 \includegraphics[width=0.8\textwidth]{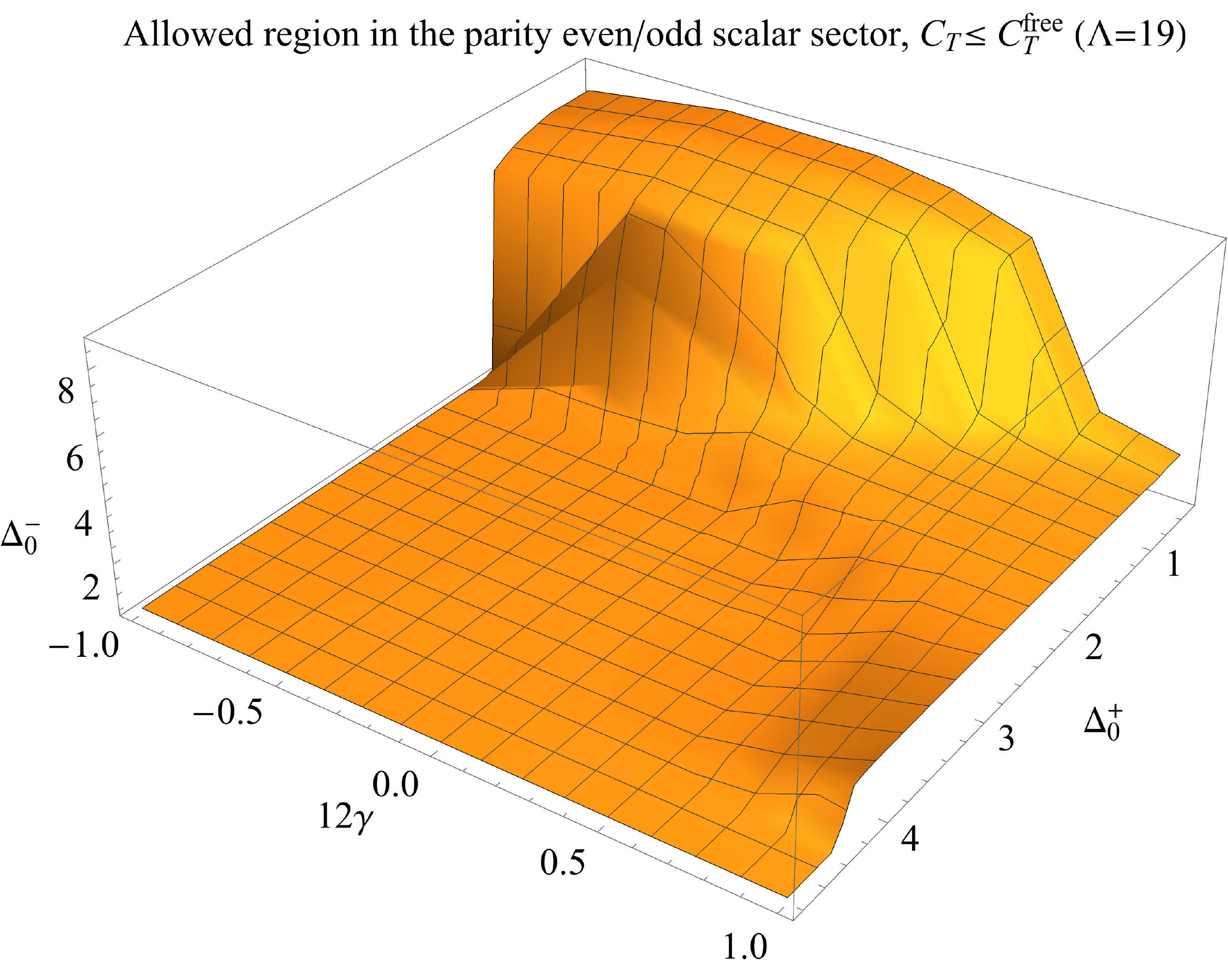}
 \end{center}
    \caption{Three dimensional view of  figure \ref{Fig:spin0evenspin0oddCTsmall}.}
\label{Fig:spin0evenspin0odd3Dplot}
\end{figure}

\begin{figure}[t]
 \begin{center}
 \includegraphics[width=0.7\textwidth]{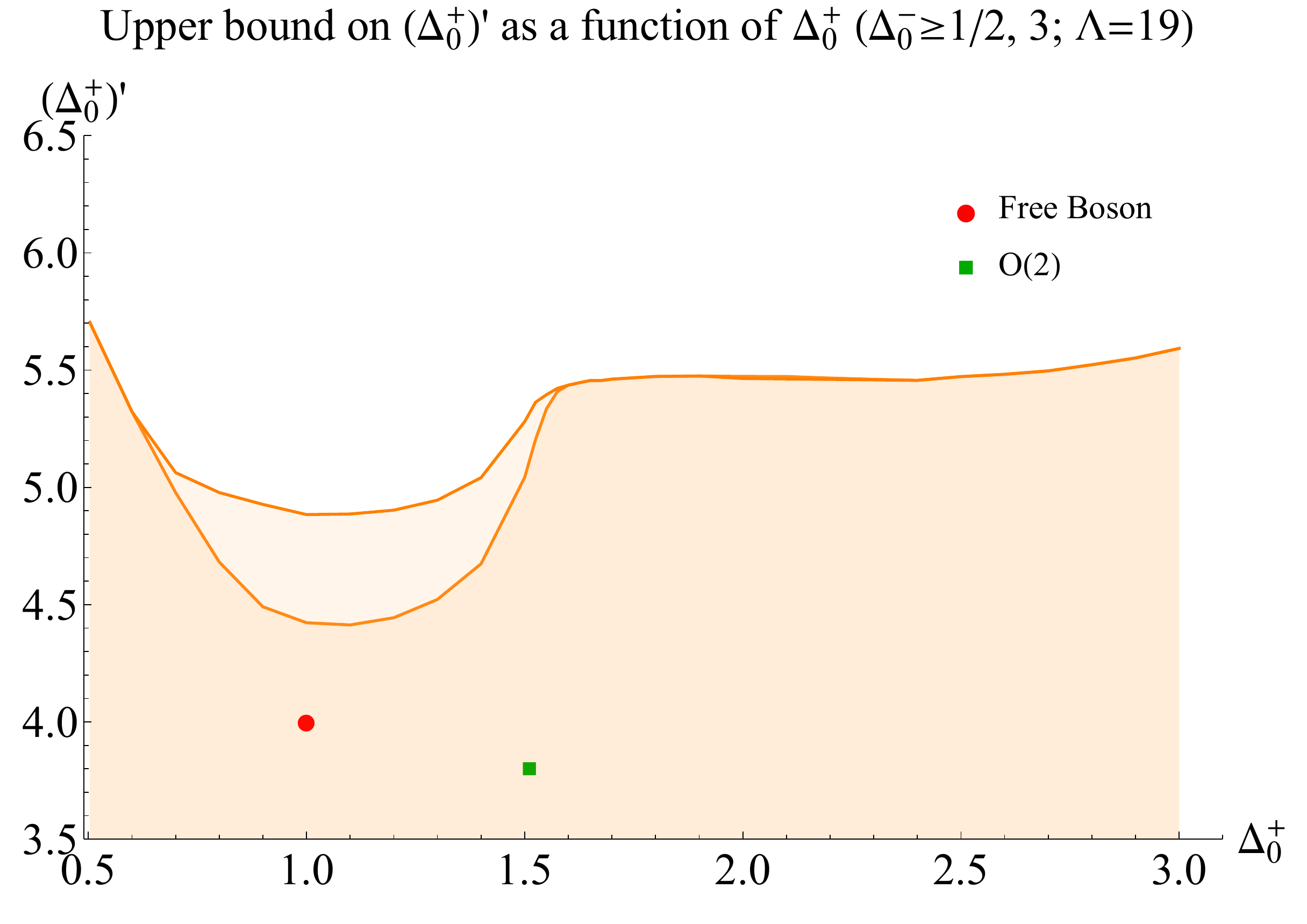}
 \end{center}
    \caption{
Upper bound on $(\Delta_0^+)'$, the dimension of the second lowest parity-even scalar operator appearing in the OPE $J\times J$ when we assume that the theory contains a scalar parity-even operator with dimension $\Delta_0^+$. 
The red dot corresponds to the free boson $((\Delta_0^+,(\Delta_0^+)')=(1,4))$, and green dot corresponds to interacting $O(2)$ theory, while points for  
 the free fermion and GFVF lie outside the range of the plot. 
Lower curve assumes  theory has no relevant scalar parity-odd operators. The bounds have been computed with $\Lambda=19$.}
\label{Fig:spin0spin0p}
\end{figure}

In the free fermion case, we find
\be
\label{eq:freefermionOPE}
J_\mu\times J_\nu \sim  \underbrace{\left(\bar{\psi} \psi\right)^2}_\text{parity-even} +\,\, \underbrace{\bar{\psi} \psi}_\text{parity-odd}+\dots , \quad \Delta_0^+=4,\, \Delta_0^-=2\,.
\ee
Finally, the GFVF is equivalent to a free photon in $AdS_4$. From the three dimensional point of view it corresponds to a conserved current with a standard 2 point function, and all higher point correlators satisfying Wick theorem. 
In this case the lightest scalar operators are given by
\be
J_\mu \times J_\nu \sim \underbrace{J_\mu J^\mu}_\text{parity-even} + \,\,   \underbrace{\epsilon^{\mu\nu\rho}J_\mu\partial_\nu J_\rho}_\text{parity-odd}+\dots , \quad \Delta_0^+=4,\, \Delta_0^-=5\,.
\ee
Notice that the GFVF is technically a so called \emph{dead-end} CFT, since it doesn't contain relevant scalar operators. On the other hand it doesn't contain a local energy momentum tensor either, since it corresponds to a U(1) gauge theory on a fixed  $AdS$ background (infinite central charge $C_T$ and no dynamical gravity). 

These solvable CFTs  are marked in figure \ref{Fig:spin0spin0odd} as described in the caption.
While the boundary of the allowed region is close to the GFVF point, it is quite far from the point of free boson theory. Instead it starts at higher values $\Delta_0^-$ and after a small plateau it  displays a kink for values of $\Delta_0^+$ seemingly in correspondence to the interacting $O(2)$ model. To our knowledge the dimension of the leading parity odd scalar in this model is not known, neither in the $\varepsilon$-expansion nor in the $1/N$ expansion. Accordingly, we conjecture that the lightest parity odd operator in \eqref{eq:freebosonOPE} in critical $O(2)$ theory acquires a positive anomalous dimension. 

One additional feature of figure \ref{Fig:spin0spin0odd} is the region extending to values $\Delta_0^+$ larger than 4 but requiring at the same time  parity odd scalars with small dimension. Let us call $\phi^-$ the parity odd scalar operator with smallest dimension. The OPE of $\phi^-$ with itself would contain a parity even scalar operator\footnote{Unless there is symmetry argument preventing this from happening, this operator must coincide with the smallest dimension parity even scalar operator entering the $J\times J$ OPE.} with dimension $\Delta_0^+$. Then, by bootstrapping the four point function $\langle\phi^-\phi^-\phi^-\phi^-\rangle$ 
we can obtain an independent bound of the form $\Delta_0^+ \leq f(\Delta_0^-)$, for some function $f$. This bounds has been already obtained in past works focused on the three dimensional Ising model \cite{ElShowk:2012ht,Kos:2014bka,El-Showk:2014dwa}. For this work we extended these results to larger values of $\Delta_0^-$. The  blue shading in figure \ref{Fig:spin0spin0odd} represents the disallowed region.
We expect that the use of mixed correlators of scalars and conserved currents will shed light on the fate of this region.

\begin{figure}[t]
 \begin{center}
 \includegraphics[width=0.7\textwidth]{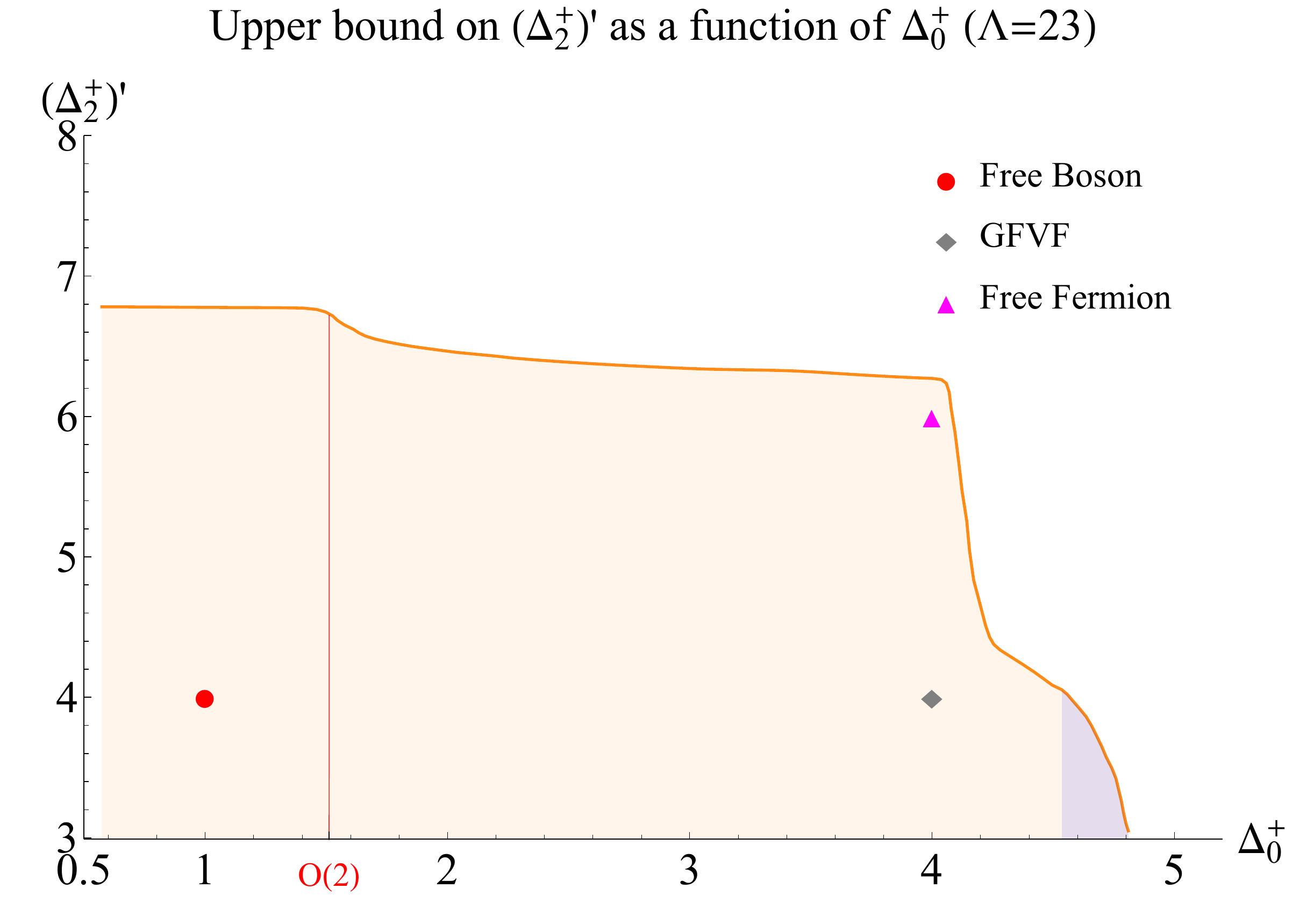}
 \end{center}
    \caption{Allowed region consistent with crossing symmetry assuming that all  parity-even scalars appearing in the OPE $J\times J$ have dimension larger than $\Delta_0^+$ and all parity-even spin-2 operators (except the energy momentum tensor) have dimensions larger than $\Delta_2^+$. No other assumption is imposed. The shaded region is allowed. Marks correspond to known CFTs: free boson $((\Delta_0^+,(\Delta_2^+)')=(1,4))$, free fermion $((\Delta_0^+,(\Delta_2^+)')=(4,6))$ and GFVF ($(\Delta_0^+,(\Delta_2^+)')=(4,4))$. The vertical red line corresponds to the central value of the allowed dimensions of the smallest dimension singlet operator in the $O(2)$ model: $\Delta_0^+=1.5117$. The shaded blue region is excluded by bootstrapping the four point function of the first parity odd scalar operator. See figure \ref{Fig:spin0spin0odd}. The bound has been computed at $\Lambda=23$.}
\label{Fig:spin0spin2}
\end{figure}

The existence of a CFTs with  large gaps in the scalar sector, namely the GFVF, shapes the bound shown in figure \ref{Fig:spin0spin0odd} for $1.6 \lesssim \Delta_0^+\lesssim 4$ and could potentially hide other theories in the bulk of the allowed region. In order to better probe this region we explored the constraints on theories with a finite value of the central charge.
To do that, we modified the conditions \eqref{eq:functional} and looked for a linear functional that satisfies the following requirements:
\be
\begin{array}{cl}
\alpha\left[ \vec{V}_{0+} + b \frac{C_T^\text{free}}{C^\text{max}_T} \vec{\tilde{\lambda}}_{JJT}(\gamma)^T \cdot \VT \cdot \vec{\tilde{\lambda}}_{JJT}(\gamma)\right]= 1,&\textrm{(normalization)}\\ 
\vec{\tilde{\lambda}}_{JJT}(\gamma)=\sqrt{\frac{3}{16}}\left(
                                   \begin{array}{c}
                                   1-12\gamma\\
                                   5-12\gamma
                                   \end{array}\right)&\\
\alpha[ \Vo] \geq 0,&\Delta\geq \Delta_0^+,\ell=0   \\ 
\alpha[ \Vp] \succeq 0,& \Delta\geq\ell+1,\ell>2\,\text{even} \\ 
\alpha[ \Vm] \ge 0,& \Delta\geq \Delta_0^+,\ell=0 \\ 
\alpha[ \Vm] \ge 0,& \Delta\geq\ell+1,\ell\geq 2 \\ 
\end{array}
\label{eq:functionalCtless}
\ee
Compared to \eqref{eq:functional} we have modified the normalization condition in order to  input a specific value of $\gamma$ and we have used \eqref{eq:lambdaofgamma}. It is straightforward to show that the bound obtained with a functional satisfying \eqref{eq:functionalCtless} only applies to CFTs with $C_T\leq C_T^\text{max}$. 

In figure \ref{Fig:spin0evenspin0oddCTsmall}, we again show the allowed region in the  plane  $(\Delta_0^+,\Delta_0^-)$ but requiring small central charge $C_T \le C_T^\text{max}=C_T^\text{free}$ and for several specific values of the parameter $\gamma$ defined in \eqref{JJOPET}. 
As expected, this excludes the GFVF which effectively has infinite central charge. 
More interestingly, one can observe that varying the parameter $\gamma$ the bounds smoothly interpolates two very different regimes. For $\gamma\simeq 1/12$, the free fermion theory drives the shape of the bound, while as we decrease $\gamma$,  the allowed region is entirely concentrated at smaller $\Delta_0^+$ but large $\Delta_0^-$. Notice also that the maximum of $\Delta_0^-$ is not reached at the free boson theory but at slightly larger values of $\gamma$ and $\Delta_0^+$.
These results are also shown as a 3D plot in figure \ref{Fig:spin0evenspin0odd3Dplot}.

In figure \ref{Fig:spin0spin0p} we show the upper bound on the dimension of the second lightest parity even scalar operator $(\Delta_0^+)'$ as a function of $\Delta_0^+$. 
We performed the analysis with and without forbidding relevant scalar parity odd operators. 
Next, in figure \ref{Fig:spin0spin2} we show the bound on  the dimension of the first non conserved spin-$2$ parity even operator $(\Delta_2^+)'$.
Notice that $\Delta_2^+=3$ because the dimension of the stress tensor is fixed.\footnote{However,  we do not exclude solutions where the OPE coefficient of the stress tensor vanishes ($C_T=\infty$).}
Interestingly, both bounds display a kink structure in the proximity of the location of the $O(2)$ model. On the other hand both the maximal allowed values of $(\Delta_0^+)'$ and $(\Delta_2^+)'$ at the kink are much larger than the ones of the free $O(2)$ model. It would be surprising if the interacting critical $O(2)$ model displayed such large anomalous dimensions. At this stage, it is unclear if the kink feature is related to the $O(2)$ model. It would be interesting to include correlations functions of charged operators in our bootstrap study to further explore this region. We leave this mixed correlator analysis for the future. 
Finally, notice  that in figure \ref{Fig:spin0spin2} the region $\Delta_0^+\gtrsim4.52$ is excluded if we also take into account the constraints coming from the four point function of the lightest parity-odd scalar appearing in $J\times J$ (see figure \ref{Fig:spin0spin0odd}).

\subsection{Central Charge bounds}
\label{sec:centralcharge}

\begin{figure}[h!]
 \begin{center}
\subfigure[]{\includegraphics[width=0.45\textwidth]{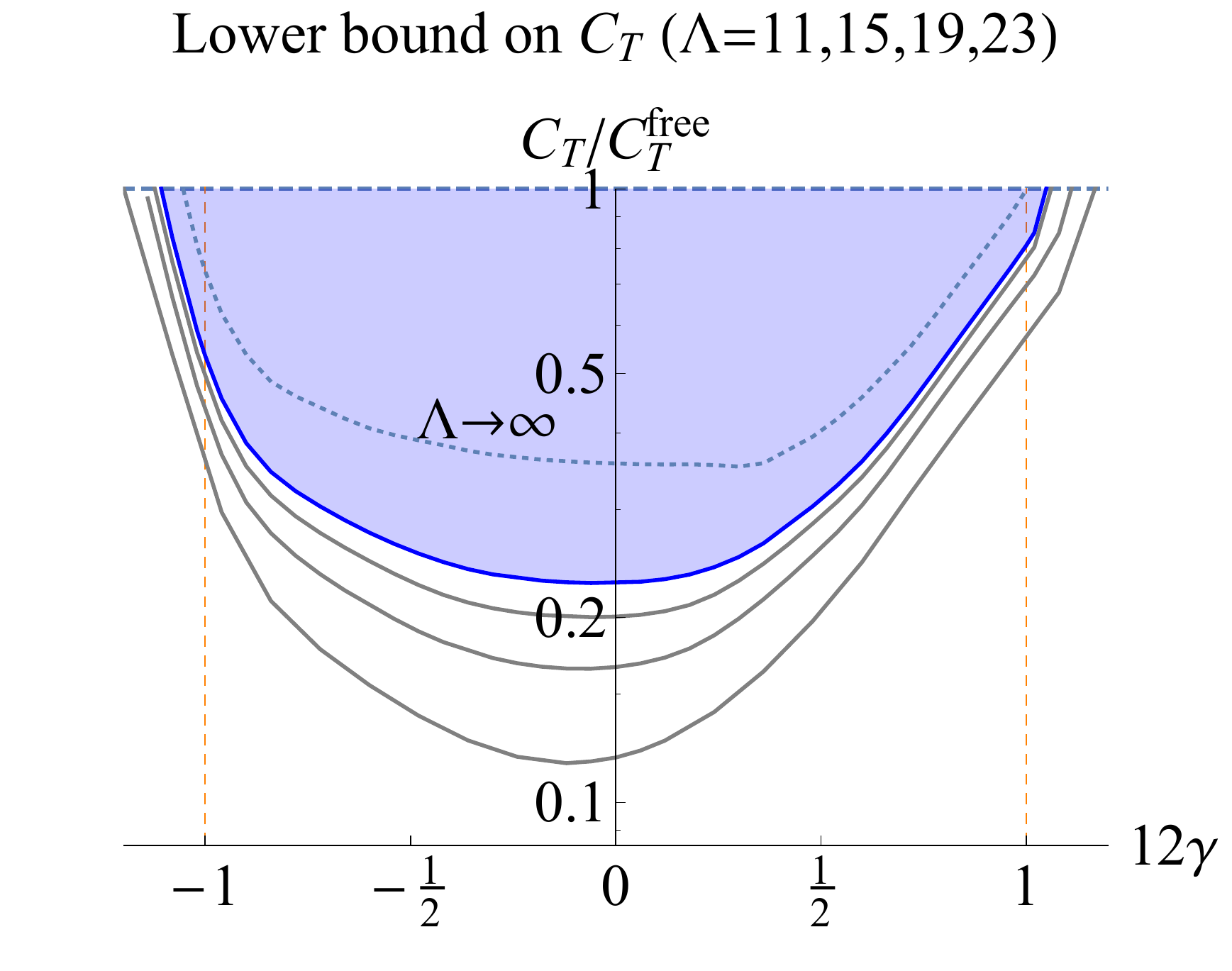}}
\subfigure[]{\includegraphics[width=0.45\textwidth]{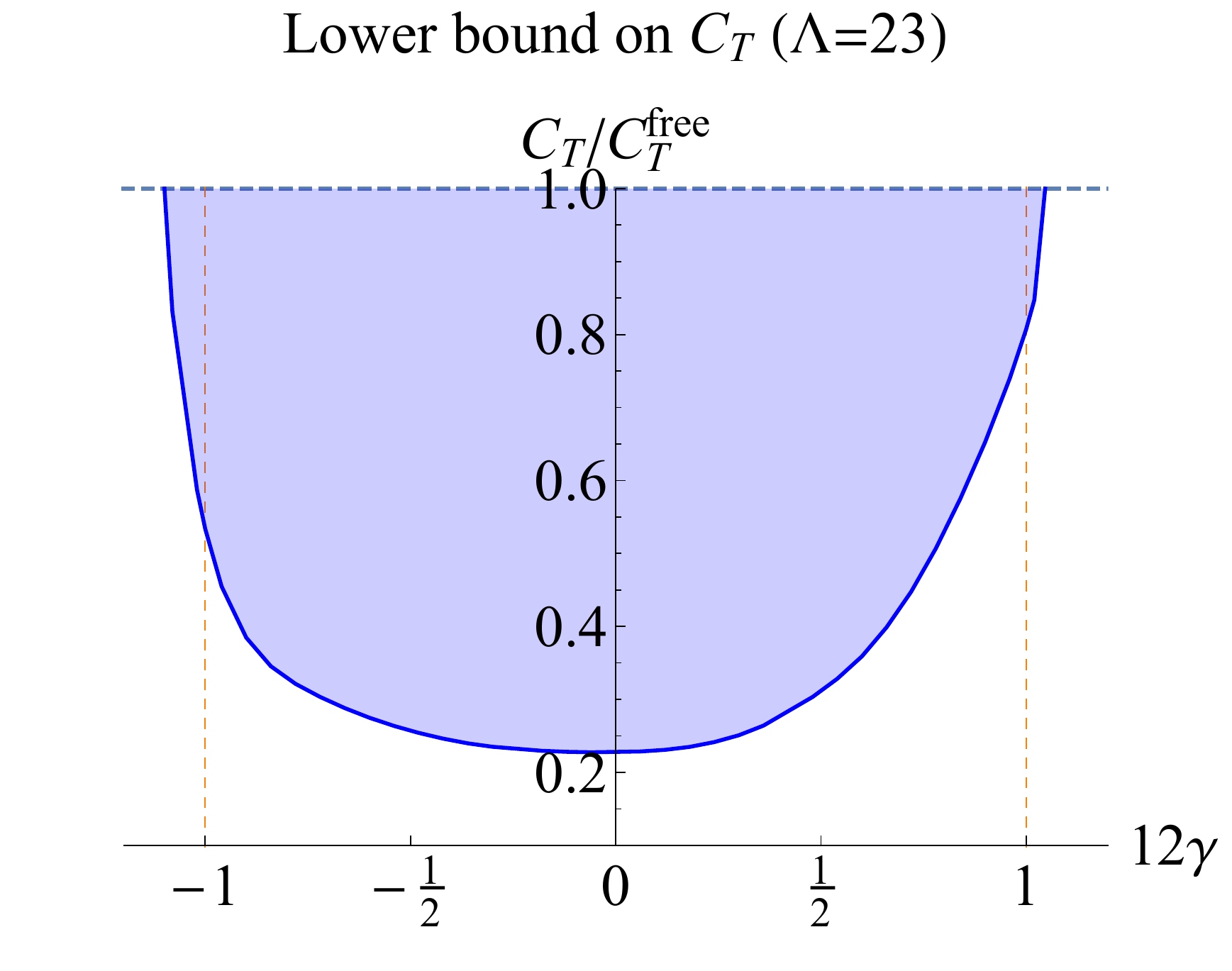}}
  \end{center}
  \caption{a) Zoom in in the region $|12\gamma|\leq 1$ of the lower bound on the central charge normalized to the central charge of a free complex boson as a function of the parameter $\gamma$ defined in \eqref{JJOPET}. The shaded region is allowed. Different curves corresponds to increasing the number of derivatives included in the numerical problem. The bounds have been computed at $\Lambda=11,15,19,23$. The dashed line corresponds to a linear extrapolation in $\Lambda^{-1}$. b) Best bound on central charge in linear scale.
  }
\label{Fig:CTvsGAMMAzoom}
\end{figure}

A well established feature of the conformal bootstrap is the possibility to place upper bounds on OPE coefficients, or equivalently a lower bound on $C_T$ \cite{Caracciolo:2009bx,Rattazzi:2010gj,Poland:2010wg}. In this section we investigate the minimal value of the central charge that a CFT with a continuous local global symmetry is allowed to have, as a function of the parameter $\gamma$.  To find such a bound, we search for a functional $\alpha$ satisfying the properties:
\be
\label{eq:ctsdp}
\begin{array}{cl}
\alpha[ \vec{\tilde{\lambda}}_{JJT}(\gamma)^T \cdot \VT \cdot \vec{\tilde{\lambda}}_{JJT}(\gamma)]= 1,&\textrm{(normalization)}\\ 
\vec{\tilde{\lambda}}_{JJT}(\gamma)=\sqrt{\frac{3}{16}}\left(
                                   \begin{array}{c}
                                   1-12\gamma\\
                                   5-12\gamma
                                   \end{array}\right)&\\
\alpha[ \Vo] \succeq 0,&\Delta\geq 1/2,\ell=0   \\ 
\alpha[ \Vp] \succeq 0,&\Delta\geq (\Delta_2^+)',\ell=2   \\ 
\alpha[ \Vp] \succeq 0,& \Delta\geq\ell+1,\ell>2\,\text{even} \\ 
\alpha[ \Vm] \ge 0,& \Delta\geq 1/2,\ell=0 \\ 
\alpha[ \Vm] \ge 0,& \Delta\geq\ell+1,\ell\geq 2 \\ 
\end{array}
\ee
Notice that compared to \eqref{eq:functional} we have eliminated the assumption of the functional $\alpha$ being positive on the identity operator contribution. As shown later, we will instead minimize $\alpha[\Vo]$.  Also, by fixing the normalization we input a specific value of $\gamma$. Here $\vec{\tilde{\lambda}}_{JJT}(\gamma)=\sqrt{C_T/C_T^{\text{free}}}\left( \tilde{\lambda}_{JJT}^{(1)},\, \tilde{\lambda}_{JJT}^{(2)}\right)$ is a two dimensional vector of OPE coefficients, with each component being a linear function of  $\gamma$, and we have used \eqref{eq:lambdaofgamma}. Finally, we introduced a gap  in the spin 2 even sector, and assume that, besides the energy momentum tensor, whose dimension saturates the unitarity bound, all the parity-even spin-2 operators satisfy $[\mathcal O_{\ell=2}]\geq(\Delta_2^+)'$.   We will come back to this assumption later. 
Applying the functional to the crossing equations \eqref{eq:crossing} and using the results of Sec.~\ref{sec:JJT} one obtains
\be
\label{eq:CTbound}
\frac{C_T^{\text{free}}}{C_T} \le -  \alpha[\VId]\,.
\ee
Therefore, the optimal bounds on $C_T$ will be set by the functional minimizing $\alpha[\Vo]$, subject to the constraints (\ref{eq:ctsdp}).

\begin{figure}[t!]
 \begin{center}
 \includegraphics[width=0.6\textwidth]{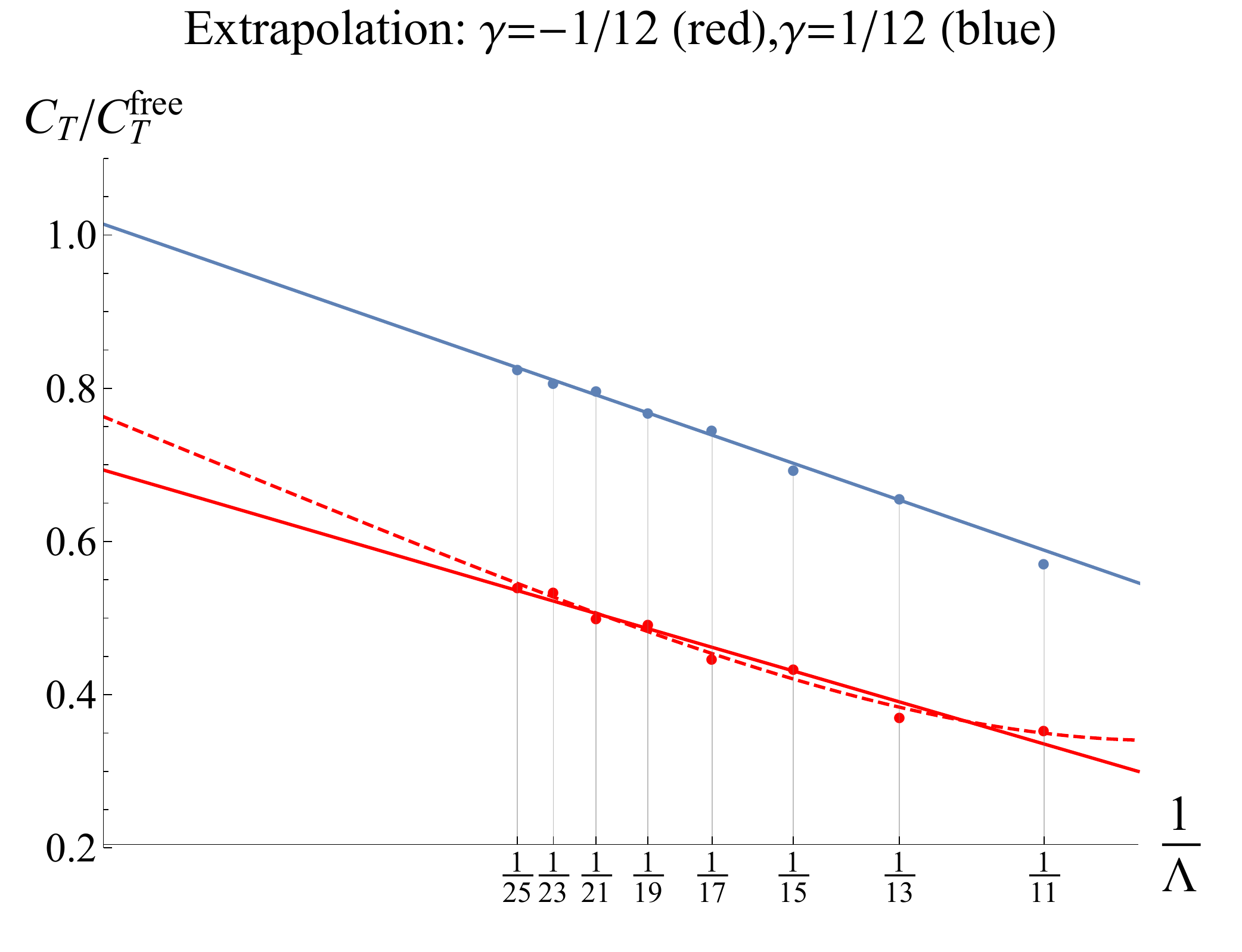}
  \end{center}
  \caption{Extrapolation of the bound on central charge normalized to $C_T^\text{free}$ as a function of the number of derivatives included in the semidefinite-programming $\Lambda$. The upper (blue) lines corresponds to a linear fit of the bounds at $\gamma=1/12$. In the limit $\Lambda\rightarrow \infty$ the extrapolation approaches the value 1. The lower curves correspond to a linear (red continuous) and quadratic (red dashed) fit of the bounds at $\gamma=-1/12$. The linear fit predicts an asymptotic bound much smaller than the free theory one. Other fits may predict a value of $C_T$ closer to $C_T^\text{free}$. This is exemplified by the quadratic fit.}
\label{Fig:CTextrapolation}
\end{figure}

In figure \ref{Fig:CTvsGAMMA}, presented in the introduction, we show our best bound on the central charge as a function of $\gamma$ and how the bound improves when increasing the numerical power $\Lambda$. As expected, inside the interval $|12\gamma|\leq1$, the bound seems to converge to a finite value, while outside it improves by a order one factor at each step. 

In figure \ref{Fig:CTvsGAMMAzoom} we display the zoomed version of the same plot. 
As discussed in Sec.~\ref{sec:JJT}, the two extremes of the interval $12|\gamma| \le 1$ are saturated by the free complex boson and the free fermion theory. In \cite{Zhiboedov:2013opa}, it was shown that when $\gamma$ assumes the extremal values, the CFT must necessarily be free, i.e.~all the correlators of the CFT must be equal to  those of a corresponding (bosonic or fermionic) free theory. One would therefore expect the bound to approach the value of the central charge of a free complex boson or free fermion given in eq.~\ref{eq:ctfree} at the extremes of the allowed interval. This doesn't appear to be the case with the current numerical power. Nevertheless we might hope to approach the optimal bound in the limit $\Lambda\rightarrow\infty$. In figure \ref{Fig:CTextrapolation} we show a linear extrapolation of the bounds computed at $\gamma=\pm1/12$ for $\Lambda=11,...,25$. For $\gamma=1/12$ a linear extrapolation (upper blue line in figure \ref{Fig:CTextrapolation})  is consistent with an asymptotic bound $C_T\ge C_T^\text{free}$. Extrapolating the bound for $\gamma=-1/12$ is trickier. Although we expect the bound to be $C_T\ge C_T^\text{free}$, the linear fit (bottom red line  in figure \ref{Fig:CTextrapolation}) clearly gives an asymptotic value smaller than $C_T^\text{free}$. Most likely, the linear extrapolation in $\Lambda$ simply does not capture the infinite number of derivatives limit.
It is plausible that the apparent convergence of the bound to a value smaller that $C_T^\text{free}$ is due to some hypothetical CFT with $C_T< C_T^\text{free}$ and $\gamma$ close to $-1/12$. With the current numerical power we cannot make a conclusive statement confirming or ruling out such a theory. 

An interesting feature of figure \ref{Fig:CTvsGAMMAzoom} is that the central charge bound is well below $C_T^{\rm free}$ not only near $12|\gamma|= 1$ but in the whole region $12|\gamma|\le 1$. Based on previous works on conformal bootstrap \cite{ElShowk:2012ht,El-Showk:2014dwa} we are keen to consider this as an indication that there might exist a number of CFTs whose central charge is smaller than the free theory one. A largely accepted lore suggests that the central charge measures the number of degree of freedom in the theory.\footnote{This is clearly the case for free theories and CFTs that are perturbatively away from a free theory.} Accordingly we expect a CFT with the central charge smaller than $C_T^\text{free}$ to have minimal possible gloabal symmetry, i.e.~only a global $U(1)$. The critical $O(2)$-model is the only known example of such a theory with $C_T\approx 0.944$. The other possible candidate, the $N=2$ Gross-Neveu model is in fact expected to have a central charge larger than $C_T^\text{free}$ (see appendix \ref{sec:reminder} for a review). The critical $O(2)$-model clearly can not explain the current shape of the bound. 
As the numerics improves, $\Lambda \to \infty$,  we expect the optimal bound to become significantly stronger and be saturated by the hypothetical new theories with $C_T\le C_T^{free}$. 

\begin{figure}[h!]
 \begin{center}
\includegraphics[width=0.7\textwidth]{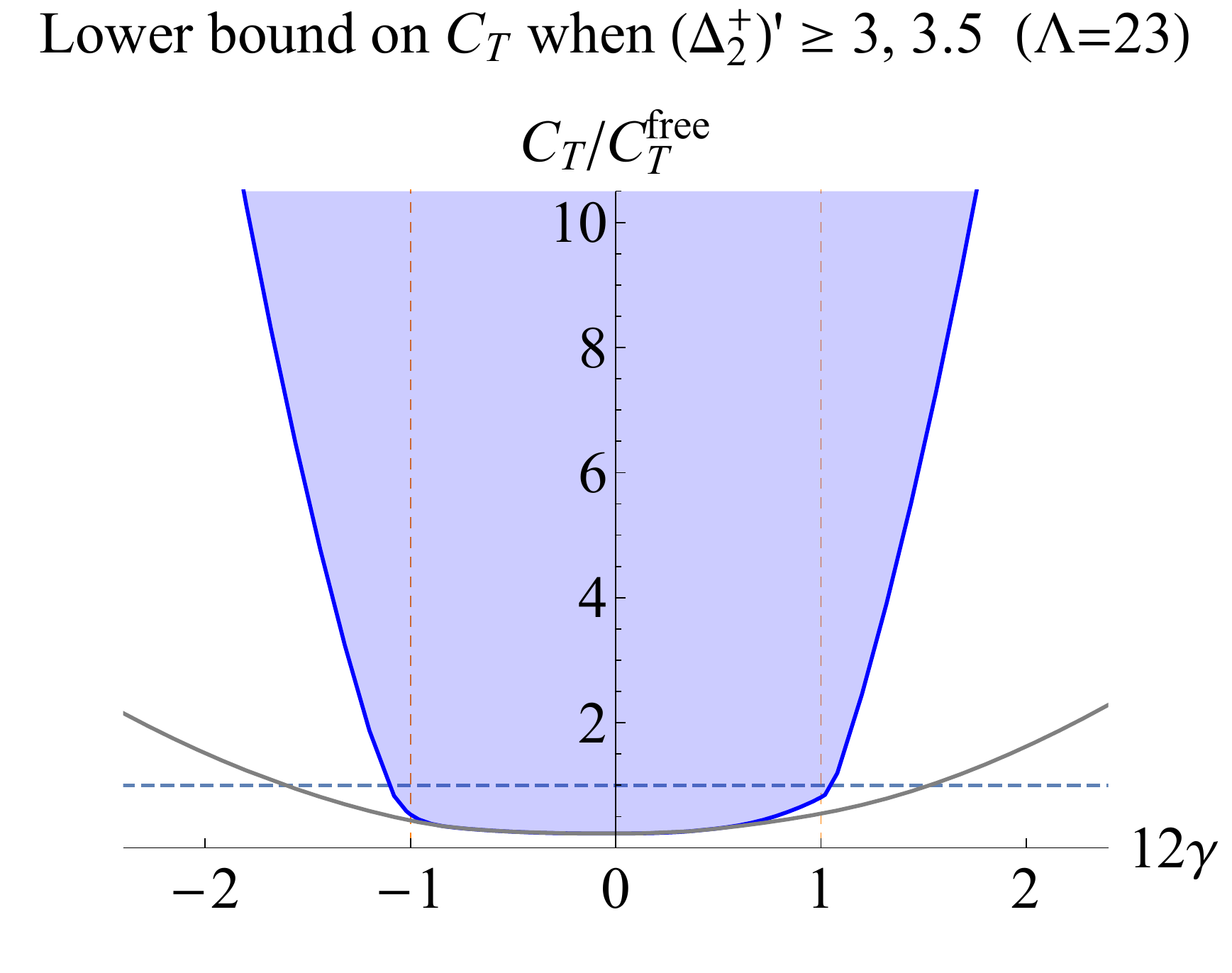}
  \end{center}
  \caption{Lower bound on the central charge as a function of the parameter $\gamma$ defined in \eqref{JJOPET}. The grey line corresponds to not imposing any gap between the energy momentum tensor and the next spin-2 parity even operator. The blue line shows how a gap $(\Delta_2^+)'=3.5$ impacts the strength of the bound. While inside the interval $|12\gamma|\leq 1$ the bound is marginally affected, the effect outside the interval is dramatic.}
\label{Fig:CTvsGAMMAGap}
\end{figure}

Let us now discuss the role of the gap in the spin-2 parity even sector. 
The key observation is that the proof of the conformal collider bound \eqref{eq:HMbound} elegantly obtained in \cite{Hofman:2016awc} relies on the assumption of the existence of a \emph{single} energy momentum tensor.
If instead a CFT possesses  several conserved spin-2 operators, the bound \eqref{eq:HMbound} must be replaced by a bound on a weighted sum over the  corresponding $\gamma$'s:
\be
-\frac1{12}\leq \sum_{i }  w_i\gamma_i \leq \frac1{12}\,.
\ee
Unfortunately, in our bootstrap analysis with a finite truncation parameter $\Lambda$, any parity even spin-2 operator of dimension close to 3 is almost indistinguishable from another stress-tensor.  This is precisely the role played by the gap $(\Delta_2^+)'$ in \ref{eq:ctsdp}: imposing a single energy momentum tensor corresponds to input a gap strictly larger than 3. In figure \ref{Fig:CTvsGAMMAGap} we show the impact of this gap on the lower bound on the central charge of the theory. 
As expected, the effect is stronger in the region disallowed by the bound \eqref{eq:HMbound} because the imposed gap on the spin-2 sector implies uniqueness of the stress tensor.
On the other hand, imposing a gap like $(\Delta_2^+)'=3.5$ probably excludes most CFT's with global symmetry bigger than $U(1)$.
For example, consider the 
OPE of two conserved currents in the $O(3)$-model:  
\bea
J_\mu^i \times J_\nu^k \supset \delta^{ik}\lambda_{JJT} T_{\mu\nu} + \lambda_{JJO} O^{ik}_{\mu\nu} 
\eea
where the spin-2 operator $ O^{ik}_{\mu\nu}$ transforms in the symmetric traceless representation of $O(3)$. When we restrict to a unique current, for instance to $i=k=3$, the operator $O_{\mu\nu}^{33}$ is a singlet of the $U(1)$ generated by $J^3_\mu$ and we expect its dimension to be perturbatively close to the unitarity bound. A similar argument holds for all $O(N>2)$ models: generically there can be more than one spin-2 operator entering the $J\times J$ OPE, whose anomalous dimension is $1/N$ suppressed. 
We expect that to properly constraint these theories one has to bootstrap the four-point functions of full set of conserved currents.

A final comment regarding the comparison between our analysis and the case of bootstrapping the stress-tensor four-point function is in order. Since the 3 point function of three stress tensors is structurally different from the one of two stress tensor and a non conserved spin-2 operator, there is no contribution in the 4 point function that could fake a second energy momentum tensor. As a consequence, the uniqueness of $T_{\mu\nu}$ is automatic and in principle there is no need to impose a gap in the spin-2 even sector. 

\subsection{Central Charge bounds with spectrum assumptions}

In this section we investigate how the bounds on the central charge change when we introduce additional assumptions on the spectrum of scalar operators or in the spin-4 parity-even sector. We therefore replace the conditions \eqref{eq:ctsdp} with the following conditions
\be
\label{eq:ctsdpWithGaps}
\begin{array}{cl}
\alpha[ \vec{\tilde{\lambda}}_{JJT}(\gamma)^T \cdot \VT \cdot \vec{\tilde{\lambda}}_{JJT}(\gamma)]= 1,&\textrm{(normalization)}\\ 
\alpha[\Vo] \geq 0,&\Delta\geq \Delta_0^+ , \ell=0   \\ 
\alpha[ \Vp] \succeq 0,&\Delta\geq (\Delta_2^+)',\ell=2   \\ 
\alpha[ \Vp] \succeq 0,&\Delta\geq \Delta_4^+,\ell=4   \\
\alpha[\Vp ]\succeq 0,&\Delta \geq\ell+1,\ell>4, \,\ell \text{ even} \\
\alpha[ \Vm] \succeq 0,&\Delta \geq\Delta_0^-,\ell=0 \\ 
\alpha[ \Vm] \succeq 0,&\Delta \geq\ell+1,\ell\geq 2 \\ 
\end{array}
\ee

\begin{figure}[t]
\centering
\subfigure[]{\includegraphics[width=0.45\textwidth]{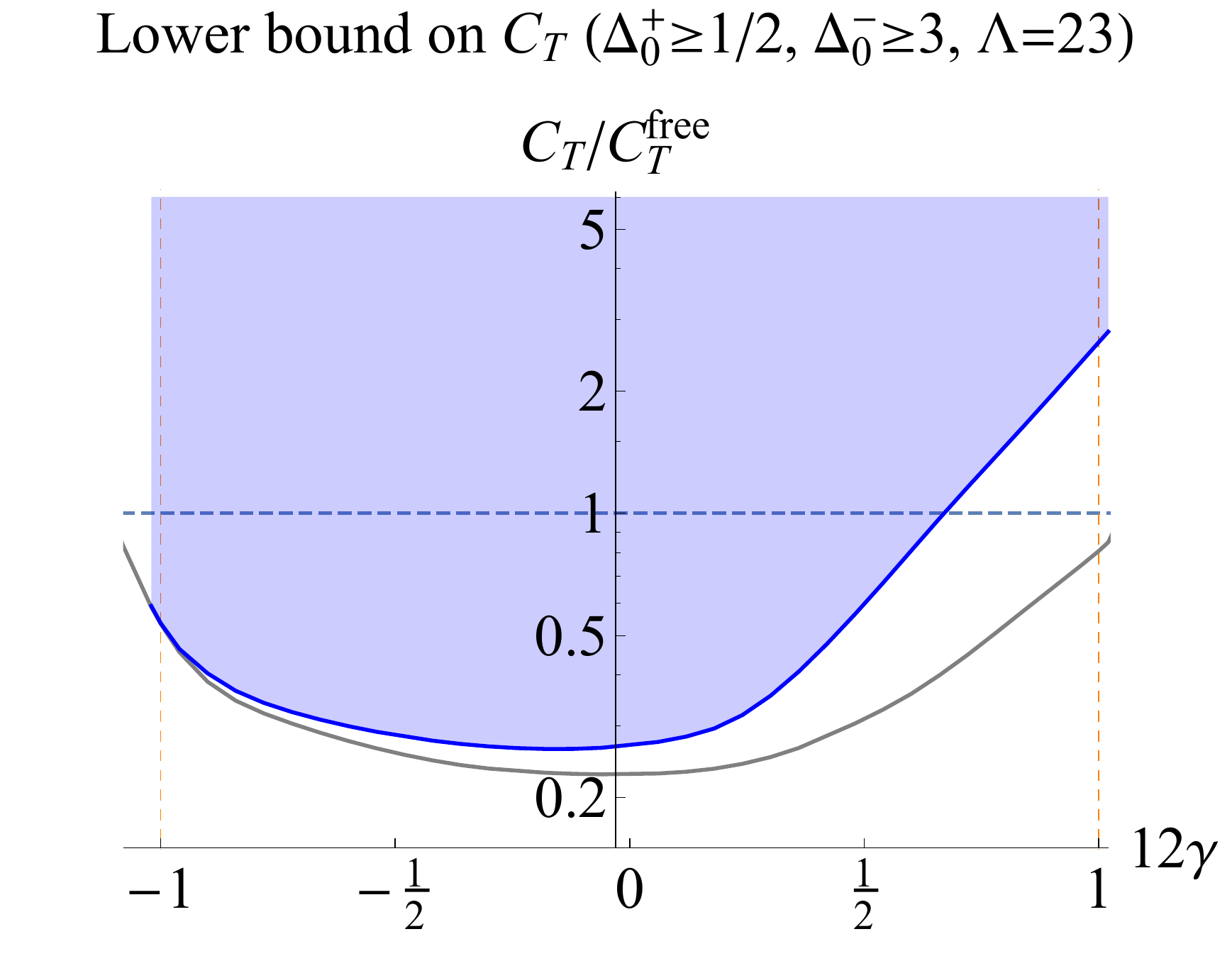}\label{Fig:noOdd}}
\subfigure[]{\includegraphics[width=0.45\textwidth]{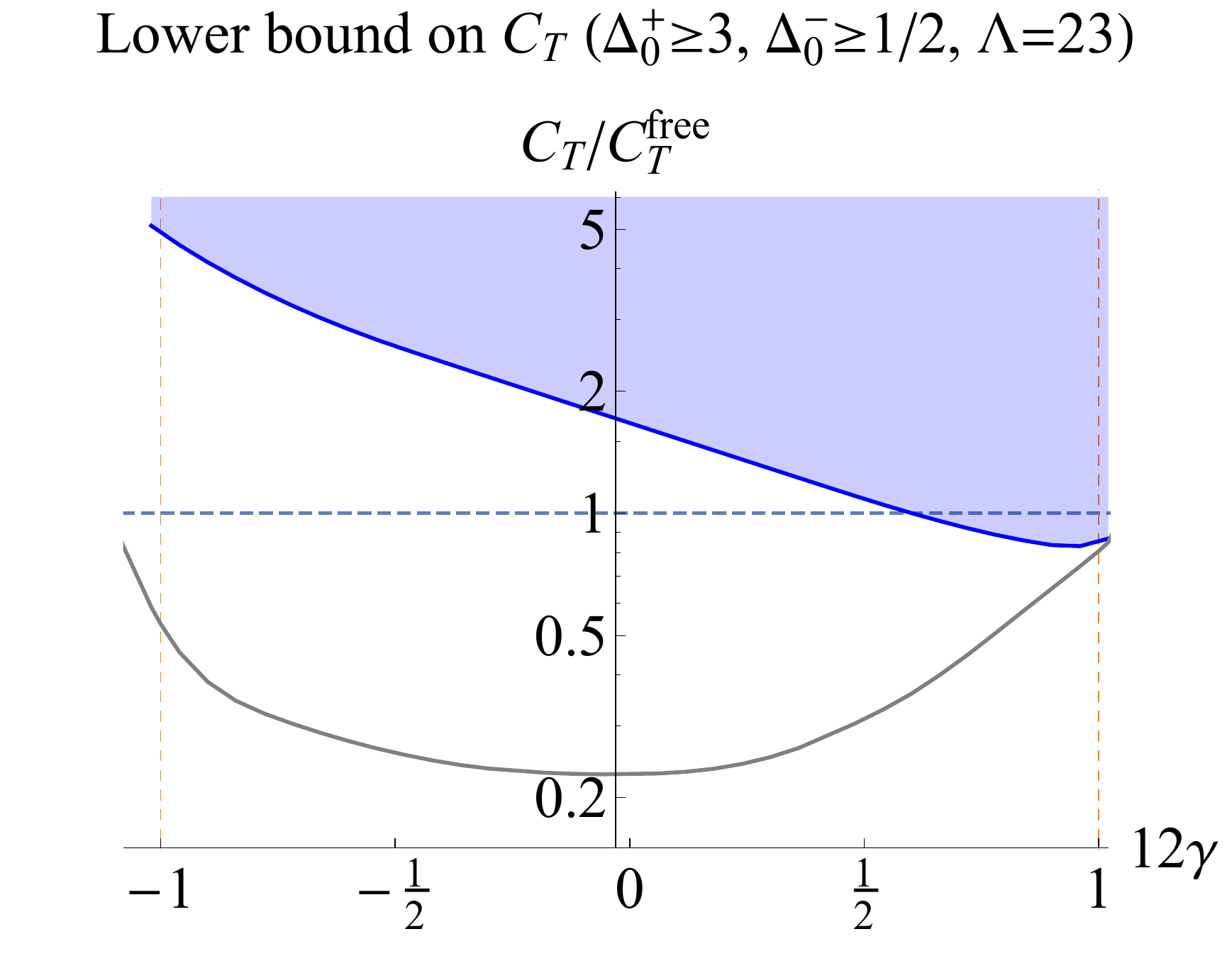}\label{Fig:noEven}}
\caption{a) Lower bound on the central charge normalized to the central charge of a free complex boson as a function of the parameter $\gamma$ defined in \eqref{JJOPET} assuming no relevant parity-odd scalars. b) Lower bound on the central charge assuming no relevant parity-even scalars. The shaded regions are excluded. The bounds have been computed at $\Lambda=23$. The grey lines on both plots correspond to the lower bound on $C_T/C_T^\text{free}$ without any assumption.}
\end{figure}

In figure \ref{Fig:noOdd} we show the impact of imposing the absence of relevant odd scalar operators in the $J\times J$ OPE. This amounts to set $\Delta_0^-=3$ while keeping all the other gaps to their minimal value consistent with unitarity. As expected, the bound on the central charge increases for positive values of $\gamma$, excluding the   free fermion theory, which is indeed ruled out by this assumption. Close to $\gamma=-1/12$ the bound  is   almost unaffected, consistent with the conjecture that the left part of the plot is driven by the free boson theory and possibly by the critical  $O(2)$-model. Notice that in this analysis we haven't made any assumption about the parity-even spectrum, and in particular no assumption about the number of relevant parity-even scalars.
A second investigation,  shown in figure \ref{Fig:noEven}, solely assumes that no relevant parity-even scalar operators are present. The impact of this assumption is more dramatic: very small room is left for theories with $C_T < C_T^\text{free}$. Although we haven't performed a careful extrapolation we believe this window will close in the limit of infinite number of derivatives $\Lambda\rightarrow\infty$.

\begin{figure}[t]
 \begin{center}
 \includegraphics[width=0.7\textwidth]{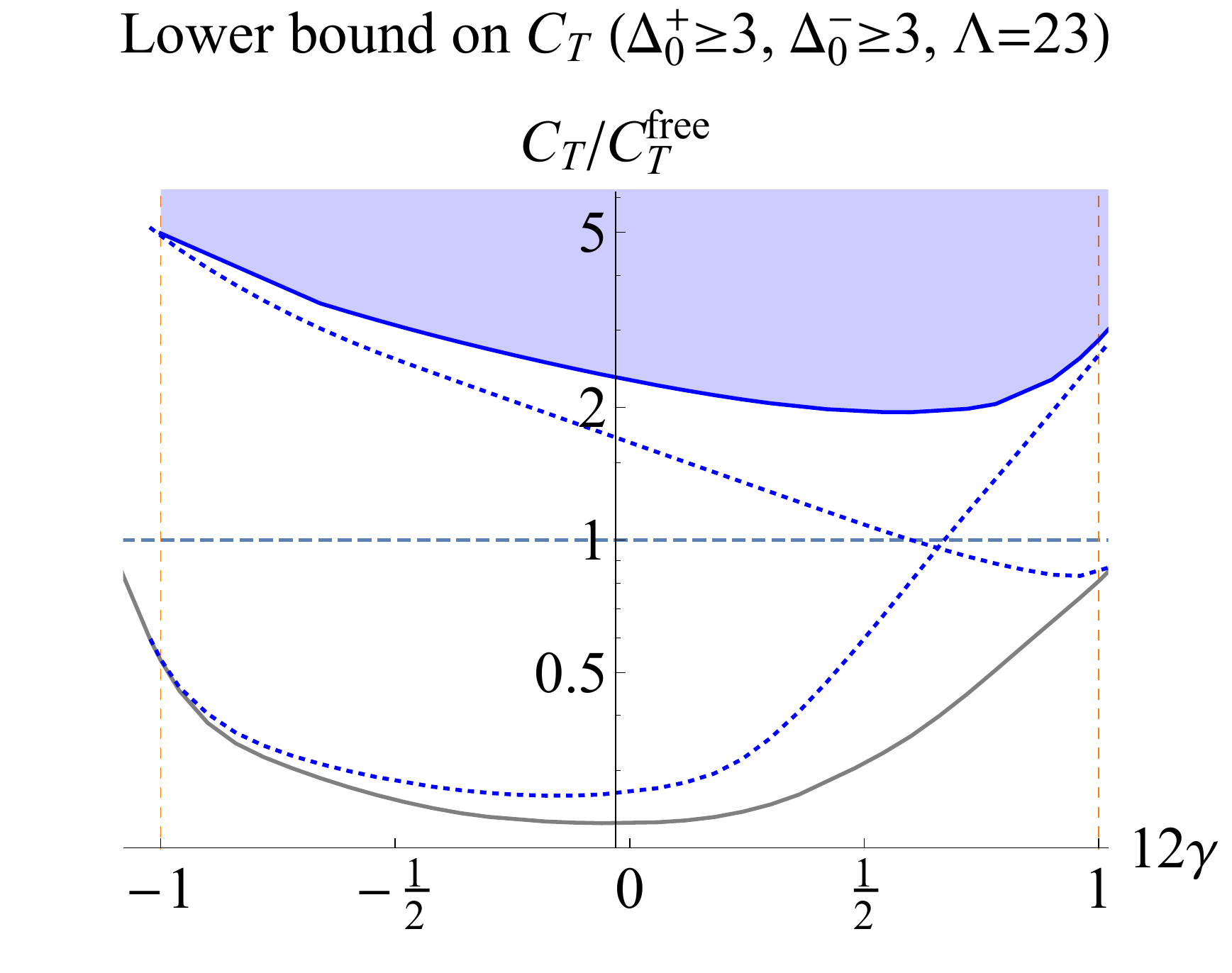}
 \end{center}
  \caption{Lower bound on the central charge normalized to the central charge of a free complex boson as a function of the parameter $\gamma$  assuming that both parity-even and parity-odd scalar operators are absent. The shaded region is allowed. The bounds have been computed at $\Lambda=19$. The dashed blue line corresponds to the bounds shown in figure \ref{Fig:noOdd} and figure \ref{Fig:noEven}. The gray line corresponds to the lower bound without any assumption. }
\label{Fig:noRelevant}
\end{figure}

Finally, in figure \ref{Fig:noRelevant} we combine both assumptions to study the central charge limits for the case of \emph{dead-end} CFTs, namely those CFTs without any relevant scalar operator. As the name suggests, these CFTs would be stable under any scalar deformation and therefore would represent an attractive point for all the renormalization group flows driven by rotation-preserving deformations. While we expect such CFTs with a large central charge (from weakly coupled abelian gauge theory in $\text{AdS}_4$), there are no known examples with small values of $C_T$. Interestingly, at present, our limits do not preclude the existence of dead-end CFTs with $C_T/C_T^\text{free}\sim O(1)$.

\begin{figure}[t]
\centering
\subfigure[]{\includegraphics[width=0.49\textwidth]{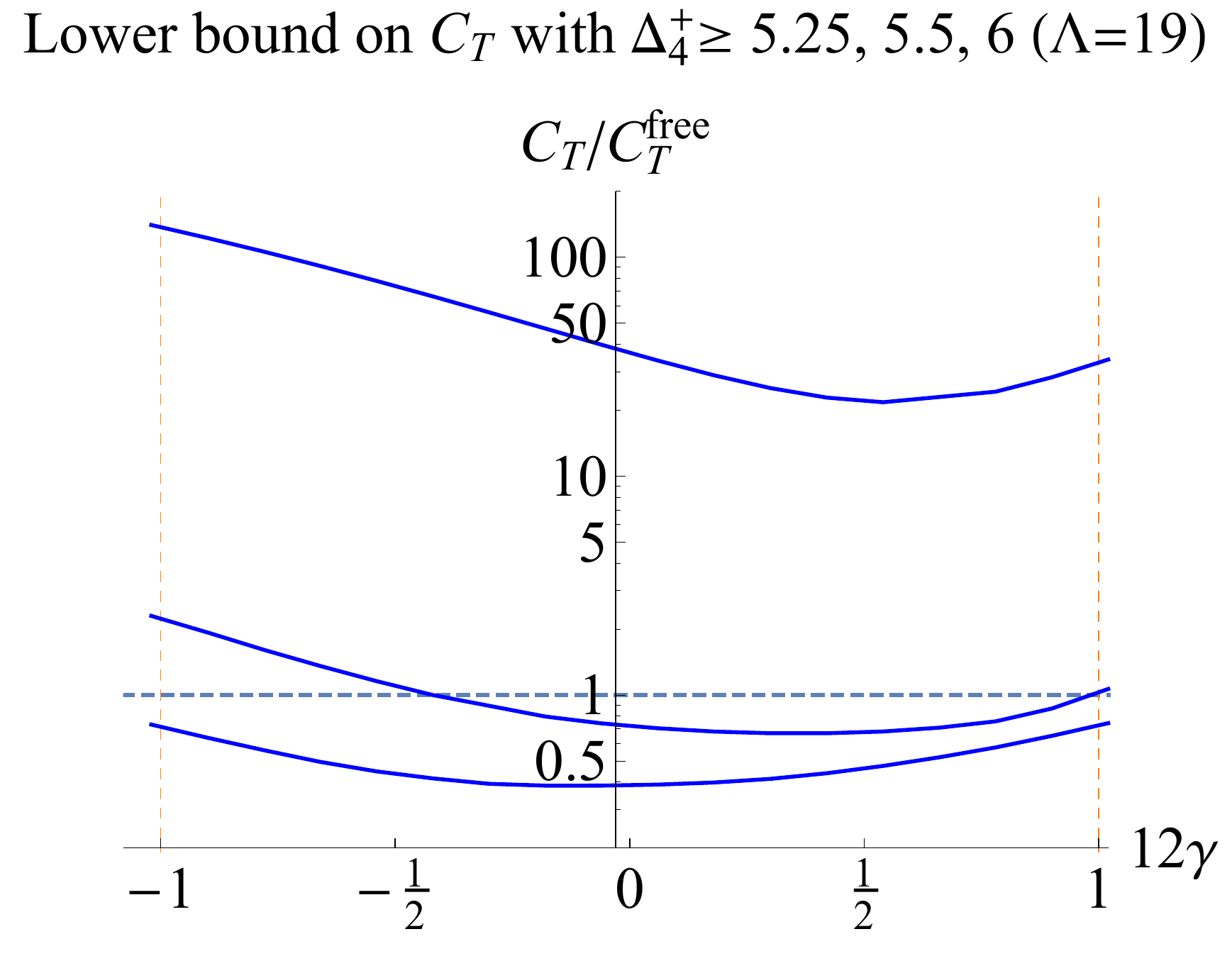}\label{Fig:gap4}}
\subfigure[]{\includegraphics[width=0.49\textwidth]{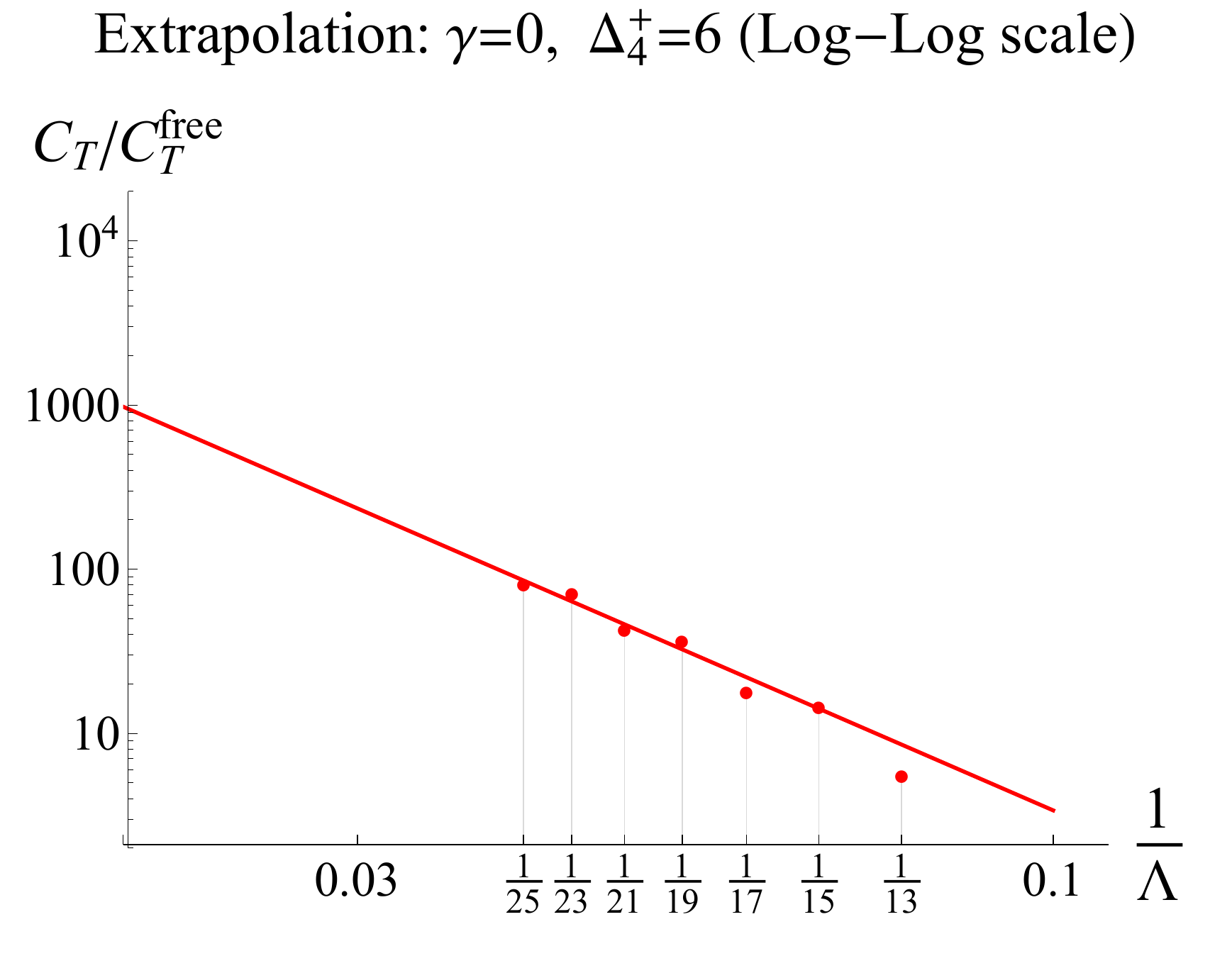}\label{Fig:gap4Extrapolation}}
\caption{a) Central charge bound as a function of $\gamma$ for several values of the gap in the spin-4 parity even sector, $\Delta_4^+=5.25,5.5,6$. As we increase the gap, solutions consistent with crossing symmetry must develop a larger central charge. b) Extrapolation of the bound on  $C_T$ for $\gamma=0,\Delta_4^+=6$. Both axis are in logarithmic scale. 
Linear fit shown as a solid red line suggests that the bound on $C_T$ diverges as $\Lambda \to \infty$.
}
\end{figure}

We now move to exploring the dependence of the central charge bound on the gap in the spin-4 parity-even sector. This can be done by tuning the parameter $\Delta_4^+$ in \eqref{eq:ctsdpWithGaps} while setting all other gaps to their minimal value consistent with unitarity. The value of the gap $\Delta_4^+$ can be considered as a knob to interpolate from free theories to holographic CFTs. Indeed, the $J\times J$ OPE in free CFTs contains a conserved spin-4 parity even operator. When going to interacting CFTs, its dimension must be lifted \cite{Maldacena:2011jn}  and the operator acquires a positive anomalous dimension. On the other hand, in holographic CFTs the lightest spin 4 operator is the ``double-trace" operator $\sim J_{(\mu_1}\partial_{\mu_2}\partial_{\mu_3}J_{\mu_4)}$ of dimension 
$6$, with corrections suppressed as $1/N$. 
As we increase the value of the gap, we exclude more and more theories, and it is natural to expect that the only solution still consistent with crossing symmetry are those which have a large central charge. This behavior is indeed realized in figure \ref{Fig:gap4}, where we show the lower bound on the central charge as a function of $\gamma$ for several values of $\Delta_4^+$. As anticipated, the bound grows with the gap. By increasing the numerical power one can presumably make the bound much stronger. In figure \ref{Fig:gap4Extrapolation} we performed an extrapolation in the number of derivatives of the central charge limit when $\Delta_4^+=6$ for the central value $\gamma=0$. 
The extrapolation suggests that $\Delta_4^+=6$ implies $C_T=\infty$, in agreement with the holographic interpretation.\footnote{Recall that the anomalous dimension $\Delta_4^+-6\sim 1/C_T$ must be  negative due to Nachtman's theorem \cite{Nachtmann:1973mr,Li:2015itl}.}

\subsection{Hunting the $O(2)$-model}

So far we have investigated bounds on the central charge under very general assumptions on the spectrum of CFTs. 
However, they do not appear to be saturated by any known CFT. The extrapolation in the number of derivatives shown in figure \ref{Fig:CTextrapolation} suggests that in this limit we can make contact with a known result, namely the free fermion theory.  On the other hand, theories such as the $O(2)$ model seem to remain in the bulk of the regions allowed by crossing symmetry. In oder to understand the reason for this  it is useful to inspect the solution of crossing along the boundary extracted with the extremal functional\footnote{We remind the reader that on the boundary of the allowed region the solution of the truncated crossing equation is unique and it is given by the zeros of the linear functional $\alpha$.} method introduced in \cite{ElShowk:2012hu} and successfully used in \cite{El-Showk:2014dwa,Simmons-Duffin:2016wlq} to extract the spectrum of the three dimensional Ising model.
We observe that all the extremal solutions contain odd operators with $\ell\geq2$ and dimension saturating the unitarity bound $\Delta_\ell^- = \ell+1$ (or very close to it). On the contrary, all known theories display a larger gap. For instance, free theories and GFVF satisfy $\Delta_\ell^- = \ell+3$ (see appendix \ref{sec:reminder}). %
Basically, the extra gap comes from the need to contract $\epsilon$-tensor indices with derivatives. 

 \begin{figure}[t]
 \begin{center}
 \subfigure[]{\includegraphics[width=0.45\textwidth]{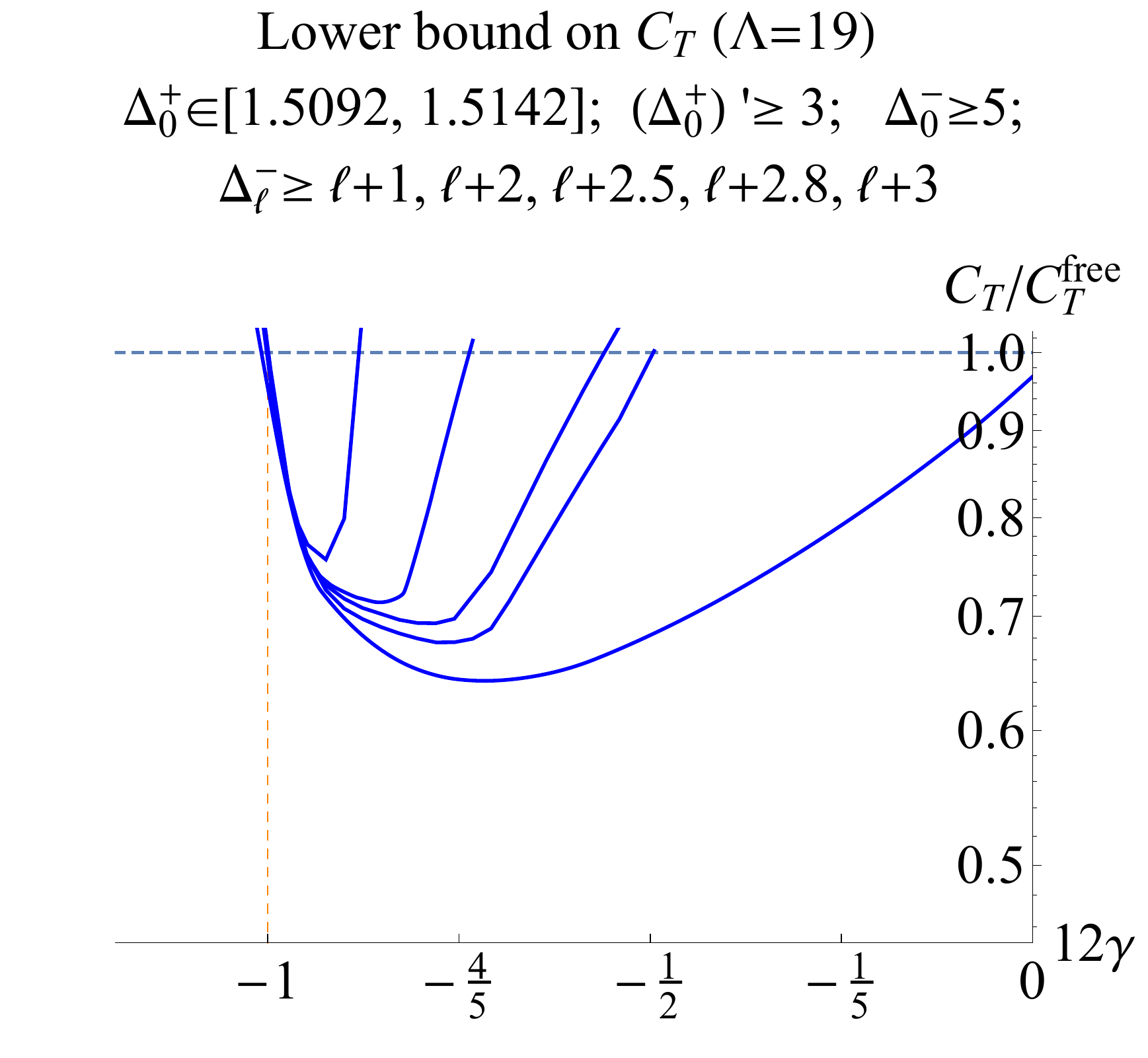}\label{Fig:noOddcurrent}}
\subfigure[]{\includegraphics[width=0.45\textwidth]{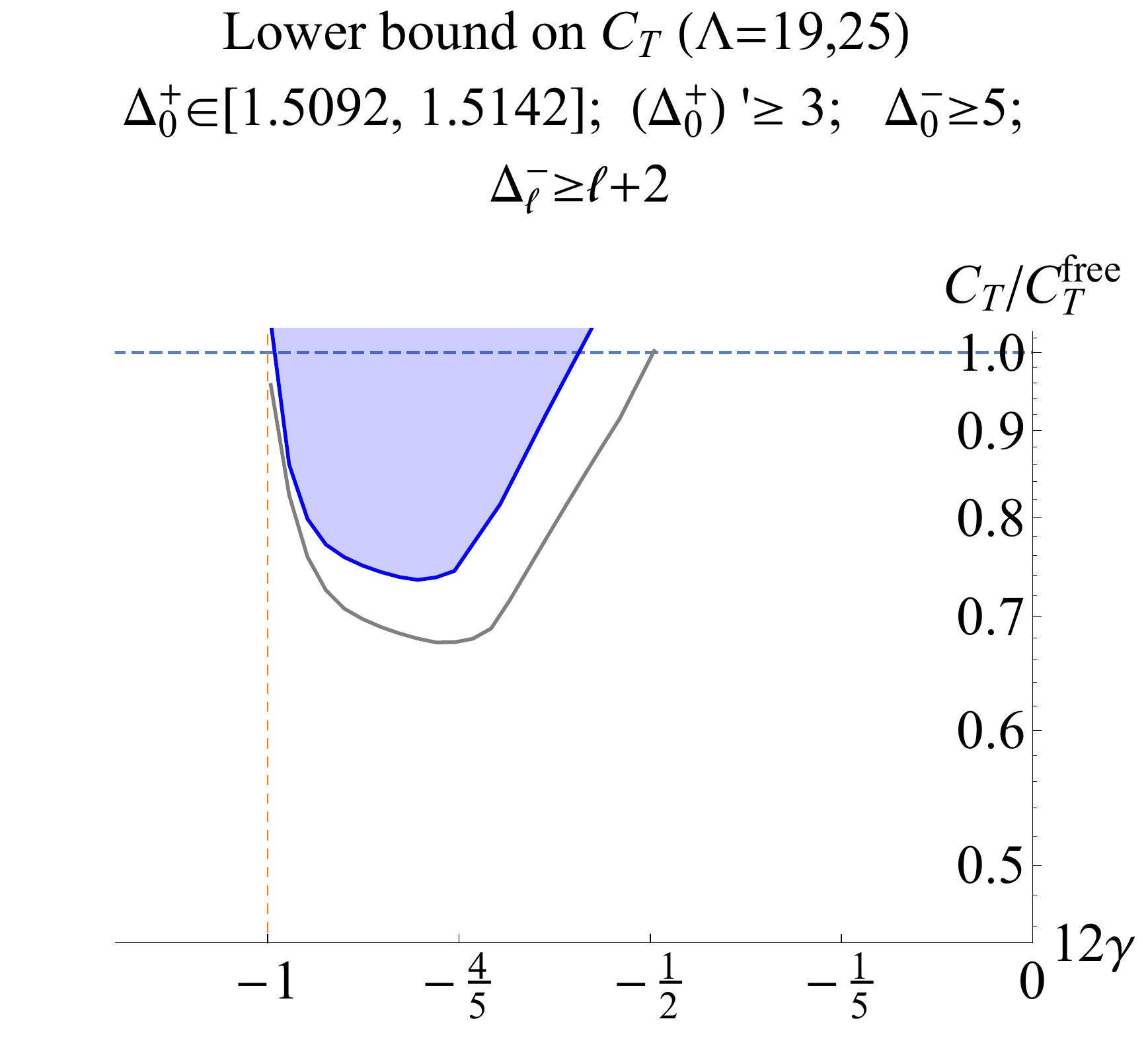}\label{Fig:noOddcurrentExtrapol}}
 \end{center}
  \caption{a) Lower bound on the central charge normalized to the central charge of a free complex boson as a function of the parameter $\gamma$. Different lines corresponds to increasing values of twist of the parity odd operators of spin $\ell\geq2$: $\tau_\text{all}^-=1,\,2,\,2.5,\,2.8,\,3$. b) Lower bounds on the central charge with $\tau_\text{all}^-=2$ for increasing numerical power: $\Lambda=19,23$. %
  }
\end{figure}

It is natural to expect that the $O(2)$ model  also displays an extra gap for all parity odd operators with spin $\ell \ge 2$. 
 Hence, in order to make contact with the $O(2)$ model, we   replace the conditions \eqref{eq:ctsdp} with the following requirements:
\be
\label{eq:ctsdpWithGaps}
\begin{array}{cl}
\alpha[ \vec{\tilde{\lambda}}_{JJT}(\gamma)^T \cdot \VT \cdot \vec{\tilde{\lambda}}_{JJT}(\gamma)]= 1,&\textrm{(normalization)}\\ 
\alpha[ \Vo] \geq 0,&\Delta\in [\Delta_\epsilon^\text{min},\Delta_\epsilon^\text{max}] (\ell=0)   \\ 
\alpha[ \Vo] \geq 0,&\Delta\geq \Delta_0^+ (\ell=0)   \\
\alpha[ \Vp] \succeq 0,&\Delta\geq \Delta_2^+,\ell=2   \\
\alpha[ \Vp] \succeq 0,&  \Delta\geq\ell+1,\ell>4, \,\ell \text{ even} \\
\alpha[ \Vm] \succeq 0,& \Delta\geq\Delta_0^-,\ell=0 \\ 
\alpha[ \Vm] \succeq 0,&  \Delta\geq\ell+\tau^-_\text{all},\ell\geq 2 \\ 
\end{array}
\ee
The novelty in the above conditions consists in raising the twist of all parity odd operators to $\tau^-_\text{all}\ge 1$, and imposing that relevant parity even scalar must be confined in a narrow interval $\Delta\in [\Delta_\epsilon^\text{min},\Delta_\epsilon^\text{max}]=[1.5092,1.5142]$, for which we take the rigorous bound from previous bootstrap studies \cite{Kos:2016ysd}. In figure \ref{Fig:noOddcurrent} we show the impact of varying $\tau^-_\text{all}$ from 1 to 3. Interestingly the bounds start developing more and more pronounced minima as we increase the value of $\tau^-_\text{all}$. In addition, the left part of the bound is insensitive to this parameter, while the right part heavily depends on it. Although from figure \ref{Fig:noOddcurrent} it would be tempting to set $\tau^-_\text{all}=3$, the  large spin analysis discussed below 
suggests that this is not possible. Nevertheless we expect that $\tau^-_\text{all}=2$ is a safe assumption. With this choice in \eqref{eq:ctsdpWithGaps}, we can obtain a rigorous bound on the parameter $\gamma$ for theories with the central charge smaller than the free theory one:
\be
\gamma \in [-0.0824,-0.0494] \qquad  \text{rigorous }\  (\text{assuming \ref{eq:ctsdpWithGaps},  }\tau^-_\text{all}=2)
\ee
The above interval has been computed at $\Lambda=25$, however, as shown in figure \ref{Fig:noOddcurrentExtrapol}, the bounds are still not converged. Using a linear extrapolation we estimate a bound
\be
\gamma \in [-0.081(1),-0.060(1)] \qquad \text{extrapolation } (\text{assuming \ref{eq:ctsdpWithGaps},  }\tau^-_\text{all}=2) 
\ee

Let us comment on the consistency of our assumption that the dimension of the leading twist parity odd operators of spin $\ell \ge 2$ in the $O(2)$ model is not too far from 3, which is the free theory value.\footnote{ Schematically, these operators have the form
$
\mathcal{O}_{\mu_1\dots \mu_\ell}^- \sim \epsilon_{\nu\rho(\mu_1}J^\nu \partial_{\mu_2}...\partial_{\mu_{\ell})} J^\rho   
$.}
The leading correction to the  dimension of these operators in the large spin expansion has been computed in \cite{Li:2015itl}  using analytic bootstrap techniques. It was found that:
\be
\delta_{\mathcal{O}_\ell^-}=\Delta_{\mathcal{O}_\ell^-}- \ell-3=-\frac{12(1-144\gamma^2)}{C_T \pi^4\ell }+\dots
 \label{eq:asymptoticDim} 
\ee
Notice that the leading correction in the above formula is negative whenever $\gamma$ satisfies the conformal collider bound. Moreover, our estimate $\gamma_{O(2)}\sim -1/12$ is compatible with the assumption of small anomalous dimension $\delta_{\mathcal{O}_\ell^-}$.

\section{Conclusions}
\label{sec:discussion}

In this work we have used the numerical conformal bootstrap to study the space of parity-even three dimensional conformal field theories with (at least) a global $U(1)$ symmetry. We did this by analyzing the four point function of identical conserved currents. We have shown that,  analogously to the case of the correlation function of 4 scalars or 4 fermions, unitarity and OPE associativity alone let us carve out the parameter space of CFTs. Inspecting the allowed values of scalar operator dimensions we found that any CFT with a conserved spin-1 current must contain both parity even and parity odd scalars. The boundary of the allowed region displays a non trivial structure with multiple features. In particular a kink appears close to the location of the $O(2)$ model, providing an upper bound on the dimension of the first parity odd scalar $\Delta_0^-\leq 7.65(1)$. A similar kink is present in the bound on the second spin-2 parity even operator. Also, we excluded the existence of dead-end CFTs with central charge smaller than twice the central charge of a free 3d Dirac fermion. We also explored bounds on the central charge with several assumptions on the CFT spectrum. In this case we observed a slower numerical converge. Nevertheless we found clear evidence of the conformal collider bounds for spin-1 currents \cite{arXiv:0803.1467,Buchel:2009sk,arXiv:1010.0443, arXiv:1210.5247}. 

The present work paves the way to many generalizations and extensions. Given the special role that the $O(2)$ model seems to play in our exclusion bounds, it is natural to expect that a mixed scalar-current bootstrap analysis will allow to precisely determine the spectrum of the theory \cite{inpreparationAVETMR}. Similarly, one could consider multiple correlators including external fermionic charged operators in order to narrow down the location of the $N=2$ Gross-Neveu model. 

As mentioned several times, the results of this work are very general and apply to CFTs with a continuous global symmetry that admits a local conserved current.\footnote{A trivial example of a theory with a global symmetry but no conserved current is a free complex field in $AdS_{d+1}$ with mass strictly larger than $-(d^2-4)/4$, which is the dual of a Generalized Free Field in $d$ spacetime-dimensions with  scaling dimension $\Delta>(d-2)/2$.} On the other hand,  by studying a single current inside a larger symmetry, we loose the ability to distinguish operators that are singlets under the entire global symmetry group from those that instead are only invariant under the specific $U(1)$ considered. As an example, spin-2 operators with dimension close to the unitarity bound but not singlet under the full global symmetry are difficult to distinguish  from the energy momentum tensor in the numerical analysis. As we have seen in Sec.~\ref{sec:centralcharge} this  dramatically affects the bound on the central charge. Hence, in order to obtain numerical evidences of the conformal collider bounds we restricted to theories with a finite gap between the  energy momentum tensor and the dimension of the next spin-2 operator.  While we expect this merely represents a technical assumption for theories with global $U(1)$ symmetry, it might not apply to CFTs with larger symmetry group. In this case, it will be important to bootstrap the full set of correlation functions $\langle J_{\mu_1}^{a_1}\,J_{\mu_2}^{a_2}\,J_{\mu_3}^{a_3}\,J_{\mu_4}^{a_4}\rangle$, with $a_i$ spanning all the generators of the global symmetry. This set up would also allow to specify the global symmetry  by inputting the group structure constants $f^{abc}$ and to put a bound on the current central charge $C_J$.
The analysis will require a minor modification of the present framework. All necessary  conformal blocks required for this analysis have been already computed in the current work. The main difference will be represented by the higher number of crossing conditions.

Throughout this work we considered parity-invariant CFTs; as a consequence, the present analysis does not apply to many Chern-Simons-matter theories. 
In order to include parity breaking effects one would need to extend our analysis in several ways. Because operators would not be classified according to their parity, all three point functions will have both parity even and parity odd tensor structures. Thus, new conformal blocks should be computed. The method explained in appendix~\ref{app:ConformalBlocks} allows to systematically perform this computation. In addition, the four point function will admit parity odd tensor structures as well. This will modify the form of conservation equations and crossing equations. 

Finally, the same investigation presented in this work can be extended to higher dimensions with minor modifications. The recurrence relation presented in appendix \ref{app:ConformalBlocks} could be generalized in order to build conformal blocks in any dimension. Alternatively, the fundamental results obtained in \cite{Elkhidir:2014woa,Echeverri:2015rwa,Echeverri:2016dun} allows us to compute the conformal blocks in four dimensions in closed form. Moreover, the analysis of the crossing equations in Sec.~\ref{sec:4pt} is valid in any spacetime dimension. This direction would be of particular interest in presence of ${\mathcal N}=1$ 4D supersymmetry. Indeed, the $U(1)$ $R$-symmetry current $\mathcal J_\mu$ is embedded in the Ferrara-Zumino supermultiplet, which also contains the energy momentum tensor as a super-descendant. The study of  $\mathcal J_\mu$ correlation functions will  provide a universal handle on all local SCFTs, allowing in principle to discover theories we currently know nothing  about \cite{inpreparationAVASAM}. 

This work represents a first exploration of an uncharted territory. Very much like 15th century navigators, we landed and explored the border of a whole new word. We created a first map of the landscape of CFTs with global symmetries which will serve as a roadmap for further investigations. We are confident that future expeditions will lead to finer understanding of this space.

\section*{Acknowledgments}
We would like to thank the organizers and participants of the workshops held in Yale in October 2016 and Princeton in March 2017 by the Simons Collaboration on ``Non-perturbative Bootstrap'' for hospitality and comments. We especially thank Simone Giombi,  Petr Kravchuk, Miguel  Paulos and Andreas Stergiou for useful discussions.
AV is supported by the Swiss National Science Foundation under grant no. PP00P2-163670. 
JP is supported by the National Centre of Competence in Research SwissMAP funded by the Swiss National Science Foundation and by the \emph{Simons collaboration on the Non-perturbative Bootstrap} funded by the Simons Foundation.
ET is supported by the Portuguese Funda\c{c}\~ao para a Ci\^encia e a
Tecnologia (FCT) through the fellowship SFRH/BD/51984/2012. ET is also partially supported by Perimeter Institute for Theoretical Physics.
 ET would like to thank FAPESP grant 2011/11973-4 for funding his
visit to ICTP-SAIFR where a part of this work was done. ET would also like to thank EPFL for hospitality.
The computations in this paper were run on the EPFL SCITAS cluster and on the CERN LXPLUS cluster.

\appendix

\section{Spectrum of Simple Theories}
\label{sec:reminder}

\subsection{Free Scalar Theories}
The simplest example of CFT in $3$ dimensions with $U(1)$ global symmetry is the theory of free complex scalar field $\varphi$. The central charge of this theory $C_T=C_T^{\rm free}$ was given in \eqref{eq:ctfree}. The global $U(1)$ current is given by the conventional expression $J_\mu=i\bar{\varphi}\partial_\mu \varphi- i\varphi\partial_\mu\bar{\varphi}$. The lightest parity even neutral scalar $\bar\varphi \varphi$ has dimension $\Delta_{0}^{+}=1$.
The lightest parity-odd scalar is more complicated. Normally, one can build a parity-odd scalar out of two vectors 
\bea
\label{odd}
\epsilon^{\mu\nu\lambda}J_\mu \partial_\nu J_\lambda\ ,
\eea
but in case of one complex field this combination vanishes. Hence the lightest parity-odd scalar has more derivatives 
\bea
i\epsilon_{\mu\nu\lambda} J_\mu (\partial_\rho \partial_\nu \varphi) (\partial_\rho \partial_\lambda \bar\varphi)
\label{PO7}
\eea
and is of dimension $\Delta_{0}^{-}=7$.

A complex field $\varphi$ can be decomposed into two real fields.  One can consider a more general case of $N$ free real fields $\varphi^i$. This theory has $O(N)$ global symmetry, $N(N-1)/2$ currents $J^{[ij]}_\mu=\varphi^{[i}\partial_\mu \varphi^{j]}$ and 
\bea
C_T={N\over 2}C_T^{\rm free}\ .
\eea
The lightest parity-even scalar $\varphi_i \varphi^i$ still has the same dimension $\Delta_{0}^{+}=1$.  When $N\ge 4$ one can combine two mutually commuting currents $J^{[ij]}_\mu$ and $J^{[kl]}_\nu$ with four distinct $i,j,k,l$ into a dimension $\Delta_{0}^{-}=5$ parity-odd scalar 
\bea
\label{PO5}
 O_{-}^{[ijkl]}=4\epsilon^{\mu\nu\rho} \phi^{[i} \partial_\mu \phi^j  \partial_\nu \phi^k   \partial_\rho \phi^{l]} =-\epsilon^{\mu\nu\rho}J^{[ij]}_\mu \partial_\nu J^{[kl]}_\rho\ .
\eea This operator is charged under full $O(N)$ but is neutral under some generators, including $J^{[ij]}_\mu$ and $J^{[kl]}_\mu$. Depending on the choice of the generator   $J_\mu=\omega_{ij}J^{ij}_\mu$ the OPE two identical $J_\mu$ will or will not include \eqref{PO5}. For example the OPE of two $J_\mu=J^{[12]}_\mu$ will remain the same as in the theory of one complex boson, with $\Delta_0^-=7$, while the OPE of two 
 $J_\mu=(J^{[12]}_\mu+J^{[34]}_\mu)/2^{1/2}$ will include \eqref{PO5}, leading to $\Delta_0^-=5$.

In the theory of a free complex boson, the four-point function of currents can be easily calculated explicitly. Using the symmetry properties of Table~\ref{4pstruct} and the crossing symmetry conditions \eqref{eq:crossingftilde} together with the definitions in  \eqref{eq:ftilde}, the vector of 43 structures in $d-$dimensions can be compactly written as:
\begin{align} 
f_s=&\left\{f_1,f_2,\hat{f}_1,f_4,f_5,f_5,f_7,f_8,f_9,
  \hat{f}_9,f_8,\hat{f}_5,\hat{f}_4,
  \hat{f}_7, \hat{f}_5,\hat{f}_5,
  \hat{f}_4,\hat{f}_7,   \hat{f}_5,f_8,f_9,  \hat{f}_9,f_8,f_7,f_5,f_5,\right.\nonumber\\
   &  \left. 
  f_4,f_{28},f_{29},f_{29},f_{31},
   \hat{f}_{29},\hat{f}_{28},f_{34},f_{29},
   \hat{f}_{29},f_{34},\hat{f}_{28},f_{29}, 
   \hat{f}_{31},\hat{f}_{29},\hat{f}_{29},f_{28}\right\}
   \label{eq:minimalF}
\end{align} 
where $ \hat{f}_{i}\equiv   u^{-1+d} v^{1-d} f_i(v,u)$. 
Finally, the 11 independent functions appearing in the above equation are:
\begin{align} 
\begin{array}{ll}
f_1= ((-2+d)^2 v^{-d/2} \left(2 u v^{d/2}+u^{d/2}
   \left(v+v^{d/2}\right)\right))/(2 u), &
  f_9= \frac{1}{2} (-2+d)^3
   u^{\frac{1}{2} (-1+d)} \sqrt{v} \\
 f_2= \frac{1}{2} (-2+d)^2
   u^{-1+\frac{d}{2}} v^{-d/2} \left(u v^{d/2}+u^{d/2} \left(v+2
   v^{d/2}\right)\right),&
    f_{28}= \frac{1}{2} (-2+d)^4
   u^{d/2} v^{1-\frac{d}{2}},\\
 f_4= \frac{1}{2} (-2+d)^3 u^{d/2}
   v^{\frac{1}{2}-\frac{d}{2}},&
    f_{29}= -\frac{1}{2} (-2+d)^4
   u^{\frac{1}{2} (-1+d)} v^{\frac{3}{2}-\frac{d}{2}},\\
 f_5= -\frac{1}{2} (-2+d)^3
   u^{\frac{1}{2} (-1+d)} v^{1-\frac{d}{2}},&
  f_{31}=
   \frac{1}{2} (-2+d)^4 u^{-1+\frac{d}{2}} v^{1-\frac{d}{2}}
   \left(v+v^{d/2}\right),\\
   f_7= \frac{1}{2} (-2+d)^3
   u^{-1+\frac{d}{2}} v^{\frac{1}{2}-\frac{d}{2}}
   \left(v+v^{d/2}\right),&
   f_{34}= \frac{1}{2} (-2+d)^4 u^{d/2}
   v^{1-\frac{d}{2}},\\
   f_8= 0\ .&
   \end{array}
      \label{freetheory}   
\end{align}
First few terms in the conformal block decomposition of \eqref{freetheory} are summarized in the table below. 
In particular it shows that the second lightest parity-odd scalar 
\bea
\bar\varphi \varphi\, \partial_\mu\bar\varphi  \partial_\mu\varphi
\eea
appearing in the OPE of two currents
has dimension $(\Delta_0^+)'=4$. This is because dimension $3$ operator $\partial_\mu\bar\varphi  \partial_\mu\varphi$ is not a primary.  Similarly, second lightest spin-2 operator $ \bar\varphi \varphi T_{\mu\nu}$
also has dimension $(\Delta_2^+)'=4$.
\bea
\setstretch{1.25}
\begin{array}{|c|c|ccc|}
 \hline
 \Delta & \ell &(\tilde{\l}_{JJ\mathcal{O}^+}^{(1)} )^2 &
 \ \  \tilde{\l}_{JJ\mathcal{O}^+}^{(1)} \tilde{\l}_{JJ\mathcal{O}^+}^{(2)} \ \ 
 & (\tilde{\l}_{JJ\mathcal{O}^+}^{(2)} )^2\\
 \hline
1 & 0 & 4 & 0 & 0 \\
 4 & 0 & 4 & 0 & 0 \\
 6 & 0 & \frac{256}{105} & 0 & 0 \\
 8 & 0 & \frac{48}{77} & 0 & 0 \\
 10 & 0 & \frac{131072}{75075} & 0 & 0 \\
 \hline
 3 & 2 & 12 & 72 & 432 \\
 4 & 2 & 0 & 0 & 768 \\
 6 & 2 & \frac{1112}{245} & -\frac{1984}{245} & \frac{117248}{245} \\
 8 & 2 & \frac{963584}{72765} & -\frac{2023424}{24255} & \frac{6815744}{8085} \\
 10 & 2 & \frac{1196480}{231231} & -\frac{3650560}{231231} & \frac{197918720}{231231} \\
 \hline
 5 & 4 & 35 & 980 & 27440 \\
 6 & 4 & 0 & 0 & \frac{442368}{7} \\
 8 & 4 & \frac{171392}{7623} & -\frac{3384320}{7623} & \frac{144711680}{7623} \\
 10 & 4 & \frac{132841472}{2147145} & -\frac{20922368}{17745} & \frac{71189430272}{2147145} \\
 \hline
 7 & 6 & \frac{231}{2} & 7623 & 503118 \\
 8 & 6 & 0 & 0 & \frac{14376960}{11} \\
 10 & 6 & \frac{6208}{55} & -\frac{3861504}{715} & \frac{263061504}{715} \\
 \hline
 9 & 8 & \frac{6435}{16} & \frac{96525}{2} & 5791500 \\
 10 & 8 & 0 & 0 & \frac{2424963072}{143} \\
 \hline
 \end{array}
\qquad 
\begin{array}{|c|c|c|}
\hline
 \Delta & \ell &  (\tilde{\l}_{JJ\mathcal{O}^-} )^2 \\
 \hline
7 & 0 & \frac{768}{5} \\
 9 & 0 & \frac{8192}{91} \\
 \hline
 5 & 2 & 768 \\
 7 & 2 & \frac{18432}{35} \\
 9 & 2 & \frac{7577600}{7007} \\
 \hline
 6 & 3 & \frac{3072}{35} \\
 8 & 3 & \frac{4096}{11} \\
 10 & 3 & \frac{507904}{1573} \\
 \hline
 7 & 4 & \frac{9728}{3} \\
 9 & 4 & \frac{41656320}{11011} \\
 \hline
 8 & 5 & \frac{737280}{539} \\
 10 & 5 & \frac{6356992}{1859} \\
 \hline
 9 & 6 & \frac{2301952}{143} \\
 \hline
 10 & 7 & \frac{214810624}{19305} \\
\hline
\end{array}
\eea 
The OPE coefficients $\tilde{c}_{JJ\mathcal O^\pm}$ in the above tables are defined in appendix \ref{app:conservedequal}.

\subsection{Critical $O(N)$ Models}
The spectrum of critical $O(N)$ models at large $N$ is in many ways similar to that one of free theories. Including leading $1/N$ corrections the central charge is given by \cite{Petkou:1994ad,Petkou:1995vu} (also see \cite{Kos:2013tga} for further references)
\bea
C_T={N\over 2}C_T^{\rm free}\left(1-{40\over 9\pi^2 N}+\dots\right)\ .
\eea 
The main difference is the dimension of the lightest parity even scalar. 
At large $N$ its dimension approaches $2$, 
\bea
\Delta_0^+=2-{32\over 3\pi^2N}+\dots
\eea
The dimensions of the parity-odd scalars are less studied. For $N=2,3$, the lightest parity odd scalar appearing in the OPE of two currents has the same quantum numbers as \eqref{PO7} and is expected to have dimension $\Delta_0^-\approx 7$. For $N\ge 4$ there is also a parity-odd operator in the representation of \eqref{PO5}. Thus for some generators $J_\mu=\omega_{ij}J^{ij}_\mu$ we still expect $\Delta_0^-\approx 7$ while for others $\Delta_0^-\approx 5$.

For small $N=2,3,\dots$ certain dimensions and central charge are known with a good precision from the conformal bootstrap and Monte-Carlo simulations \cite{Calabrese:2002bm,Kos:2013tga,Kos:2015mba,Kos:2016ysd}. We report some of them in the table below: 
\begin{center}
\vspace{-1.cm}
\begin{eqnarray}
\begin{tabular}{| l | c | c| c |}
\hline
  $N$ & 2 & 3 & 4 \\
\hline
  $C_T/C_T^{\rm free}$ &  0.944 &  1.416 & 1.892 \\
\hline
  $\Delta_\epsilon$ &  1.51124(22) &  1.5939(10) & 1.6649(35) \\
\hline
  $\Delta_{\epsilon'}$ &  3.795(9) & 3.782(12) & 3.774(12)  \\
  \hline
  $\Delta_T$ &  1.237(4) & 1.211(3) & 1.189(2)  \\
\hline
\end{tabular}
\end{eqnarray}
\end{center}
Here $\epsilon$,$\epsilon'$ are the first and second singlet scalar operators appearing in the OPE $\phi_a\times\phi_b$, while $T_{ab}$ is the leading scalar operator transforming in the tensor traceless representation of $O(N)$. Under a given $U(1)\subset O(N)$, $T_{ab}$ decomposes into neutral and charged components. The neutral ones are allowed to enter the OPE of the conserved current associated with the $U(1)$. This means that 
\begin{align}
&\Delta_0^+ = \Delta_\epsilon, \qquad (\Delta_0^+)' = \Delta_{\epsilon'} \qquad \text{for } O(2)\nonumber\\
&\Delta_0^+ = \Delta_T, \qquad (\Delta_0^+)' = \Delta_{\epsilon} \qquad \text{for } O(N\geq 3)\nonumber
\end{align}

\subsection{Free Fermion Theories}
In three dimensions, a free Dirac fermion $\psi$ is invariant under a global $U(1)$ symmetry. This theory has the same central charge as a free complex scalar, $C_T=C_T^{\rm free}$. The lightest parity-odd scalar $\bar\psi \psi$ has dimension $\Delta_0^-=2$, while lightest parity-even scalar $(\bar\psi \psi)^2$ has dimension  $\Delta_0^+=4$. The four point function of the conserved current $J_\mu=\bar \psi \sigma_\mu \psi$ can be easily calculated explicitly. The four point function contains two distinct contributions $f_s = f_s^\text{disc} - f_s^\text{con}/\Upsilon$
where the disconnected piece $ f_s^\text{disc}$ is given by  \eqref{disconnected},
while the connected one is given below. Also, $\Upsilon=\Tr \mathds{1}$ denotes the trace of the identity in $\gamma$-matrix algebra in $d-$dimensions, $\mathds{1}=(\gamma^1)^2$ ($\Upsilon=2$ in 3 dimensions).  Following the same conventions as in \eqref{eq:minimalF},  we have:
\begin{equation}
\begin{array}{l}
f^\text{con}_1= u^{-1+\frac{d}{2}} v^{-d/2}
   \left(-u^{1+\frac{d}{2}}+u^{d/2} (1+v)+(-1+v)
   \left(-1+v^{d/2}\right)-u \left(1+v^{d/2}\right)\right),\\
  f^\text{con}_2= u^{-1+\frac{d}{2}} v^{-d/2} \left(u+u^{1+\frac{d}{2}}-u
   v^{d/2}-u^{d/2} (1+v)+(-1+v) \left(1+v^{d/2}\right)\right),\\
   f^\text{con}_4= -2
   u^{d/2} v^{\frac{1}{2}-\frac{d}{2}}
   \left(-1+u^{d/2}+v^{d/2}\right),\\
   f^\text{con}_5=
   2 u^{\frac{1}{2} (-1+d)}
   v^{1-\frac{d}{2}} \left(-1+u^{d/2}+v^{d/2}\right),\\
   f^\text{con}_7= -2
   u^{-1+\frac{d}{2}} v^{\frac{1}{2}-\frac{d}{2}} \left(u^{d/2}
   (1+v)+(-1+v) \left(-1+v^{d/2}\right)\right),\\
   f^\text{con}_8= 0,\\
   f^\text{con}_9=-2
   u^{\frac{1}{2} (-1+d)} v^{\frac{1}{2}-\frac{d}{2}}
   \left(1+u^{d/2}-v^{d/2}\right),\\
   f^\text{con}_{28}=4 u^{d/2} v^{1-\frac{d}{2}}
   \left(-1+u^{d/2}+v^{d/2}\right),\\
   f^\text{con}_{29}= -4 u^{\frac{1}{2} (-1+d)}
   v^{\frac{3}{2}-\frac{d}{2}} \left(-1+u^{d/2}+v^{d/2}\right),\\
   f^\text{con}_{31}=
  4 u^{-1+\frac{d}{2}} v^{1-\frac{d}{2}} \left(u^{d/2} (1+v)+(-1+v)
   \left(-1+v^{d/2}\right)\right),\\
   f^\text{con}_{34}=4 u^{d/2} v^{1-\frac{d}{2}}
   \left(-1+u^{d/2}+v^{d/2}\right)\ .
\end{array}
\label{eq:freetheoryF}
\end{equation}
A first few terms in the conformal block decomposition of \eqref{eq:freetheoryF} are summarized in the table below. 
\bea
\setstretch{1.25}
\begin{array}{|c|c|ccc|}
 \hline
 \Delta & \ell & (\tilde{\l}_{JJ\mathcal{O}^+}^{(1)} )^2 &
 \ \  \tilde{\l}_{JJ\mathcal{O}^+}^{(1)} \tilde{\l}_{JJ\mathcal{O}^+}^{(2)} \ \ 
 & (\tilde{\l}_{JJ\mathcal{O}^+}^{(2)} )^2\\
 \hline
 4 & 0 & 4 & 0 & 0 \\
 6 & 0 & \frac{256}{105} & 0 & 0 \\
 8 & 0 & \frac{48}{77} & 0 & 0 \\
 10 & 0 & \frac{131072}{75075} & 0 & 0 \\
\hline
 3 & 2 & 0 & 0 & 192 \\
 6 & 2 & \frac{296}{35} & -\frac{2752}{35} & \frac{34304}{35} \\
 8 & 2 & \frac{676864}{72765} & -\frac{446464}{24255} & \frac{3948544}{8085} \\
 10 & 2 & \frac{182080}{21021} & -\frac{1602560}{21021} & \frac{24248320}{21021} \\
 \hline
 5 & 4 & 0 & 0 & 2240 \\
 6 & 4 & 0 & 0 & 36864 \\
 8 & 4 & \frac{278912}{7623} & -\frac{7470080}{7623} & \frac{247930880}{7623} \\
 10 & 4 & \frac{98435072}{2147145} & -\frac{1516617728}{2147145} & \frac{54330294272}{2147145} \\
 \hline
 7 & 6 & 0 & 0 & 16632 \\
 8 & 6 & 0 & 0 & \frac{10076160}{11} \\
 10 & 6 & \frac{116544}{715} & -\frac{6226944}{715} & \frac{386924544}{715} \\
 \hline
 9 & 8 & 0 & 0 & 102960 \\
 10 & 8 & 0 & 0 & \frac{167903232}{13} \\
\hline
 \end{array}
\qquad
\begin{array}{|c|c|c|}
\hline
 \Delta & \ell & (\tilde{\l}_{JJ\mathcal{O}^-} )^2 \\
 \hline
 2 & 0 & 32 \\
 7 & 0 & \frac{768}{5} \\
 9 & 0 & \frac{8192}{91} \\
 \hline
 5 & 2 & 768 \\
 7 & 2 & \frac{18432}{35} \\
 9 & 2 & \frac{7577600}{7007} \\
 \hline
 6 & 3 & \frac{3072}{35} \\
 8 & 3 & \frac{4096}{11} \\
 10 & 3 & \frac{507904}{1573} \\
 \hline
 7 & 4 & \frac{9728}{3} \\
 9 & 4 & \frac{41656320}{11011} \\
 \hline
 8 & 5 & \frac{737280}{539} \\
 10 & 5 & \frac{6356992}{1859} \\
 \hline
 9 & 6 & \frac{2301952}{143} \\
 \hline
 10 & 7 & \frac{214810624}{19305} \\
\hline
 \end{array}
\eea
The OPE coefficients $\tilde{c}_{JJ\mathcal O^\pm}$ in the above tables are defined in appendix \ref{app:conservedequal}.

\subsection{QED$_3$}
A theory of $N_f$ Dirac fermions $\psi^i$ in 3d coupled to a $U(1)$ gauge field $A_\mu$  flows to a non-trivial IR fixed point if  $N_f$ is sufficiently large. This theory has global $SU(N_f)$ flavor symmetry, with the currents $J^{a}$, with $a=1,2,\dots,N_F^2-1$. Flavor symmetry might be spontaneously broken for small $N_f$ by chiral condensate. Besides this there is a topological $U(1)$, with the topological current  $J^{\rm top} \propto \star F$. The operators charged under this $U(1)$ are monopole operators.
When $N_f$ is odd the theory is not parity-invariant \cite{Borokhov:2002ib}. Accordingly we consider only even $N_f$ such that the effective number of Majorana fermions $N=2N_f$ is a multiple of four. 
For large $N$, the central charge is given by \cite{Giombi:2016fct}
\bea
C_T={N\over 2}C_T^{\rm free} \left(1+{4192-360 \pi^2\over 45\pi^2 N}+\dots\right)
\eea 
For minimal possible value $N=4$ this gives $C_T\approx 2.72 C_T^{\rm free}$. This result is valid only if there  is no spontaneous symmetry breaking.  

Identifying the lightest parity even and odd scalars appearing in the OPE of two currents requires consideration. Since monopole operators are charged under topological $U(1)$ they are excluded from the OPE of both $J^{\rm top}\times J^{\rm top}$ and $J^a\times J^a$.  First, we consider OPE of two $J^{\rm top}$ which contains only $SU(N_f)$ singlets. For large $N$, the lightest parity-odd singlet scalar $\bar\psi_i \psi^i$ has dimension \cite{Chester:2016ref, Chester:2016wrc}
\bea
\Delta_0^-=\Delta_0=2+{256\over 3\pi^2 N}+\dots
\eea
while lightest parity-even scalar is a combination of $(\bar\psi_i \psi^i)^2$ and $F_{\mu\nu}^2$ of dimension
\bea
\Delta_0^+=\Delta'_{0,-}
=4+{128(2-\sqrt{7})\over 3\pi^2 N}+\dots
\eea

The OPE of two flavor currents include all fields charged in representations appearing in the product of two adjoints. In this case the lightest parity-odd scalar is in adjoint representation of $SU(N_f)$, $(O_{n=1})^i_j=\psi^i\bar\psi_j$. At leading order it has dimension \cite{Chester:2016ref, Chester:2016wrc}
\bea
\Delta_0^-=\Delta_1=2-{128\over 3\pi^2 N}+\dots
\eea
which is smaller than $\Delta_0$. Similarly, the lightest parity-even operator is $(O_{n=2})^{[ij]}_{[kl]}=\psi^i \psi^j \bar\psi_k \bar\psi_l$ of dimension
\bea
\Delta_0^+=\Delta_2=4-{128\over \pi^2 N}+\dots
\eea
which is smaller than 
$\Delta'_{0,-}=4+{128(2-\sqrt{7})\over 3\pi^2 N}+\dots$ and the dimension of the adjoint operator $O_{1,-}'$ made out of four $\psi$'s, $\Delta'_{1,-}=4+{16(25-\sqrt{2317})\over 3\pi^2 N}+\dots$. 

We see that in both cases at large $N$, $\Delta_0^+\approx 4$ and $\Delta_0^-\approx 2$, and from this point of view QED$_3$ is similar to the free fermion theory.

\subsection{Gross-Neveu Models}
One Dirac spinor can be decomposed into two Majorana spinors. A theory of $N\ge 2$ free Majorana fermions has $O(N)$ symmetry, while the dimension of lightest parity even and odd scalars remain the same for all $N$. Upon adding a parity-odd scalar field $\phi$ with quartic interaction which couples to Majorana fermions via Yukava coupling $\bar\psi_i \phi \psi^i$, the theory flows into an interacting fixed point characterized by $O(N)$ symmetry. %
The lowest parity-odd scalar $\phi$ has dimension \cite{Gracey:1992cp,Derkachov:1993uw,Gracey:1993kc} (also see \cite{Iliesiu:2015qra} for further references)
\bea
\Delta_0^-=1-{32\over 3\pi^2 N}+\dots
\eea
The lightest parity-even scalar $\phi^2$ has dimension
\bea
\Delta_0^+=2+{32\over 3\pi^2 N}+\dots
\eea
while the central charge is given by \cite{Diab:2016spb}
\bea
\label{CGN}
C_T={N\over 2}C_T^{\rm free}\left(1+{8\over 9\pi^2 N}+\dots\right) \ .
\eea
Below we compare $C_T$ and $\Delta_0^+,\Delta_0^-$ for small $N$
found using leading $1/N$ expansion and Pade extrapolation of $\epsilon$-expansion in $d=2,4$ \cite{Diab:2016spb,Fei:2016sgs}\footnote{To calculate $C_T$ for small $N$ we follow Pade approximation procedures developed in \cite{Diab:2016spb}. Namely we employ ${\rm Pade}_{[4,1]}$ or ${\rm Pade}_{[1,4]}$  choosing the one which has no poles in the interval $2<d<4$.  Namely ${\rm Pade}_{[1,4]}$ for $N=3,4$ and  ${\rm Pade}_{[4,1]}$ for $N=2$.} and bootstrap techniques \cite{Poland3DON}.

\begin{center}
\begin{eqnarray}
\begin{tabular}{| l | c | c| c |}
\hline
  $N$ & 2 & 3 & 4 \\
\hline
\hline
  $N/2+4/(9\pi^2)$ & 1.045 &  1.545 & 2.045 \\
\hline
  $C_T/C_T^{\rm free}$\ \, Pade &  1.190 &  1.486 & 2.029 \\
\hline
\hline
  $1-32/(3\pi^2 N)$  &  0.46 &  0.64 & 0.73 \\
\hline
  $\Delta_0^-$\ \ \ \, Pade &  0.656 &  0.688 & 0.753 \\
\hline
$\Delta_0^-$\ \ \ \, Bootstrap &  0.660 &  0.724 & 0.772 \\
\hline
\hline
 $2+32/(3\pi^2 N)$  &  2.54 &  2.36 & 2.27 \\
\hline
  $\Delta_0^+$\ \ \ \, Pade &  1.75 &  2.285 & 2.148 \\
\hline
  $\Delta_0^+$\ \ \ \, Bootstrap &  2.14 &  2.17 & 2.25 \\
\hline
\end{tabular}
\end{eqnarray}
\end{center}
It is worth noting that, similar to critical bosonic $O(N)$ theories,  central charge of Gross-Neveu models even for small $N$ is substantially close to the free theory counterpart.

\subsection{Generalized Free Vector Field}
The generalized free vector field (GFVF) is a theory of a conserved   current $J_\mu$ (of dimension $d-1$) with the standard two-point function and all higher-point correlation functions satisfying Wick theorem. In particular the four-point function of currents $\langle JJJJ \rangle$  includes only the disconnected piece (all other components are zero), 
\bea
\label{disconnected}
f_1=1\ ,\quad f_2=u^{d-1}\ ,\quad f_3=u^{d-1} v^{1 - d}\ .
\eea
This theory contains no stress-energy tensor, i.e.~$C_T^{-1}=0$. The only operators present in the spectrum are those build of $J_\mu$. In particular the lightest parity even scalar $J^\mu J^\mu$ has dimension $\Delta_0^+=4$ and parity-odd scalar given by \eqref{odd}  has dimension $\Delta_0^-=5$. 
GFVF is dual to $U(1)$ gauge theory in $AdS_4$ in the limit of zero Newton constant, when only disconnected Witten diagrams contribute. 
In the table below, we list some OPE coefficients that we obtained from the conformal block expansion of $\langle JJJJ \rangle$ in $d=3$ dimensions.
\bea
\setstretch{1.25}
\begin{array}{|c|c|ccc|}
 \hline
 \Delta & \ell &  (\tilde{\l}_{JJ\mathcal{O}^+}^{(1)} )^2 &
 \ \  \tilde{\l}_{JJ\mathcal{O}^+}^{(1)} \tilde{\l}_{JJ\mathcal{O}^+}^{(2)} \ \ 
 & (\tilde{\l}_{JJ\mathcal{O}^+}^{(2)} )^2\\
  \hline
4 & 0 & \frac{8}{3} & 0 & 0 \\
 6 & 0 & \frac{512}{315} & 0 & 0 \\
 8 & 0 & \frac{512}{385} & 0 & 0 \\
 10 & 0 & \frac{262144}{225225} & 0 & 0 \\
\hline
 4 & 2 & 0 & 0 & 512 \\
 6 & 2 & \frac{2048}{245} & -\frac{11776}{245} & \frac{161792}{245} \\
 8 & 2 & \frac{1998848}{218295} & -\frac{3473408}{72765} & \frac{18219008}{24255} \\
 10 & 2 & \frac{6103040}{693693} & -\frac{33095680}{693693} & \frac{626524160}{693693} \\
\hline
 6 & 4 & 0 & 0 & \frac{294912}{7} \\
 8 & 4 & \frac{278528}{7623} & -\frac{6619136}{7623} & \frac{235667456}{7623} \\
 10 & 4 & \frac{294649856}{6441435} & -\frac{476053504}{585585} & \frac{164580294656}{6441435} \\
\hline
 8 & 6 & 0 & 0 & \frac{10485760}{11} \\
 10 & 6 & \frac{1048576}{6435} & -\frac{17825792}{2145} & \frac{29360128}{55} \\
 \hline
 10 & 8 & 0 & 0 & \frac{1879048192}{143} \\
 \hline
 \end{array}
\qquad
\begin{array}{|c|c|c|}
\hline
 \Delta & \ell & (\tilde{\l}_{JJ\mathcal{O}^-} )^2\\
 \hline
 5 & 0 & \frac{256}{3} \\
 7 & 0 & \frac{512}{5} \\
 9 & 0 & \frac{131072}{1001} \\
 \hline
 5 & 2 & 512 \\
 7 & 2 & \frac{24576}{35} \\
 9 & 2 & \frac{6553600}{7007} \\
 \hline
 6 & 3 & \frac{24576}{175} \\
 8 & 3 & \frac{65536}{231} \\
 10 & 3 & \frac{2097152}{4719} \\
 \hline
 7 & 4 & \frac{8192}{3} \\
 9 & 4 & \frac{47185920}{11011} \\
 \hline
 8 & 5 & \frac{2621440}{1617} \\
 10 & 5 & \frac{16777216}{5577} \\
 \hline
 9 & 6 & \frac{2097152}{143} \\
 \hline
 10 & 7 & \frac{234881024}{19305} \\
\hline
 \end{array}
\eea
The OPE coefficients $\tilde{c}_{JJ\mathcal O^\pm}$ in the above tables are defined in appendix \ref{app:conservedequal}.

\section{Relations between parity odd structures}
\label{sec:ind3pt}

Parity odd conformally invariant three point functions can be construct using the $\epsilon$-tensor. 
In $d=3$,  there are six parity odd building blocks:
\beq
\epsilon_{ij}\equiv    \epsilon(Z_i,Z_j,P_1,P_2,P_3)\,\qquad \widetilde\epsilon_{ij}\equiv   \epsilon(P_i,P_j,Z_1,Z_2,Z_3) \qquad(i,j=1,2,3)\,, 
\eeq
However, not all of them are  independent.
To see this we use  the following identity
\begin{align}
&0=\det \left(
\begin{array}{cccccc}
(P_1 \cdot \Xi) & (P_2 \cdot\Xi) & (P_3 \cdot\Xi) &  (Z_1  \cdot  \Xi) &  (Z_2 \cdot\Xi) & (Z_3\cdot \Xi)  \\
\Bigl( P_1 \Bigr)_{\!\!A} & \Bigl( P_2 \Bigr)_{\!\!A} & \Bigl( P_3 \Bigr)_{\!\!A} & \Bigl( Z_1 \Bigr)_{\!\!A}& \Bigl( Z_2 \Bigr)_{\!\!A}&\Bigl( Z_3 \Bigr)_{\!\!A}
\end{array}
\right) 
\label{eq:det6by6}
\end{align}
where $\Xi$ is an arbitrary 5 dimensional vector. 
The determinant vanishes automatically because the first row of the matrix is a linear combination of the other 5 rows.
By choosing for instance $\Xi=P_1$ one gets
\bea
&&-(P_1 \cdot P_2)\epsilon(P_1,P_3,Z_1,Z_2,Z_3)+(P_1 \cdot P_3)\epsilon(P_1,P_2,Z_1,Z_2,Z_3)\\
&&-(P_1 \cdot Z_2)\epsilon(P_1,P_2,P_3,Z_1,Z_3)+(P_1 \cdot Z_3)\epsilon(P_1,P_2,P_3,Z_1,Z_2)=0
\eea
Similarly one can get two more equations by choosing $\Xi=P_2,P_3$. All together these relations allow to express $\widetilde\epsilon_{ij}$ in terms of linear combination of $\epsilon_{ij}$ only. 

In addition, one can find linear relations involving only the three $\epsilon_{ij}$. 
This follows immediately if we choose $\Xi$ orthogonal to the three $P$'s.
This is achieved with
\be
\Xi^A = \left[ Z_1^A (P_1 \cdot X^+)- P_1^A (Z_1 \cdot X^+) \right] (P_2\cdot P_3) + (X^-)^A (Z_1\cdot X^-)\,,
\ee
where
\be
(X^\pm)^A = P_2^A (P_1\cdot P_3) \pm P_3^A (P_1\cdot P_2)\,.
\ee
One can easily check that
\begin{align}
\Xi \cdot Z_1 &= -2 (P_1\cdot P_2)(P_1\cdot P_3)(P_2\cdot P_3) V_{1,23}^2 \\
\Xi \cdot Z_2 &= 2 (P_1\cdot P_2)(P_1\cdot P_3)(P_2\cdot P_3) (H_{12} + V_{1,23}V_{2,31}) \\
\Xi \cdot Z_3 &= 2 (P_1\cdot P_2)(P_1\cdot P_3)(P_2\cdot P_3) (H_{13} + V_{1,23}V_{3,12}) 
\end{align}
and conclude that \eqref{eq:det6by6} reduces to
\be
V_{1,23}^2 \epsilon_{23}- (H_{12} + V_{1,23}V_{2,31}) \epsilon_{13}- (H_{13} + V_{1,23}V_{3,12}) \epsilon_{12}=0\,.
\ee
Similarly, we can find another 2 equations by permuting the 3 points. These relations were taken into account in the construction of  conformal invariant three point functions in section \ref{sec:3pt}.

\section{Basis for four point function}
\label{sec:4ptbasis}

Conformal invariant tensor structures for four point functions are constructed using the building blocks $H_{ij}$ and $V_{i,jk}$ defined in equation \eqref{eq:HV}.
However, not all combinations are linearly independent. In fact, it is sufficient to use the following set:
\be
\left\{H_{12},H_{13},H_{14},H_{23},H_{24},H_{34},V_{1,23},V_{1,43},V_{2,34},V_{2,14},V_{3,21},V_{3,41},V_{4,21},V_{4,32}   \right\}
\ee
All others can be expressed in terms of a linear combination of the above. For instance the following identity holds:
\bea
&&\sqrt{-2(P_i \cdot P_j) (P_i \cdot P_k) (P_j \cdot P_k)} (P_i \cdot P_l) V_{i,jk} + \sqrt{-2(P_i \cdot P_j) (P_i \cdot P_l) (P_j \cdot P_l)} (P_i \cdot P_k) V_{i,kj}\nonumber\\
&&+ \sqrt{-2(P_i \cdot P_l) (P_i \cdot P_k) (P_l \cdot P_k)} (P_i \cdot P_j) V_{i,kl} =0\,.
\eea
Out of the above list one can construct 43 tensor structure. These are listed in Table.~\ref{4pstruct}. 
\newpage
\begin{table}[h!]
\begin{center}
$
{\fontsize{10.7}{12.8}\selectfont 
\begin{array}{|c|c|c|c|c|}
\hline
s & Q_s & 1234 \rightarrow 3412 & 1234 \rightarrow 2143 & \text{Indep}\\
\hline
  1 & H_{12} H_{34} & 1 & 1 & \color{red} 1 \\
 2 & H_{13} H_{24} & 2 & 2 & \color{red} 2 \\
 3 & H_{14} H_{23} & 3 & 3 & \color{red} 3 \\
 4 & H_{12} V_{3,21} V_{4,12} & 27 & 4 & \color{red} 4 \\
 5 & H_{12} V_{3,21} V_{4,32} & 26 & 6 & \color{red} 5 \\
 6 & H_{12} V_{3,41} V_{4,12} & 25 & 5 & 5 \\
 7 & H_{12} V_{3,41} V_{4,32} & 24 & 7 & \color{red} 7 \\
 8 & H_{13} V_{2,14} V_{4,12} & 11 & 20 & \color{red} 8 \\
 9 & H_{13} V_{2,14} V_{4,32} & 9 & 21 & \color{red} 9 \\
 10 & H_{13} V_{2,34} V_{4,12} & 10 & 22 & \color{red} 10 \\
 11 & H_{13} V_{2,34} V_{4,32} & 8 & 23 & 8 \\
 12 & H_{14} V_{2,14} V_{3,21} & 19 & 16 & \color{red} 12 \\
 13 & H_{14} V_{2,14} V_{3,41} & 17 & 17 & \color{red} 13 \\
 14 & H_{14} V_{2,34} V_{3,21} & 18 & 18 & \color{red} 14 \\
 15 & H_{14} V_{2,34} V_{3,41} & 16 & 19 & 12 \\
 16 & H_{23} V_{1,23} V_{4,12} & 15 & 12 & 12 \\
 17 & H_{23} V_{1,23} V_{4,32} & 13 & 13 & 13 \\
 18 & H_{23} V_{1,43} V_{4,12} & 14 & 14 & 14 \\
 19 & H_{23} V_{1,43} V_{4,32} & 12 & 15 & 12 \\
 20 & H_{24} V_{1,23} V_{3,21} & 23 & 8 & 8 \\
 21 & H_{24} V_{1,23} V_{3,41} & 21 & 9 & 9 \\
 22 & H_{24} V_{1,43} V_{3,21} & 22 & 10 & 10 \\
 23 & H_{24} V_{1,43} V_{3,41} & 20 & 11 & 8 \\
 24 & H_{34} V_{1,23} V_{2,14} & 7 & 24 & 7 \\
 25 & H_{34} V_{1,23} V_{2,34} & 6 & 26 & 5 \\
 26 & H_{34} V_{1,43} V_{2,14} & 5 & 25 & 5 \\
 27 & H_{34} V_{1,43} V_{2,34} & 4 & 27 & 4 \\
 28 & V_{1,23} V_{2,14} V_{3,21} V_{4,12} & 43 & 28 & \color{red} 28 \\
 29 & V_{1,23} V_{2,14} V_{3,21} V_{4,32} & 39 & 30 & \color{red} 29 \\
 30 & V_{1,23} V_{2,14} V_{3,41} V_{4,12} & 35 & 29 & 29 \\
 31 & V_{1,23} V_{2,14} V_{3,41} V_{4,32} & 31 & 31 & \color{red} 31 \\
 32 & V_{1,23} V_{2,34} V_{3,21} V_{4,12} & 42 & 36 & \color{red} 32 \\
 33 & V_{1,23} V_{2,34} V_{3,21} V_{4,32} & 38 & 38 & \color{red} 33 \\
 34 & V_{1,23} V_{2,34} V_{3,41} V_{4,12} & 34 & 37 & \color{red} 34 \\
 35 & V_{1,23} V_{2,34} V_{3,41} V_{4,32} & 30 & 39 & 29 \\
 36 & V_{1,43} V_{2,14} V_{3,21} V_{4,12} & 41 & 32 & 32 \\
 37 & V_{1,43} V_{2,14} V_{3,21} V_{4,32} & 37 & 34 & 34 \\
 38 & V_{1,43} V_{2,14} V_{3,41} V_{4,12} & 33 & 33 & 33 \\
 39 & V_{1,43} V_{2,14} V_{3,41} V_{4,32} & 29 & 35 & 29 \\
 40 & V_{1,43} V_{2,34} V_{3,21} V_{4,12} & 40 & 40 & \color{red} 40 \\
 41 & V_{1,43} V_{2,34} V_{3,21} V_{4,32} & 36 & 42 & 32 \\
 42 & V_{1,43} V_{2,34} V_{3,41} V_{4,12} & 32 & 41 & 32 \\
 43 & V_{1,43} V_{2,34} V_{3,41} V_{4,32} & 28 & 43 & 28\\
 \hline
 \end{array}}
 $
\end{center}
\caption{\footnotesize{
The third and fourth column show how the structures map to each other under the special permutations that preserve the cross ratios. The 43 structures split into 19 multiplets under these permutations. In the last column, we show in red the label of the 19 independent   functions $f_s(u,v)$ that multiply each multiplet. }}
\label{4pstruct}
\end{table}%
\newpage
\section{Simplifying crossing}
\label{ap:ftilde}

The functions $\tilde{f}_s$ are defined by the following linear map, 
\be
\tilde{f}_{\tilde s}(u,v) = \sum_{s=1}^{43} \, \Mat_{\tilde s s} \, f_{s}(u,v) \ , \qquad \quad
\tilde s=1,\dots,19\ .
\ee
Here $\Mat$ is a $19 \times 43$ matrix defined by
\beq
\begin{array}{l l}
\tilde{f}_{1}=  f_{28}-f_{33}  \qquad \qquad& \tilde{f}_{8}=  f_{34} \\
\tilde{f}_{2}=  f_{29}-f_{32}  \qquad& \tilde{f}_{9}=  f_{28}+f_{33}  \\
\tilde{f}_{3}= f_7-f_{14}  \qquad& \tilde{f}_{10}=  f_{29}+f_{32}  \\
\tilde{f}_{4}=  f_4-f_{13}  \qquad& \tilde{f}_{11}= f_7+f_{14}  \\
\tilde{f}_{5}=  f_5-f_{12} \qquad& \tilde{f}_{12}= f_4+f_{13}  \\
\tilde{f}_{6}=  f_9-f_{10}  \qquad& \tilde{f}_{13}=  f_5+f_{12}  \\
\tilde{f}_{7}= f_1-f_3 \qquad& \tilde{f}_{14}=  f_9+f_{10}  \\
\tilde{f}_{19}=  f_{31}-f_{40}  \qquad& \tilde{f}_{15}=   f_8 \\
& \tilde{f}_{16}= f_1+f_3  \\
& \tilde{f}_{17}=  f_2 \\
& \tilde{f}_{18}=  f_{31}+f_{40}  \\
\end{array} 
\label{eq:ftilde}
\eeq
where we suppressed the arguments $(u,v)$ from all functions $\tilde{f}_s$ and $f_s$.

\section{Conformal blocks}
\label{app:ConformalBlocks}

In this appendix we explain how to obtain the recurrence relation for the  conformal blocks of four external vector operators in $d=3$ defined in section \ref{sec:CBs}.
In particular we will show how to compute all the ingredients of formulae (\ref{RecRelSpin+}-\ref{RecRelSpin-})
by using the  conventions and ideas introduced in \cite{arXiv:1509.00428}.

\subsection{Conventions for $JJJJ$}
\label{App:conventions}
In this section we define our conventions for the conformal blocks. We are interested in finding all the CBs for four generic external vectors
\begin{align}\label{CBwithSPIN3d}
\langle J_1 J_2 J_3 J_4 \rangle=
{\sum_{\Ocal^+} \sum_{p,q=1}^5} %
\raisebox{1.7em}{$\xymatrix@=6pt{{J_1}\ar@{-}[rd]& & & &J_3 \ar@{-}[ld]   \\  
& *+[o][F]{\mbox{\tiny \emph{p}}}  \ar@{=}[rr]^{\displaystyle \Ocal^+ } & & *+[o][F]{\mbox{\tiny \emph{q}}}  &  \\	
J_2 \ar@{-}[ru]& & & &J_4 \ar@{-}[lu]}$}
+
{\sum_{\Ocal^-} \sum_{\pp,\qq=1}^4} %
\raisebox{1.7em}{$\xymatrix@=6pt{{J_1}\ar@{-}[rd]& & & &J_3 \ar@{-}[ld]   \\  
& *+[o][F]{\mbox{\tiny \pp}}  \ar@{=}[rr]^{\; \;\; \displaystyle \Ocal^- } & & *+[o][F]{\mbox{\tiny \qq}}  &  \\
J_2 \ar@{-}[ru]& & & &J_4 \ar@{-}[lu]}$}
\ . 
\end{align}
where the operators  $J_i$ have spin $1$ and dimension $\D_i$  and $\Ocal^\pm$ are operators with spin $\ell$ dimension $\D$ and parity $\pm$.

As a first step we explain our convention for the labels $p,q$ of the OPE. We define the leading OPE in terms of a linear combination of tensor structures
\be \label{OPEwithSPIN}
\Ocal^\pm(x,z)J_1(0,z_1)\sim \frac{J_2(0,\partial_{z_2})}{(x^2)^{\a_\pm}} \sum_{q}
c_{12\Ocal^\pm}^{(q)} \; t^{(q)}_{\ell \pm}( x,z,z_1,z_2)  \ ,
\ee
where $z_{\m}$ are null polarization vectors. Here  $\sim$ means that we are considering only the channel of the OPE in which $\Ocal^\pm \times J_1$ exchanges the operator $J_2$ (therefore omitting all the other possible exchanged primaries), and  
taking into account only the leading term of the OPE for $x^\m \rightarrow 0$ (therefore omitting all the contribution of the descendant operators). We also define 
\beq
\a_+\equiv \frac{\D+\D_1-\D_2+\ell+2}{2}\,,\qquad
\a_-\equiv \frac{\D+\D_1-\D_2 +\ell+1}{2}\,.
\eeq
 The various  OPE coefficients $c_{12\Ocal^\pm}^{(q)}$  multiply the tensor structures   $ t^{(q)}_{\ell\pm}(x,z,z_2,z_3)$, which are  Lorentz invariant and  satisfy
\begin{align}
t^{(q)}_{\ell+}(\m x, \l z, \l_1 z_1,\l_2 z_2)&= \m^{\ell+2} \l^\ell \l_1 \l_2 \; t^{(q)}_{\ell+}( x,z,z_1,z_2) \  , \\
t^{(q)}_{\ell \,-}(\m x, \l z, \l_1 z_1,\l_2 z_2)&= \m^{\ell+1} \l^\ell \l_1 \l_2 \; t^{(q)}_{\ell\, -}( x,z,z_1,z_2) \ .
\end{align}
The sum over $q$ in (\ref{OPEwithSPIN}) runs from one to five for parity even operators, since we can build the following five structures
\begin{align}
\begin{split}
t_{\ell+}^{(1)}( x,z,z_1,z_2)&\equiv (x\cdot z)^{\ell} (z_1 \cdot z_2) x^2\ ,\\
t_{\ell+}^{(2)}( x,z,z_1,z_2)&\equiv (x\cdot z)^{\ell} (x\cdot z_1) (x \cdot z_2)\ ,\\
t_{\ell+}^{(3)}( x,z,z_1,z_2)&\equiv (x\cdot z)^{\ell-1} (z\cdot z_1) (x \cdot z_2) x^2\ ,\\
t_{\ell+}^{(4)}( x,z,z_1,z_2)&\equiv (x\cdot z)^{\ell-1}(z\cdot z_2) (x \cdot z_1) x^2\ ,\\
t_{\ell+}^{(5)}( x,z,z_1,z_2)&\equiv (x\cdot z)^{\ell-2} (z\cdot z_1) (z \cdot z_2) x^4\ .
\end{split}
\label{t_five_allowed}
\end{align}
Notice that for $\ell=0$ only $t^{(1)}$ and $t^{(2)}$ survive and for $\ell=1$ all are  allowed except $t^{(5)}$. 
These structures are related by a simple linear transformation to the structures of the main text. To be more precise, the basis
\be
T^{(q)}_+=\left\{V_{1,23} V_{2,31} V_{3,12}^\ell,H_{12} V_{3,12}^\ell,H_{13} V_{2,31} V_{3,12}^{\ell-1},H_{23} V_{1,23} V_{3,12}^{\ell-1},H_{13} H_{23} V_{3,12}^{\ell-2}\right\} 
\ee
can be related to $t^{(q)}_{\ell\, +}\rightarrow\sum_{q'=1}^5 [\mathfrak{M}_{+}]_{q q'}  \ T^{(q')}_+$ by
\be
\mathfrak{M}_{+}=\left(
\begin{array}{ccccc}
 0 & -1 & 0 & 0 & 0 \\
 1 & 0 & 0 & 0 & 0 \\
 0 & 2 & -1 & 0 & 0 \\
 0 & 0 & 0 & 1 & 0 \\
 0 & 0 & 0 & -2 & 1 \\
\end{array}
\right) \ .
\ee

Similarly for parity odd operators with generic spin there are four allowed tensor structures which can be build by using the three dimensional epsilon tensor (see appendix \ref{Parity_Odd_Structures_in_three_dimensions})
\be \label{structuresParityOdd}
\begin{array}{l} 
t^{(1)}_{\ell\, -}= \e(x,z_1,z_2)  (x \cdot z)^{\ell}\\
t^{(2)}_{\ell\, -}=\e(x,z,z_1) (x \cdot z_2) (x \cdot z)^{\ell-1}\\
t^{(3)}_{\ell\, -}= \e(x,z,z_2) (x \cdot z_1) (x \cdot z)^{\ell-1}\\
t^{(4)}_{\ell\, -}=[\e(x,z,z_1) (z \cdot z_2)+\e(x,z,z_2) (z \cdot z_1)] (x \cdot z)^{\ell-2} x^2 \ .
\end{array}
\ee
Again it is clear that for $\ell=0$ there is only  $t^{(1)}$ and for $\ell=1$ only $t^{(1)},t^{(2)},t^{(3)}$. 
The basis in embedding space
\be
T^{(q)}_-=\left\{V_{2,31} \epsilon _{13} V_{3,12}^{\ell-1},V_{1,23} \epsilon _{23} V_{3,12}^{\ell-1},H_{23} \epsilon _{13} V_{3,12}^{\ell-2},H_{13} \epsilon _{23} V_{3,12}^{\ell-2}\right\}
\ee
can be related to $t^{(q)}_{\ell\, -}\rightarrow\sum_{q'=1}^4 [\mathfrak{M}_{-}]_{q q'}  \ T^{(q')}_-$ for $\ell\geq2$ by means of the following matrix
\be
\mathfrak{M}_{-}=
\left(
\begin{array}{cccc}
 0 & 4 & 0 & 0 \\
 0 & 0 & -1 & 0 \\
 -2 & -2 & 2 & -2 \\
 \frac{1}{2} & \frac{1}{2} & \frac{3}{2} & -\frac{1}{2} \\
\end{array}
\right) \ .
\ee

\subsection{Null States}
In this section we write all the possible primary descendant states that can be exchanged when the external operators are all vectors.

In $d=3$ the only irreducible representations of the rotation group are   traceless and symmetric tensors of spin $\ell$.
We consider such a primary state of spin $\ell$ and we contract it with null polarization vectors $z_{\m}$  as follows
\be
|\Ocal;z \rangle \equiv z_{\m_1} \dots z_{\m_\ell} \Ocal^{\mu_1 \dots \mu_\ell}(0)|0\rangle\ \equiv \Ocal(z,0)|0\rangle\ .
\ee
It is possible to recover the expressions with the indices 
\be
\Ocal^{\mu_1 \dots \mu_\ell}(x)= \frac{1}{\ell! (\frac{d}{2}-1)_\ell} D_z^{\mu_1} \cdots D_z^{\mu_\ell} \Ocal(z,x) \ ,
\ee
by acting with the differential operator $D_z$ of \cite{SpinningCC,Sofia},
\be
D_z^\m=\left( \frac{d}{2}-1+z \cdot \partial_z \right) \partial_z^\m -\frac{\, z^\m}{2} 
\  \partial_z  \cdot \partial_z \ .
\ee

The  primary descendant states are of four possible types which are additionally  labeled  by an integer $n$ ($n$ runs over all positive integers for type {\I} and type \II, and over a finite set for type {\III} and \IV).
We define each descendant state $|\Ocal_A; z \rangle$ by the action of an  operator $\mathcal{D}_A$ (built as a linear combination of many $P^\m$, the generators of the translations) on a  primary state $| \Ocal ;z  \rangle$
\be
|\Ocal_A; z \rangle = \Dcal_A  | \Ocal ;z  \rangle \ .
\label{DA}
\ee
The operators $\mathcal{D}_A$ can be fixed by asking that $K_\mu |\Ocal_A; z \rangle=0$ when  $\D=\D^\star_A$. 
This is simply the requirement that when $\D=\D^\star_A$, the descendants $|\Ocal_A; z \rangle$ become primaries.
The operators $\mathcal{D}_A$  can be written in the following compact form \cite{arXiv:1509.00428}
\begin{align}
\Dcal_{\I,n}&\equiv c_{\I,n} \ (z\cdot P)^n  \ , \label{DI}\\
\Dcal_{\II,n}& \equiv c_{\II,n}\ (D_z \cdot P)^n  \ ,  \label{DII}\\
\Dcal_{\III,n}&\equiv c_{\III,n}\ \Vcal_{\frac{1}{2}} \ \Vcal_{\frac{3}{2}} \cdots \Vcal_{n-{\frac{1}{2}}}  \ ,  \label{DIII} \\
\Dcal_{\IV,n}&\equiv c_{\IV,n} \ \Ecal  \ \Vcal_{1}\ \Vcal_{2} \cdots \Vcal_{n-1}   \ , \label{DIV}
\end{align}
where $\Vcal_j$ and $\Ecal$ are defined by
\begin{align}  \label{Vcalj}
\Vcal_j&=(j+\ell)(\ell-j)  P^2-2  (P\cdot z) (P\cdot D_z) \ , \\ 
 \Ecal&= \epsilon_{\m \n \s} P^\m z^\n D_z^\s \ .
\end{align}
The coefficients $c$ are unimportant normalization constants that we are free to set to any value. For convenience, we choose
\be
c_{\I,n}=1=c_{\IV,n} \ , \quad   c_{\II,n}=\frac{1}{\left(\frac{1}{2}-\ell\right)_n (-\ell)_n} \ , \quad c_{\III,n}=\frac{1}{\left(\ell-n+\frac{1}{2}\right)_{2 n}} \ .
\ee

We want to stress that the operators of   type $\I,\II$ and $\III,$ do not change the parity of the state on which they are applied, while the one of type $\IV$  does, namely
\be
\Dcal_A | \Ocal^\pm \rangle =
\left\{ \begin{array}{l l}
  | \Ocal^\pm_A \rangle \qquad & A=(\I,n),(\II,n),(\III,n) \\
   | \Ocal^\mp_A \rangle \qquad & A=(\IV,n) \\
\end{array}
\right. \ .
\ee
Using the definitions (\ref{DI}-\ref{DIV}) and  the commutation relations of the conformal algebra, we can compute the residue $Q_A$ at the pole $\D^\star_A$ of the inverse of the norm of the primary descendant states 
\be
\langle \Ocal_A; z|\Ocal_A; z \rangle ^{-1}=\frac{Q_A}{\D-\D^\star_A}
\langle \Ocal; z|\Ocal; z \rangle^{-1} + O((\D-\D^\star_A)^0)\ .
\ee
The result for the four types is
\begin{align}
\label{QA}
\begin{split}
Q_{\I,n}&=-\frac{2^{-n}}{(n-1)! n!} \ , \\
Q_{\II,n}&=\frac{(-2)^{-n}}{(n-1)! n!} \frac{ (2 n-2 \ell-1)}{(2 \ell+1)} \frac{ (-\ell)_n}{ (\ell-n+1)_n}\ , \\
Q_{\III,n}&=-\frac{(-4)^{-2n}}{(n-1)! n!}  \frac{
   \left(\ell-n+\frac{1}{2}\right)}{ \left(\ell+n+\frac{1}{2}\right)} \frac{1}{ \left(\frac{1}{2}-n\right)_{2 n}}  \ , \\
Q_{\IV,n}&=\frac{2}{(2 n-1)! (2 n-2)!} \frac{1}{(1-2 \ell)^2} \frac{1}{(-\ell-n+1)_{2 n-2} (\ell-n+1)_{2 n}}\ .
\end{split}
\end{align}

\subsection{The Residue $R_A$}
\label{RA}
The residue $R_A$ is obtained using formula (\ref{RA=MAQAMA}), where $Q_A$ are defined in (\ref{QA}).

The matrices $M_A$ can be defined by the action of differential operator $\Dcal_A$ on the tensor structures appearing in the leading OPE \eqref{OPEwithSPIN}. For the first three types we have
\be
\Dcal_{A}\frac{t^{(q)}_{\ell\pm}(x,z, z_1,z_2)}{(x^2)^{\a_\pm}} =\sum_{q'} \left(M_{\pm  A}^{(L)}\right)_{q q'} \frac{t_{\ell_A\pm}^{(q\rq{})}(x,z,  z_1,z_2)}{(x^2)^{\a_{\pm A}}}  \qquad {\scriptsize A= \left\{ \begin{array}{l} \I,n \\ \II,n  \\ \III,n \end{array} \right.}\ ,
\ee
where the matrices $M_{+A}$ are $5 \times 5$ while the  $M_{-A}$ are $4 \times 4$. The exponent $\a_{\pm A}$ is equal to $\a_{\pm}$ where we replace $\ell \rightarrow \ell_A$ and $\D \rightarrow \D +n_A$. Moreover, we set $\D=\D^\star_A$.

The type {\IV} is slightly different since it changes the parity of the primary state, therefore
\be
\Dcal_{A}\frac{t^{(q)}_{\ell \pm }(x,z, z_1, z_2)}{(x^2)^{\a_\pm}} =\sum_{q'} \left(M_{\pm A}^{(L)}\right)_{q q'}  \frac{t_{\ell_A \mp}^{(q')}(x,z, z_1, z_2)}{(x^2)^{\a_{\mp A}}}  \qquad  A= \IV,n\ .
\ee
In this case  $M_{+\IV}$ is a rectangular matrix $5 \times 4$ while $M_{-\IV}$ is $4 \times 5$. 

The definitions given here can be directly used to compute $M$. However there can be better strategies to implement this computation. One possible strategy is to act with each \emph{building block differential operator} contained in $\Dcal_A$ (namely $P\cdot z$, $P\cdot D_z$, $V_j$ and $\Ecal$) on the set of tensor structures, in order to obtain a correspondent  \emph{building block matrix} that rotates the tensor structures. The full result can be computed as products of these building block matrices, as detailed in \cite{arXiv:1509.00428}. 
A new strategy is explained in appendix \ref{MA_New}, where we show how to obtain $M_A$ for the types $(\I,n),(\II,n),(\III,n)$ in a closed form by doing a trivial computation.

It is worth commenting that by direct computation one can check that $M_{\IV,n}$
 vanishes for $n>2$ which means that only two poles of type $(\IV,n)$ contribute, namely $\D=0,1$ as mentioned in appendix \ref{app:RecGeneralD}. 

\subsection{Conformal Block at Large $\D$}
\label{largedelta}

In this section we explain how to compute $h_\infty$, the large $\D$ limit of the conformal blocks. To do so, we are going to solve the Casimir differential equation at the leading order  for large $\D $ with the appropriate initial condition for $G_{\D \ell \pm}^{(p,q)}$  when $x_{12}, x_{34} \rightarrow 0$. 

The Casimir equation can be schematically expressed as
\be
\frac{1}{2}(J_1+J_2)^2 \ G_{\D \ell \pm}^{(p,q)}(\{P_i;Z_i\})= c_{\D \ell} \, G_{\D \ell \pm}^{(p,q)}(\{P_i;Z_i\}) \ , \label{casimir eq spin}
\ee
where $J_i^{AB} \equiv i ( P^A_i \partial^B_{P_i}- P^B_i \partial^A_{P_i}+ P\leftrightarrow Z )$.
We consider the leading order in $\D$ of (\ref{casimir eq spin}) and we substitute  the definitions (\ref{def:CBsExp}) and $g^{(p,q)}_{\D \ell \pm,s}(r, \eta)=(4r)^{\D} h^{(p,q)}_{\D \ell \pm,s}(r, \eta)$. The result is a set of $43$ coupled first order differential equations for the functions $h_s$
\be  \label{GenericCoupledPDEs}
\partial_r h_s= \sum_{t=1}^{43} \mathcal{M}_{s\, t}(r,\eta) h_{t} \ ,
\ee 
where $\mathcal{M}$ is a $43 \times 43$ matrix of explicitly known rational functions of $r$ and $\eta$ and where we dropped all the labels of $h_s$ (which will be reintroduced when we will fix the initial condition of the Casimir equation).  Since the $43$ equations \eqref{GenericCoupledPDEs} are of the first order, we have $43$ independent solutions $h^{(s')}_{s}$, labeled by $s'=1\dots 43$. We then use the   ansatz
\begin{align} \label{hss}
h^{(s')}_{s}(r,\eta)&\equiv \Acal(r,\eta) P_s^{ (s\rq{})}(r,\eta)  \ , 	\\
\Acal(r,\eta)&\equiv \frac{ \left(1-r^2\right)^{-3-\frac{d}{2}}}{ \left(1+r^2-2 r \eta \right)^{\frac{1}{2}-\D_{12}+\D_{34}} \left(1+r^2+2 r \eta \right)^{\frac{5}{2}+\D_{12}-\D_{34}} } \ ,
\end{align}
to obtain a set of differential equations for the $43$ functions $P_s^{ (s')}$. This ansatz, inspired by the solution of the scalar Casimir at large $\D$, has the property of eliminating completely the dependence on $\D_{12}=\D_1-\D_2$ and  $\D_{34}=\D_3-\D_4$ from the differential equation.
Moreover it turns out that we can easily fix all the functions $P_s^{(s')}(r,\eta)$ since they are simply polynomials in $r$ (of maximal degree $12$) and $\eta$.   Notice also that we are leaving $d$ unfixed: in fact the solution that we find works in any dimension. We can further choose a basis such that 
\be
P_s^{(s')}(0,\eta)=\delta_s^{s'} \ , \qquad
s,s\rq{}=1,\dots, 43\,.
\ee  
The functions $h_\infty$ can then be written as a linear combination of the $43$ functions $h_s^{(s')}$ as follows
\be
h^{(p,q)}_{\infty \ell \pm ,s}(r,\eta)=\sum_{s'=1}^{43} h_s^{(s')}(r,\eta) f^{(p,q)}_{\ell \pm, s'}(\eta),
\ee
where the functions $f$ are constants of integration that can be fixed by imposing the correct initial condition for the differential equation.
In particular with our conventions $f^{(p,q)}_{\ell \pm, s}(\eta) \equiv h^{(p,q)}_{\infty \ell \pm,s}(0,\eta)$. Therefore we can fix them by studying 
the OPE limit (namely $x_2\to x_1$ and $x_4 \to x_3$ which also imply $r \rightarrow 0$) of $G_{\D \ell \pm}^{(p,q)}$. 
As explained in \cite{arXiv:1509.00428}, by studying the OPE limit of  $G_{\D \ell \pm}^{(p,q)}$ we obtain the following equation, that can be used to define the functions $f$,
\begin{align}
 \frac{ t^{(p)}_{\ell \pm}( \hat x_{12}, 
   I(x_{24})\cdot D_z ,I(x_{12}) \cdot  z_1, z_2)t^{(q)}_{\ell \pm}(\hat x_{34},z,I(x_{34}) \cdot z_3,z_4) 
}{\ell! (h-1)_\ell}
 \approx  \sum_{s=1}^{43} f^{(p,q)}_{\ell\pm,s}(\eta) Q_s\ .\label{smallr:eq.forf}
\end{align}
Here $Q_s$ are the 43 four point function tensor structures in the OPE limit   $x_2\to x_1$ and $x_4 \to x_3$. Our choice of structures $Q_s$ is such that the 43 structures remain finite and linearly independent in this limit. The contractions with the tensor $I$ denote $I(x) \cdot z = z-2x (x\cdot z)/x^2 $ and  %
$ \eta\approx - \hat x_{12} \cdot I(x_{24}) \cdot \hat x_{34} $.
 
As a last remark we want to stress that from this computations, all the functions $h^{(p,q)}_{\infty \ell +, s}$ were found in generic dimensions. However the leading term of the blocks $h^{(p,q)}_{\infty \ell -, s}$ is by construction related to the three dimensional case. Nevertheless to generalize it to any dimension it straightforward. To that end it is sufficient to replace $f^{(p,q)}_{\ell -,s}$ with the OPE limit  of the higher dimensional conformal blocks for the exchange of operators in the SO(d) representations $(\ell,1)$, $(\ell,2)$ and $(\ell,1,1)$ \cite{Costa:2016hju}.

\subsection{Tensor structures for equal and conserved currents}
\label{app:conservedequal}
In this section we obtain the matrices $m_\pm$ which are needed in order to obtain the blocks of conserved equal currents, according to (\ref{def:CB_conserved_equal}).
\subsubsection{Parity even}
In the OPE of two identical operators  $\Ocal_1=\Ocal_2$ there is a smaller set of allowed tensor structures in comparison to the one defined in section \ref{App:conventions}. The invariance under the exchange of $\Ocal_1$ and $\Ocal_2$ can be easily formulated in terms of the OPE by asking that the a linear combination $\sum_{q=1}^5 c^{(q)} t_{\ell+}^{(q)}$ of the OPE structures is invariant under the map
\be \label{O1=O2MAP}
\begin{array}{c l}
x &\to - x \\
z_1 &\to I(x)\cdot z_2 \\
z_2 &\to I(x)\cdot z_1
\end{array}
\ .
\ee
This automatically gives the following set of constraints on the coefficients $c^{(q)}$,
\be
\label{cqSystemForEqualOp}
\left\{
\begin{array}{l}
c^{(1)} \left((-1)^\ell-1\right)=0 \\
c^{(2)} \left((-1)^\ell-1\right)+2
   \left(c^{(3)}+c^{(4)}+2 c^{(5)}\right) (-1)^\ell=0
   \\
 \left(c^{(4)}+2 c^{(5)}\right) (-1)^\ell+c^{(3)}=0 \\
 \left(c^{(3)} +2 c^{(5)}\right) (-1)^\ell+c^{(4)}=0 \\
 c^{(5)} \left((-1)^\ell-1\right)=0 \\
\end{array}
\right. \ .
\ee
Solving this set of constraints one can define a new set of allowed tensor structures. From (\ref{cqSystemForEqualOp}) it is clear that  for $\ell=0$ we have just two possible structures, for even $\ell>0$ we have four, while for odd $\ell$ we always have one. 
 
   When $\Ocal_1=J_1$ is a conserved current, we have
\be
\partial_{z_1}\cdot \partial_{x_1} J_1(x_1,z_1)=0 \qquad 
\Longrightarrow 
\qquad
 \label{conservationOPE}
\partial_{z_1}\cdot \partial_{x_1} \frac{\sum_{q=1}^5 c^{(q)} t^{(q)}_{\ell+}(-x_1,z,z_1,z_2)}{(x_1^2)^{\a_+}}=0 \ .
\ee
The conservation condition applied to the OPE provides a constraint on the allowed combinations of the OPE coefficients,  
\be \label{conservedk}
\left\{
\begin{array}{l}
 -2 (\alpha_+ -1) c^{(1)}-2 (\alpha_+ -1)
   c^{(3)}+c^{(2)} (-2 \alpha_+ +d+\ell+1)=0 \\
 -2 (\alpha_+ -2) c^{(5)}+c^{(4)} (-2 \alpha_+ +2
   h+\ell+1)+c^{(1)} \ell+c^{(3)}=0 \\
\end{array}
\right. \ .
\ee
This implies that for $\ell=0$ there exists only one allowed structure, for $\ell=1$ there are two, and for $\ell>1$ there are always three. 

To find all the allowed structures for equal  conserved currents we can solve simultaneously the systems (\ref{cqSystemForEqualOp}) and (\ref{conservedk}). 
We decide to define the basis $\tilde t$ of the tensor structures for conserved and equal currents as the following linear combination of the basis of generic external vectors
\be
\label{ConservedEqualBasis+}
\tilde t^{(\tilde{p})}_{\D \ell+}(x,z,z_1,z_2)=\sum_{p=1}^5 (m_+)_{\tilde{p} p} \  t^{(p)}_{\ell +}(x,z,z_1,z_2) \ , \qquad (\tilde{p}=1,2) \ ,
\ee
where
\be
\label{m+}
m_+=\left(
\begin{array}{ccccc}
 (2-\Delta ) (\ell+\Delta ) & (\Delta -\ell) (\ell+\Delta ) & 2 \ell (\Delta -2) & 0 & -\ell (\Delta -2) \\
 \ell-\Delta +2 & 0 & -\ell+\Delta -2 & \Delta -\ell & \ell-\Delta +1 \\
\end{array}
\right) \ .
\ee
In particular for odd values of $\ell$ there are no allowed structures while in the case $\ell=0$ only $\tilde t^{(1)}$ is allowed. Instead for all even  $\ell>0$ both structures are allowed. The OPE coefficients $\tilde c$ in the basis $\tilde t$ are related to the ones of section \ref{App:conventions} by 
\be
c^{(p)}_{12\Ocal^+} =\sum_{\tilde p=1}^2 \tilde \l^{(\tilde p)}_{12\Ocal^+}   (m_+)_{\tilde p p} \ .
\ee

\subsubsection{Parity Odd Structures in three dimensions}
\label{Parity_Odd_Structures_in_three_dimensions}
In equation (\ref{structuresParityOdd}) we defined the leading OPE of two spin one operators $J_1$ and $J_2$ with a pseudo-tensor $\Ocal^-$ of spin $\ell$ and dimension $\Delta$.
Notice that we did not include the structures 
\be
\e(z,z_1,z_2) x^2 (x \cdot z)^{\ell-1} \ , 	\qquad \e(x,z,z_1) (z \cdot z_2) (x \cdot z)^{\ell-2} x^2\ ,
\ee
since they can be written as linear combinations of the previous structures $t_{\ell-}$. In fact we have the following two identities
\begin{align} \label{detepsilon}
&0=\det \left(
\begin{array}{cccc}
x^2& x\cdot z& x\cdot z_1& x \cdot z_2\\
\Bigl( x \Bigr)_{\!\!\m} & \Bigl( z \Bigr)_{\!\!\m}& \Bigl( z_1 \Bigr)_{\!\!\m}&\Bigl( z_2 \Bigr)_{\!\!\m}
\end{array}
\right) =
\begin{array}{l}
\\
x^2 \e(z,z_1,z_2)-(x\cdot z) \e(x,z_1,z_2) \\
+(x\cdot z_1) \e(x,z,z_2)-(x\cdot z_2) \e(x,z,z_1) \ ,
\end{array}
\\
&0=\det \left(
\begin{array}{cccc}
z\cdot x &  0 \; \ &z\cdot z_1& z\cdot z_2\\
\Bigl( x \Bigr)_{\!\!\m} & \Bigl( z \Bigr)_{\!\!\m}& \Bigl( z_1 \Bigr)_{\!\!\m}&\Bigl( z_2 \Bigr)_{\!\!\m}
\end{array}
\right)=
\begin{array}{l}
\\
(z \cdot x) \e(z,z_1,z_2)+(z\cdot z_1) \e(x,z,z_2) \\
-(z\cdot z_2) \e(x,z,z_1) \ ,
\end{array}
\end{align}
which reduce the space of six possible tensor structures to just four independent ones.
We also remind that in the case of $\ell=0$ there is only one structure $t^{(1)}_{\ell\, -}$, while for $\ell=1$ we have three possible structures $t^{(1)}_{\ell\, -}, t^{(2)}_{\ell\, -},t^{(3)}_{\ell\, -}$.

In the case of equal operators we need to find the linear combinations of (\ref{structuresParityOdd}) that are invariant under the map (\ref{O1=O2MAP}). We obtain that  for $\ell=0,1$ we can only have a single structure, while for $\ell>1$ there are two.
For two different conserved currents one would have  one single structure for $\ell=0$, two for $\ell=1$ and three for $\ell>1$. For conserved equal currents in three dimensions  we obtain just one structure $\tilde t_{\D \ell \, -}$ that takes two different forms for $\ell$ even and $\ell$ odd, 
\be
\label{ConservedEqualBasis-}
\tilde t_{\D \ell -}(x,z,z_1,z_2) = \sum_{p} \; (m_{\ell-})_{p} \;  t^{(p)}_{\ell-}(x,z,z_1,z_2) 
\ee
with
\be
\label{m-}
m_{\ell-}= 
\left\{
\begin{array}{l l}
(\Delta -3,\ell,\ell,0)   & \  \ell \mbox{ even,} \\
(0,\Delta -\ell-3,\Delta +\ell+1,1-\Delta)  & \ \ell>1, \mbox{ odd.}
\end{array}
\right.
\ee
For the special case $\ell=1$ there are no allowed tensor structures. For $\ell=0$ instead $\tilde t_{\D \ell -}$ is still allowed.
We define the OPE coefficient $\tilde c$ in the basis $\tilde t$ by
\be
c^{(p)}_{12\Ocal^-} =\tilde \l_{12\Ocal^-} (m_-)_{p}  \ .
\ee
\subsection{Comments on the recurrence relation}
In this appendix we explain some new technical developments on the conformal block recurrence relations which were obtained as a part of this work.  Findings presented in this section will be useful for the task of computing the full set of conformal blocks needed to implement the numerical conformal bootstrap in any spacetime dimension and for any ``low spin''
external operator.

\subsubsection{New strategy to compute the matrices $M_A$}
\label{MA_New}
The matrices $M_A$ are not trivial to compute using the strategy proposed in appendix \ref{RA}. This is due to the fact that the first three types $(\I,n)$, $(\II,n)$, $(\III,n)$ have the label $n$ which  in principle can take infinitely many values. The label $n$ is also related to the degree of the differential operator which one has to apply to the structures, therefore it may look very nontrivial to obtain closed form results for $M_A$.
However there is a simple way to compute the matrices $M_A$ in a closed form, which we are going to explain in this section. 
For this purpose we define the conformal blocks $\hat G^{(p,q)}_{\Ocal}$ in the basis of the differential operators 
\be
\label{GtoGhat}
\hat G^{(p,q)}_{\Ocal}
\equiv
 {\mathfrak D}_{\rm left}^{(p)}{\mathfrak D}_{\rm right}^{(q)} G_{\Ocal}
 \ee
 where $G_{\Ocal}$ is the scalar block. The blocks in the basis used in this paper are related to $\hat G^{(p,q)}_{\Ocal}$ as
 \be
 \hat G^{(p,q)}_{\Ocal}=\sum_{p'q'} (\mathfrak n_\Ocal)_{pp'}(\mathfrak n_\Ocal)_{qq'} G^{(p',q')}_{\Ocal} \ ,
 \ee
 where $\mathfrak n_\Ocal$ are matrices of coefficients independent of the cross-ratios.
  In the differential basis the residues at the poles are diagonal in the labels $p$ and $q$,
 \begin{align}
\hat G^{(p,q)}_{\Ocal}&= {\mathfrak D}_{\rm left}^{(p)}{\mathfrak D}_{\rm right}^{(q)} \frac{m^{(L)}_A Q_A m^{(R)}_A}{\D-\D_A^\star} G_{\Ocal_A}  + O\left((\D-\D^\star_A)^0 \right)\nonumber \\
 &=\frac{ m^{(L , p)}_A Q_A  m^{(R , q)}_A}{\D-\D_A^\star} {\mathfrak D}_{\rm left}^{(p)}{\mathfrak D}_{\rm right}^{(q)} G_{\Ocal_A}  + O\left((\D-\D^\star_A)^0 \right) \nonumber \\
 &=\frac{ r^{(p,q)}_A}{\D-\D_A^\star} \hat G^{(p,q)}_{\Ocal_A} + O\left((\D-\D^\star_A)^0 \right) \ ,
 \label{decoupling}
\end{align}
where $m^{(L)}_A Q_A m^{(R)}_A \equiv r_A$  is the residue of the scalar block and $ r^{(p,q)}_A\equiv  m^{(L,p)}_A Q_A  m^{(R,q)}_A$.
The coefficient $ m^{(p)}_A$ is morally the same as $m_A$ since the differential part of the operator ${\mathfrak D}^{(p)}$ clearly does not act on the residue $r_A$. However   the operators ${\mathfrak D}^{(p)}$ also act by shifting $\D_{12}$ (or $\D_{34}$) by some units.  Therefore the coefficient  $ m^{(p)}_A$ is trivially obtained by implementing such shifts on $m_A$.

Knowing $m^{(p)}_A$, it is trivial to obtain $M_A$ in a closed form, just by performing a change of basis from the differential basis to the one that we need. For example for the parity even conformal blocks we got
\be
\left( M^{(L)}_{+ A} {\phantom{\Big|}} \right)_{p q} = \sum_{p'=1}^5 \left(\mathfrak n_{\D_A^\star, \ell,+}^{-1}\right)_{p p'}
\;
 m^{(L,p')} \;
  \left(\mathfrak n_{\D_A, \ell_A, +} {\phantom{\Big|} }\right)_{p' q} \ ,
 \qquad {\scriptsize A= \left\{ \begin{array}{l} \I,n \\ \II,n  \\ \III,n \end{array} \right.}
\ee
where 
\begin{align}
m^{(L,1)}_A=m^{(L,2)}_A=m^{(L,5)}_A=m^{(L)}_A& \ , \\
m^{(L,3)}_A=m^{(L)}_A \Big|_{\D_{12}\rightarrow \D_{12}-2}\ , \qquad \ \ & m^{(L,4)}_A=m^{(L)}_A \Big|_{\D_{12}\rightarrow \D_{12}+2} \ ,
\end{align}
amd where $m^{(L)}_A$ are the scalar coefficients  \cite{arXiv:1509.00428},
\bea
m^{(L)}_{\I,n}&\equiv&(-2 i)^n \left(\frac{-n+\Delta
   _{12}+1}{2}\right)_n \ , \\
m^{(L)}_{\II,n}&\equiv &\frac{i^n \left(\frac{-n+\Delta
   _{12}+1}{2} \right)_n (2 h+\ell-n-2)_n}{(h+\ell-n-1)_n} \ , \\
   m^{(L)}_{\III,n}&\equiv &
   \frac{(-4)^n (h-n-1)_\ell \left(\frac{-h-\ell-n+\Delta
   _{12}+2}{2} \right)_n \left(\frac{h+\ell-n+\Delta _{12}}{2}
   \right)_n}{(h+n-1)_\ell}  \ .
\eea
The $5 \times 5$ matrix $\mathfrak n_{\D, \ell, +}$ implements the change of basis from the differential operator basis to the  basis (\ref{t_five_allowed}) that is used in this paper, 
\be
\label{LeadOPEtoDifferential}
\mathfrak n_{\D, \ell, +}\equiv \left(
\begin{array}{ccccc}
 \ell-\beta  & 2 \beta  (\Delta-\beta ) & (\beta -1) \ell & \ell
   (\beta -\Delta ) & \frac{(1-\ell) \ell}{2}  \\
 -1 & 0 & 0 & 0 & 0 \\
 1-\beta  & 2 (\beta -1) \beta  & (1-\beta ) \ell & (1-\beta)
   \ell& \frac{(\ell-1) \ell }{2} \\
 \beta -\Delta -\ell+1 & 2 (\beta -\Delta ) (\beta -\Delta +1) &
   \ell (\Delta-\beta  ) & \ell (\Delta-\beta  ) & \frac{ (\ell-1) \ell }{2}
  \\
 \beta -\Delta  & 2 \beta  (\Delta-\beta  ) & \ell (\beta
   -\Delta ) & (\beta -1) \ell & \frac{(1-\ell) \ell}{2} \\
\end{array}
\right) \ ,
\ee
 where $\b=(\D_{12}+\D+\ell)/2$. 
Similarly one can obtain the matrices for the parity odd conformal blocks. With this method we cannot fix the residue of the fourth type, since the scalar block does not have any pole of  type $\IV$. However these are just two new poles which can be easily investigated by a direct computation.

\subsubsection{Simplifications of the recurrence relations}
\label{DecouplingRecRel}
In this appendix we mention some interesting ways to simplify the recurrence relations which were not adopted in this work, but could be very useful for future investigations.

The first remarkable simplification of the recurrence relation comes from equation (\ref{decoupling}). Using the basis of the differential operators it is possible to define a fully diagonal recurrence relation of the form
\be
h^{(p,q)}_{\D \ell,s}(r,\eta)=
h^{(p,q)}_{\infty \D \ell,s}(r,\eta)+ \sum_A (4r)^{n_A} \frac{r^{(p,q)}_{\ell A}}{\D-\D^\star_A}  h^{(p,q)}_{\D_A \ell_A,s}(r,\eta) \, .
\label{eq:recinDeltaDiagonal}
\ee
Here $r^{(p,q)}_A$ is a constant which is obtained by shifting the scalar constant $r_A$.
From (\ref{decoupling}) it is also clear that the labels $A$, $\ell_A$, $n_A$ and $\D_A$ appearing in this recurrence relation would be the same as in  the scalar relation \cite{arXiv:1509.00428}.
The form (\ref{eq:recinDeltaDiagonal}) is much more convenient to compute  complicated conformal blocks (for example, for    four external stress tensors) since it decouples all the recurrence relations. However, in order to obtain such a beautiful result, one has to pay a price.
In fact in this basis function $h_\infty$ is a polynomial in $\D$. The order of such polynomial depends on the choice of the external operators and it is related to the matrix that changes basis from the usual $\D$-independent three point functions basis (for example the leading OPE basis  (\ref{t_five_allowed}) used in this paper) to the differential one (for example \eqref{GtoGhat} that can be obtained by implementing the change of basis \eqref{LeadOPEtoDifferential} which is $\D$-dependent). The fact that $h_\infty$ is a polynomial in turn implies that to obtain $h_\infty$ we need to solve the Casimir equation at sub-leading orders in large $\D$. 

Similarly one could implement the recurrence relation for the conformal blocks directly in the basis of conserved equal currents (\ref{ConservedEqualBasis+}). This was already successfully tried for one external conserved current in general dimension \cite{arXiv:1509.00428}. This approach would also give rise to important simplifications since we would just have $5$ coupled recurrence relations instead of $41$ (an even more drastic simplification would appear in the case of four external stress tensors). However, also in this case, the behaviour at large $\Delta$ of the blocks would change. In particular since the coefficients of (\ref{m+}) 
are  polynomial of order two in $\D$ (and we have to contract twice these matrices with the conformal blocks), we would need to solve the Casimir equation at four subleading orders in large $\D$. 

It is also interesting to notice that  one could implement the recurrence relation for the conformal blocks in the basis of equal vectors obtained by solving for example  (\ref{cqSystemForEqualOp}). This basis is very convenient since it does not depend on $\D$ and therefore it does not modify the behaviour at large $\D$ of the conformal blocks. It is therefore very simple to implement such a change and it would introduce a significant simplification. 

In this work we decided to stick to the most conservative recurrence relation valid for any external vector even if we had all the ingredients necessary to apply any of the previous simplifications (in fact we managed to solve the Casimir equation to many subleading orders in large $\D$). We did not further analyzed the simplified recurrence relations since the actual algorithm was already efficient enough for the computations done in this work. However it would be very interesting to obtain very efficient formulas for this and for more complicated conformal blocks valid in any dimensions. To do so it may be worth to implement the ideas mentioned in this section.
\subsubsection{A new implementation of the recurrence relation}
\label{appendix:cbnewmethod}
\begin{figure}[h!]
\graphicspath{{Fig/}}
\def\svgwidth{10 cm}
\centering
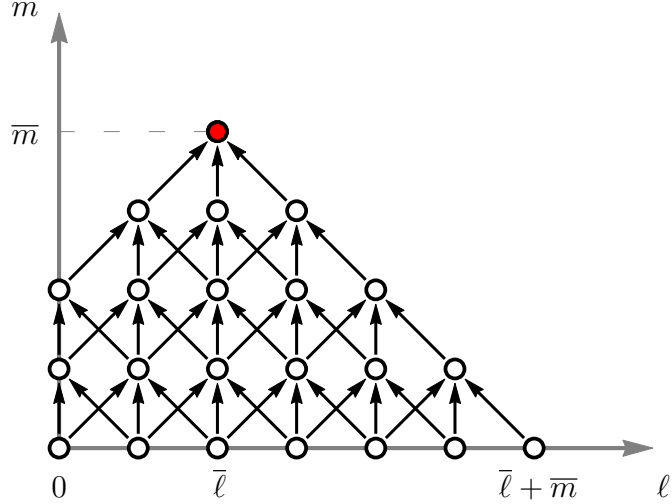 
\caption{ \label{fig:RecRelSpin}
 This plot shows which information is used to compute the block $h^{(p,q)}_{\D \bar \ell,s}[\bar m]$, represented as the red dot. In order to build this  block one needs to know all the  $h^{(p',q')}_{\D \ell,s}[m]$ for $\ell=\max[0,\bar \ell-\bar m], \dots , \bar \ell+\bar m$ and $m=\bar m - |\ell -\bar \ell|$. 
}
\end{figure}

In this section we want to comment on a new way to implement the recurrence relations, which was adopted in this work. We will exemplify the method for the case of four external vectors in three dimensions, but one can implement it also in any other case.
 It is convenient to introduce the notation
\be
h^{(p,q)}_{\D \ell \pm,s}(r ,\eta) \equiv \sum_{m=0}^\infty  r^m h^{(p,q)}_{\D \ell \pm,s}[m](\eta)\ ,
\ee
where $h^{(p,q)}_{\D \ell \pm,s}[m](\eta)$ are the coefficient of the expansion in $r$ of the conformal blocks, which are just functions of $\eta$. We shall drop the dependence on $\eta$ in the following formulae  for convenience.
The recurrence relation can be easily casted in the following $r-$independent form\footnote{Formula (\ref{recrelm}) is a more compact way to write 
(\ref{RecRelSpin+}-\ref{RecRelSpin-}). We hope not to confuse the reader with this change of notation. The simbol $\pm_A=\pm$ for the first three types, while  $\pm_{\IV,n}=\mp$.
}
\be
\label{recrelm}
h^{(p,q)}_{\D \ell \pm,s}[m]=
h^{(p,q)}_{\infty \ell \pm,s}[m]+ \sum_A\sum_{p',q'}  (4)^{n_A} \frac{(R_{\pm A})_{pp'qq'}}{\D-\D^\star_A}  h^{(p',q')}_{\D_A \ell_A \pm_A ,s}[m-n_A] \, .
\ee
Since $n_A>1$ it is clear how to use the formula to build $h[\bar m]$ from the knowledge of  $h[m]$ with $m<\bar m$. However it is interesting to notice that the blocks $h^{(p,q)}_{\D \ell \pm,s}[m]$ for $m<\bar m$ are called in the recurrence relation with different spins $\ell_A$. In particular in order to know $h^{(p,q)}_{\D \bar \ell +,s}[\bar m]$ (or $h^{(p,q)}_{\D \bar \ell -,s}[\bar m]$)
one needs to know\footnote{We put primes in the label $p$ and $q$ to stress that we need all the blocks for any $p$ and $q$, since (\ref{recrelm}) couples those labels. On the other hand $s$ is diagonal so we can study one vale of $s$ at the time.} $h^{(p',q')}_{\D \ell \pm,s}[\bar m-1]$ with $\ell=\bar \ell-1,\bar \ell,\bar \ell+1$, which can be computed knowing $h^{(p'',q'')}_{\D \ell \pm,s}[\bar m-2]$ with $\ell=\bar \ell-2,\bar \ell-1,\bar \ell,\bar \ell+1,\bar \ell+2$ and so on, until $h^{(p''',q''')}_{\D \ell \pm,s}[0]$ with $\ell=\max[0,\bar \ell-\bar m], \dots , \bar \ell+\bar m$. Therefore formula (\ref{recrelm}) can be conveniently implemented as a recurrence relation in spin. In figure \ref{fig:RecRelSpin} it is shown how to build $h^{(p,q)}_{\D \bar \ell \pm,s}[\bar m]$ using the information of the conformal blocks at different spins and lower value of $m$. 

\begin{figure}[h]
\graphicspath{{Fig/}}
\def\svgwidth{12 cm}
\centering
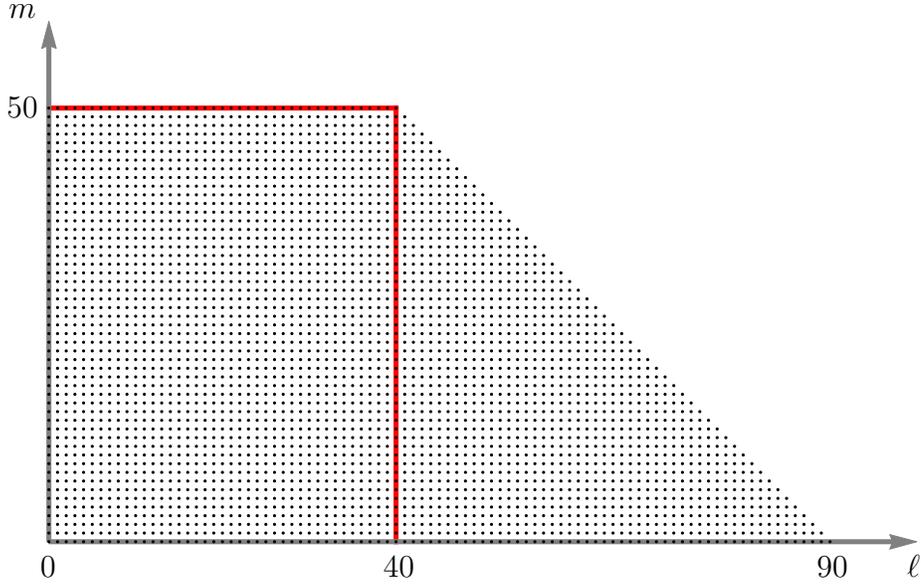 
\vspace{0.15 cm}
\caption{ \label{fig:n50l40}
Schematical representation of the computation of the blocks performed in this paper.
}
\end{figure}

The code that we implemented uses this strategy to compute all the blocks with various spins at once. Given the inputs of $\bar m$ and $\bar \ell$, it starts by computing all the blocks for $m=0$ and $\ell=0,\dots,\bar \ell+ \bar m$, then it computes all the blocks with $m=1$ and $\ell=0,\dots,\bar \ell+ \bar m-1$, and so on, until one obtain the blocks with $m=\bar m$ and $\ell=0,\dots,\bar \ell$. 
The output of the algorithm is therefore a list of all the blocks with $m\leq \bar m$ and $\ell\leq \bar \ell$, but it also generates blocks with higher spins and lower $m$.
With this algorithm we were able to obtain the blocks with $\bar m=50$ and $\bar \ell=40$, as showed in figure \ref{fig:n50l40}. 
Each dot in the picture corresponds to $41$ exact functions of $\D$ and $\eta$: $h^{(p,q)}_{\D \ell +,s}[m]$ for $p,q=1,\dots 5$ and $h^{(\pp,\qq)}_{\D \ell -,s}[m]$ for $\pp,\qq=1,\dots 4$ for a given structure $s$. This algorithm should be repeated $19$ times, which correspond to the number of structures $s$ that we considered.

\subsubsection{The recurrence relation in any dimension}
\label{app:RecGeneralD}
In this section we review how to write recurrence relations for conformal blocks in any dimension $d$ \cite{arXiv:1509.00428, Costa:2016xah}. We do this in order to comment on some subtleties involving the dependence on $d$ of the recurrence relations.

From representation theory we know the full set of poles $\D^\star_A$ of a bosonic conformal block $G_{\D \ell}^{(p,q)}$ for the exchange of the most generic $SO(d)$ representation $\ell=(l_1,\dots,l_{[\frac{d}{2}]})$.
In odd dimensions all the poles are simple and the residues at the poles $\D^\star_A$ are linear combinations of conformal blocks $G_{\D_A,\ell_A}^{(p',q')}$ associated to the exchange of a (null) primary descendant in the $SO(d)$ representation $\ell_A=(\D_A,l_{1 A},\dots,l_{[\frac{d}{2}] A})$, namely
\be
G_{\D \ell}^{(p,q)}(x_i) \sim \frac{1}{\D-\D^\star_A }\sum_{p\rq{},q\rq{}} (R_{\ell A})_{p p\rq{} q q\rq{}} G_{\D_A\ell_A}^{(p\rq{},q\rq{})}(x_i)  \ .
\ee
In the following table we write the full set of poles $\D^\star_A$ and the labels associated to the primary descendants $\D_A,\ell_A$ 
\be \label{AllTheLabelsd}
\begin{array}{|l | ccc|}
\hline
\phantom{\Big(}	\qquad \quad\quad	A								&\D^\star_A 					&n_A 										&l_{i A}	 \\ 
\hline
\phantom{\Big(}  \I_k, \; \;n: n\in [1,l_{k-1}-l_k]				& k-l_k-n 			 		& n 	  	   		&\quad l_i + n \d_{i k} \quad \\ 
\phantom{\Big(} \II_k,  \;n: n\in [1,l_{k}-l_{k+1}\,]  \;\; 	&      \quad   d+l_k-k-n \quad & \quad n \quad  			&\quad l_i - n \d_{i k} \quad  \\
\phantom{\Big(} \III,     \, \,   n: n \in [1,\infty) \; 				& \frac{d}{2}-n  						& 2n       			 	 &l_i     \\ 
\phantom{\Big(} \IV,     \, \,   n: n \in [1,l_{[\frac{d}{2}]}] \; 				& \frac{d+1}{2}-n  					& 2n  -1     			 	 &l_i     \\ 
\hline
\end{array}
\ee
where $k=1,\dots ,[\frac{d}{2}]$, $n$ is an integer and $\D_A=\D^\star_A+n_A$. 
 We can then reconstruct the full conformal block by summing over all the poles in $\D$ and over the regular part.
In radial coordinates this can be done by writing 
\be
g^{(p,q)}_{\D \ell,s}(r,\eta)=(4r)^\D h^{(p,q)}_{\D \ell,s}(r,\eta) \,,
\ee
where 
\be
h^{(p,q)}_{\D \ell,s}(r,\eta)=
h^{(p,q)}_{\infty \ell,s}(r,\eta)+ \sum_A\sum_{p',q'}  (4r)^{n_A} \frac{(R_{\ell A})_{pp'qq'}}{\D-\D^\star_A}  h^{(p',q')}_{\D_A \ell_A,s}(r,\eta) \,,
\label{eq:recinDelta}
\ee
and the regular part $h^{(p,q)}_{ \infty \ell,s}(r,\eta) \equiv \lim_{\D \rightarrow \infty } h^{(p,q)}_{\D \ell,s}(r,\eta)$. 

We would like to point out the difference between the behavior of the recurrence relations when $d$ is an odd integer and when it is a continuous analytic variable. The even-dimensional case is obtained as a limit of general $d$.

First we want to comment on the types $\I_k, \II_k$. We want to show that there may be more poles associated to these types when we do not fix the spacetime dimension. In fact the label $k$ is bounded to be in the set $1,\dots ,[\frac{d}{2}]$. While for any concrete integer $d$ this range is finite, when $d$ is not fixed this range is virtually infinite.
When $d$ is not fixed the range only truncates because of the choice of the external operators. 
In fact, at a given pole of  type $A$, the residue is a conformal block $h_{\Ocal_A}$. 
However $\Ocal_A$ may be labeled by an $SO(d)$ representation which cannot possibly couple to the external operators, namely $\Ocal_A \notin (\Ocal_1\times \Ocal_2) \cap (\Ocal_3\times \Ocal_4)$.
This phenomenon is particularly relevant for the types $(\I_k,n)$ (or $(\II_k,n)$) when $k>1$. In fact in these cases the  $SO(d)$ representation of $\Ocal_A$ is different from the one of $\Ocal$, since the Young tableau associated to $\Ocal_A$ has $n$ more boxes (or $n$ less boxes) in its $k-$th row.
As an example, if we consider any choice of  traceless and symmetric external operators in any dimension, the label $k$ is bounded to take the values $k=1,2,3$.
When $d$ is fixed to an odd integer instead we have a finite number of $k$ independently of the choice of the external operators. For example in three dimensions the types $\I_k$ and $\II_k$ exist just for $k=1$ and similarly for $d=5$ we only have $k=1,2$.

The reverse phenomenon happens for the type $\IV$. In generic dimensions the type $(\IV,n)$ cannot appear since $n$ runs in the set $[1,l_{[\frac{d}{2}]}]$. 
In fact, since $d$ is not specified, the label $l_{[\frac{d}{2}]}$ of the exchanged operator has to vanish for any choice of the external operators.
Instead, for $d$ integer, $l_{[\frac{d}{2}]}$ is well defined and the poles of type $\IV$ may appear depending on the choice of the external operators.

The fact that some poles exist only in integer dimensions and that others exist only when $d$ is not integer may seem paradoxical, since we claim that the recurrence relations are analytic functions of $d$.
The resolution of the paradox comes from the fact that the poles of type $\I_k, \II_k$ which disappear at integer values of $d$, are exactly replaced by the poles of type $\IV$ which exist only at $d$ integer.

For example in the case of $4$ external vectors in generic $d$, besides the poles of  type $\I_1$, $\II_1$ and $\III$ (which are present also in the scalar case), there are also new poles coming from type $\I_2, n=1,2$, $\II_2, n=1,2$, $\I_3, n=1$, $\II_3, n=1$. We can read their position from the table \eqref{AllTheLabelsd}. Once we set the value of $d=3$,  we  obtain three new poles at positions $\D=0,1,2$.
On the other hand when we consider the table \eqref{AllTheLabelsd} directly in three dimensions, we  predict $\ell$  new poles of type $\IV$ at the positions $\D=1,0,-1,\dots, -\ell+2$. 
One can check that, rather magically, the residues $R$ correspondent to  type $\IV$ vanish for $n>2$ giving  poles only at $\D=1,0$. In a similar way one can prove that the pole at position $\D=2$ predicted in general dimension has a vanishing residue when $d=3$.

\section{Details about numerical implementation}
\label{sec:numdetails}

Let us describe the numerical implementation one step at a time.
\begin{enumerate}
 \item As explained in appendix \ref{app:ConformalBlocks}, conformal blocks are naturally computed in a basis with 43 tensor structures. However, in three dimensions there exist two linear relations among the 43 tensor structures, the first task we need to perform is to express the three dimensional crossing equations in terms of the 43 (19 for equal currents) original functions. 
 
In our basis of tensor structures, the two linear relations read:
 {\footnotesize{
 \begin{align}
&\left(u^2+(-1+v)^2-2 u (1+v)\right) Q_2+\left(-u^2-(-1+v)^2+2 u (1+v)\right) Q_3+4 u v Q_8-2 \sqrt{u} \sqrt{v} (-1+u+v) Q_9\nonumber\\
&-2 \sqrt{u} \sqrt{v} (-1+u+v) Q_{10}+4 u v Q_{11}+2  u \sqrt{v} (1-u+v) Q_{12}+2 \sqrt{u} (1+u-v) v Q_{13}\nonumber\\
&+2 \sqrt{u} \left(1+u^2-v-u (2+v)\right) Q_{14}+2 u \sqrt{v} (1-u+v) Q_{15}+2 u \sqrt{v} (1-u+v) Q_{16}+2 \sqrt{u} (1+u-v) v Q_{17}\nonumber\\
&-2 \sqrt{u} \left(-1-u^2+v+u (2+v)\right) Q_{18}+2 u \sqrt{v} (1-u+v) Q_{19}+4 u v Q_{20}-2 \sqrt{u} \sqrt{v} (-1+u+v) Q_{21}\nonumber\\
&-2 \sqrt{u} \sqrt{v} (-1+u+v) Q_{22}+4 u v Q_{23}-4 u^2 v Q_{28}+4 u^{3/2} v^{3/2} Q_{29}+4 u^{3/2} v^{3/2} Q_{30}\nonumber\\
&-4 u (-1+v) v Q_{31}+4 (-1+u) u^{3/2} \sqrt{v} Q_{32}-4 u (1+u) v Q_{33}-4 (-1+u) u v Q_{34}+4 u^{3/2} v^{3/2} Q_{35}\nonumber\\
&+4 (-1+u) u^{3/2} \sqrt{v} Q_{36}-4 (-1+u) u v Q_{37}-4 u (1+u) v Q_{38}+4 u^{3/2} v^{3/2} Q_{39}\nonumber\\
&-4 u \left(1-2 u+u^2-v\right) Q_{40}+4(-1+u) u^{3/2} \sqrt{v} Q_{41}+4 (-1+u) u^{3/2} \sqrt{v} Q_{42}-4 u^2 v Q_{43}=0\nonumber\\
&\nonumber\\
&\left(u^2+(-1+v)^2-2 u (1+v)\right) Q_1+\left(-u^2-(-1+v)^2+2 u (1+v)\right) Q_3+2 u (-1+u-v) \sqrt{v} Q_4\nonumber\\
&-2 \sqrt{u} (1+u-v) v Q_5-2 \sqrt{u} (1+u-v) v Q_6-2 \sqrt{v} \left((-1+v)^2-u (1+v)\right) Q_7+2 u \sqrt{v} (1-u+v) Q_{12}\nonumber\\
&+2 \sqrt{u} (1+u-v) v Q_{13}+2 \sqrt{u} \left(1+u^2-v-u (2+v)\right) Q_{14}+2 u \sqrt{v} (1-u+v) Q_{15}+2 u\sqrt{v} (1-u+v) Q_{16}\nonumber\\
&+2 \sqrt{u} (1+u-v) v Q_{17}+2 \sqrt{u} \left(1+u^2-v-u (2+v)\right) Q_{18}+2 u \sqrt{v} (1-u+v) Q_{19}\nonumber\\
&-2 \sqrt{v} \left((-1+v)^2-u (1+v)\right)Q_{24}-2 \sqrt{u} (1+u-v) v Q_{25}-2 \sqrt{u} (1+u-v) v Q_{26}+2 u (-1+u-v) \sqrt{v} Q_{27}\nonumber\\
&+4 u v (1-u+v) Q_{28}+4 \sqrt{u} (1+u-v) v^{3/2} Q_{29}+4 \sqrt{u} (1+u-v) v^{3/2} Q_{30}+4 v \left(1-(2+u) v+v^2\right) Q_{31}\nonumber\\
&+4 u^{3/2} (-1+u-v) \sqrt{v} Q_{32}-4 u (1+u-v) v Q_{33}+4 u v (-u+v) Q_{34}+4 \sqrt{u} (1+u-v) v^{3/2} Q_{35}\nonumber\\
&+4 u^{3/2} (-1+u-v) \sqrt{v} Q_{36}-4 u (u-v) v Q_{37}-4 u (1+u-v) v Q_{38}+4 \sqrt{u} (1+u-v) v^{3/2} Q_{39}\nonumber\\
&-4 u \left(1+u^2-u (2+v)\right) Q_{40}+4 u^{3/2} (-1+u-v)  \sqrt{v} Q_{41}\nonumber\\
&+4 u^{3/2} (-1+u-v) \sqrt{v} Q_{42}+4 u v (1-u+v) Q_{43}=0
\label{eq:linearRelations}
 \end{align}}}

In this project we choose to eliminate structures $Q_{31}$ and $Q_{40}$ using this two identities. This is motivated  by their invariance under the  permutations $1234\leftrightarrow 3412$ and  $1234\leftrightarrow 2143$, as shown in Table~\ref{4pstruct}, and by the fact that they do not mix under crossing (see eq.~\eqref{eq:crossingftilde} and appendix \ref{ap:ftilde}). Also, inverting \eqref{eq:linearRelations} in terms of this pair does not introduce any singularity at the crossing symmetric point $u=v=1/4$.
\footnote{On the contrary, this choice is not optimal when performing the conformal block decomposition by matching powers of $r$: this is because it would introduce additional singularities at $r=0$. For such an exercise it is more convenient to invert eq.~\eqref{eq:linearRelations} for $Q_{37}$ and $Q_{38}$. }

Finally we plug in the expression of $Q_{31}$ and $Q_{40}$ in the tensor structure expansion of the four point function \eqref{eq:4pJJJJ} to obtain the expression of the 17 independent functions in terms of the original 19:

 \beq
  \sum_{s=1}^{43}  f_s(u,v) \,Q_s =
  \sum_{s=1}^{41}  f_s^{3D}(u,v) \,Q_s \,,
\eeq
Note that, despite the index $s$ on the rhs of the above expression runs from $1$ to $41$, there are only 17 distinct functions $f_s^{3D}(u,v)$.

\item Next, we pass to the basis that diagonalizes crossing symmetry. This is done by defining the 17 functions $\tilde{f}_{s}$ shown in \eqref{eq:ftilde}.

\item As described in \cite{Kos:2014bka}, the problem of finding $\alpha$ satisfying \eqref{eq:functional} can be transformed into a semidefinite program. 
The form of the functional $\alpha$ is given in \eqref{eq:functionaldef}. The first step is to compute the derivatives of the vectors $V_+$,$V_{\ell +}$ and $V_{\ell -}$ defined in \eqref{eq:Vdef}.
To do this, we started directly from the explicit form of the conformal blocks as a power series in the variable $r$ defined in \eqref{radcoord}. 

\item Once we take the derivatives and set $r=3-2\sqrt{2}$ and $\eta=1$, these expressions reduce to rational approximations for conformal blocks in the variable $\Delta$.  Keeping only the polynomial numerator in these rational approximations, \eqref{eq:functional} becomes a ``polynomial matrix program" (PMP), which can be solved with \texttt{SDPB} \cite{Simmons-Duffin:2015qma}.  
We use \texttt{Mathematica} to compute and store tables of derivatives of conformal blocks.  Another \texttt{Mathematica} program reads these tables, computes the polynomial matrices corresponding to the $\vec V$'s, and uses the package \texttt{SDPB.m} to write the associated PMP to an \texttt{xml} file.  This \texttt{xml} file is then used as input to \texttt{SDPB}. Our settings for \texttt{SDPB} are given in Table \ref{tab:parameters}.

\end{enumerate}

As discussed in Sec.~\ref{sec:conservation}, the minimal crossing constraints consist in 5 bulk equations, 5 boundary equations and one constraint at a point. Once one considers derivatives of the crossing equations at a given point, the conservation equations \eqref{eq:conservation} (and their derivatives at the crossing symmetric point $u=v=1/4$) simply become a set of linear relations between various derivatives of the functions  $\tilde{f}_{s}$. We explicitly checked that the set of derivatives included in the numerical bootstrap has maximal rank, i.e. there is no linear dependence induced by the conservation equations. Also, we explicitly checked that the system made by the conservation equations and their derivatives at the crossing symmetric point can be used to determine neglected components.  

Because the functions involved have definite symmetric properties under $z\rightarrow \bar{z}$, the number of non-vanishing derivatives included for a given $\Lambda$ is:
 \be
 \dim(\vec \alpha) = 5\frac{\lfloor \frac{\Lambda+2}{2}\rfloor(\lfloor \frac{\Lambda+2}{2}\rfloor+1)}{2} + 5 \lfloor \frac{\Lambda+2}{2}\rfloor + 1.
 \ee

The degree of the numerator and denominator is controlled by the order of the conformal blocks expansion in $r$, or equivalently by the number of poles kept in the recursion relations (\ref{RecRelSpin+}-\ref{RecRelSpin-}). 
Contrary to previous conformal bootstrap works \cite{Kos:2014bka}, we used the obtained expressions as they are, without employing any further approximation. Approximations might be useful to push to higher number of derivatives.

Finally, we must choose which spins to include in the PMP.  
We have chosen the number of spins to depend on $\Lambda$ as follows
\bea
S_{\L=11} &=& \{0,\dots,24\} \cup \{29,30\},\nn\\
S_{\L=15} &=& \{0,\dots,34\} \cup \{39,40\},\nn\\
S_{\L=19} &=& \{0,\dots,40\} \cup \{49,50\},\nn\\
S_{\L=23} &=& \{0,\dots,40\} \cup \{44,45,49,50,59,60\}.
\eea

\begin{table}[!htb]
\centering
\begin{tabular}{|l|c|c|c|c|}
\hline
$\Lambda$ & 11 & 15 & 19 & 23 \\
order & 30 & 40 & 40 & 50 \\
spins & $S_{\L=11}$ & $S_{\L=15}$ & $S_{\L=19}$ & $S_{\L=23}$\\
\texttt{precision} & 448 & 576 & 768 & 896\\
\texttt{findPrimalFeasible} & True & True & True & True\\
\texttt{findDualFeasible} & True & True & True & True\\
\texttt{detectPrimalFeasibleJump} & True & True & True & True\\
\texttt{detectDualFeasibleJump} & True & True & True & True\\
\texttt{dualityGapThreshold} & $10^{-30}$ & $10^{-30}$ & $10^{-30}$ & $10^{-70}$ \\
\texttt{primalErrorThreshold} & $10^{-30}$ & $10^{-30}$ & $10^{-40}$ & $10^{-70}$ \\
\texttt{dualErrorThreshold} & $10^{-30}$ & $10^{-30}$ & $10^{-40}$ & $10^{-70}$ \\
\texttt{initialMatrixScalePrimal} & $10^{40}$ & $10^{50}$ & $10^{50}$ & $10^{60}$\\
\texttt{initialMatrixScaleDual} & $10^{40}$ & $10^{50}$ & $10^{50}$ & $10^{60}$\\
\texttt{feasibleCenteringParameter} & 0.1 & 0.1 & 0.1 & 0.1 \\
\texttt{infeasibleCenteringParameter}  & 0.3 & 0.3 & 0.3 & 0.3\\
\texttt{stepLengthReduction} & 0.7 & 0.7 & 0.7 & 0.7\\
\texttt{choleskyStabilizeThreshold}  & $10^{-40}$ & $10^{-40}$ & $10^{-100}$ & $10^{-120}$ \\
\texttt{maxComplementarity} & $10^{100}$ & $10^{130}$ & $10^{160}$ & $10^{180}$\\
\hline
\end{tabular}
\caption{\texttt{SDPB} parameters for the computations of scaling dimension bounds in this work. For $C_T$   bounds we need to set all of the Boolean parameters in the table to False. In addition to that, we used $\texttt{dualityGapThreshold}=10^{-6}$, while all the rest of the parameters were kept at the same values as for the dimension bounds.}
\label{tab:parameters}
\end{table}

\newpage

\bibliographystyle{utphys}
\bibliography{JJJJbiblio}
\end{document}

%% file: Plots/Conservation.pdf_tex
\begingroup%
  \makeatletter%
  \providecommand\color[2][]{%
    \errmessage{(Inkscape) Color is used for the text in Inkscape, but the package 'color.sty' is not loaded}%
    \renewcommand\color[2][]{}%
  }%
  \providecommand\transparent[1]{%
    \errmessage{(Inkscape) Transparency is used (non-zero) for the text in Inkscape, but the package 'transparent.sty' is not loaded}%
    \renewcommand\transparent[1]{}%
  }%
  \providecommand\rotatebox[2]{#2}%
  \ifx\svgwidth\undefined%
    \setlength{\unitlength}{1008.71386719bp}%
    \ifx\svgscale\undefined%
      \relax%
    \else%
      \setlength{\unitlength}{\unitlength * \real{\svgscale}}%
    \fi%
  \else%
    \setlength{\unitlength}{\svgwidth}%
  \fi%
  \global\let\svgwidth\undefined%
  \global\let\svgscale\undefined%
  \makeatother%
  \begin{picture}(1,1)%
    \put(0,0){\includegraphics[width= \unitlength]{./Plots/Conservation.pdf}}%

\put(0.98 ,0.33){\color[rgb]{0,0,0}{\makebox(0,0)[lb]{\smash{$u$}}}}
\put(0.1 ,1.1){\color[rgb]{0,0,0}{\makebox(0,0)[lb]{\smash{$v$}}}}
\put(0.90 ,0.9){\color[rgb]{0,0,0}{\makebox(0,0)[lb]{\smash{$y$}}}}
\put(0.87 ,0.07){\color[rgb]{0,0,0}{\makebox(0,0)[lb]{\smash{$t$}}}}

\put(0.36 ,0.2){\color[rgb]{0,0,0}{\makebox(0,0)[lb]{\smash{$\frac{1}{4}$}}}}
\put(0.11 ,0.48){\color[rgb]{0,0,0}{\makebox(0,0)[lb]{\smash{$\frac{1}{4}$}}}}
  \end{picture}%
\endgroup%

%% file: Plots/subrep3.pdf_tex
\begingroup%
  \makeatletter%
  \providecommand\color[2][]{%
    \errmessage{(Inkscape) Color is used for the text in Inkscape, but the package 'color.sty' is not loaded}%
    \renewcommand\color[2][]{}%
  }%
  \providecommand\transparent[1]{%
    \errmessage{(Inkscape) Transparency is used (non-zero) for the text in Inkscape, but the package 'transparent.sty' is not loaded}%
    \renewcommand\transparent[1]{}%
  }%
  \providecommand\rotatebox[2]{#2}%
  \ifx\svgwidth\undefined%
    \setlength{\unitlength}{1008.71386719bp}%
    \ifx\svgscale\undefined%
      \relax%
    \else%
      \setlength{\unitlength}{\unitlength * \real{\svgscale}}%
    \fi%
  \else%
    \setlength{\unitlength}{\svgwidth}%
  \fi%
  \global\let\svgwidth\undefined%
  \global\let\svgscale\undefined%
  \makeatother%
  \begin{picture}(1,1)%
    \put(0,0){\includegraphics[width= \unitlength]{./Plots/subrep3.pdf}}%

\put(0.59 ,0.567){\color[rgb]{0,0,0}{\makebox(0,0)[cc]{\smash{$\Ocal_{\!\!\!\, A}$}}}}

\put(0.3 ,0.43){\color[rgb]{0,0,0}{\makebox(0,0)[lb]{\smash{$\Ocal$}}}}

\put(0.395,0.06){\color[rgb]{0,0,0}{\makebox(0,0)[cb]{\smash{$\D=\D^\star_A$}}}}
\put(0.575,0.06){\color[rgb]{0,0,0}{\makebox(0,0)[lb]{\smash{$\D_A$}}}}
\put(0.07 ,0.51){\color[rgb]{0,0,0}{\makebox(0,0)[lb]{\smash{$\ell_A$}}}}
\put(0.085 ,0.375){\color[rgb]{0,0,0}{\makebox(0,0)[lb]{\smash{$\ell$}}}}



  \end{picture}%
\endgroup%

%% file: Plots/RecRelSpin.pdf_tex
\begingroup%
  \makeatletter%
  \providecommand\color[2][]{%
    \errmessage{(Inkscape) Color is used for the text in Inkscape, but the package 'color.sty' is not loaded}%
    \renewcommand\color[2][]{}%
  }%
  \providecommand\transparent[1]{%
    \errmessage{(Inkscape) Transparency is used (non-zero) for the text in Inkscape, but the package 'transparent.sty' is not loaded}%
    \renewcommand\transparent[1]{}%
  }%
  \providecommand\rotatebox[2]{#2}%
  \ifx\svgwidth\undefined%
    \setlength{\unitlength}{1008.71386719bp}%
    \ifx\svgscale\undefined%
      \relax%
    \else%
      \setlength{\unitlength}{\unitlength * \real{\svgscale}}%
    \fi%
  \else%
    \setlength{\unitlength}{\svgwidth}%
  \fi%
  \global\let\svgwidth\undefined%
  \global\let\svgscale\undefined%
  \makeatother%
  \begin{picture}(1,0.80)%
    \put(0,0){\includegraphics[width= \unitlength]{./Plots/RecRelSpin.pdf}}%

\put(0.9 ,0.04){\color[rgb]{0,0,0}{\makebox(0,0)[lb]{\smash{$\ell$}}}}

\put(0.69 ,0.04){\color[rgb]{0,0,0}{\makebox(0,0)[lb]{\smash{$\bar \ell+\bar m$}}}}
\put(0.31,0.04){\color[rgb]{0,0,0}{\makebox(0,0)[lb]{\smash{$\bar \ell$}}}}
\put(0.095,0.04){\color[rgb]{0,0,0}{\makebox(0,0)[lb]{\smash{$0$}}}}

\put(0.06 ,0.68){\color[rgb]{0,0,0}{\makebox(0,0)[cc]{\smash{$m$}}}}
\put(0.06 ,0.51){\color[rgb]{0,0,0}{\makebox(0,0)[cc]{\smash{$\bar m$}}}}

  \end{picture}%
\endgroup%

%% file: Plots/n50l40.pdf_tex
\begingroup%
  \makeatletter%
  \providecommand\color[2][]{%
    \errmessage{(Inkscape) Color is used for the text in Inkscape, but the package 'color.sty' is not loaded}%
    \renewcommand\color[2][]{}%
  }%
  \providecommand\transparent[1]{%
    \errmessage{(Inkscape) Transparency is used (non-zero) for the text in Inkscape, but the package 'transparent.sty' is not loaded}%
    \renewcommand\transparent[1]{}%
  }%
  \providecommand\rotatebox[2]{#2}%
  \ifx\svgwidth\undefined%
    \setlength{\unitlength}{1008.71386719bp}%
    \ifx\svgscale\undefined%
      \relax%
    \else%
      \setlength{\unitlength}{\unitlength * \real{\svgscale}}%
    \fi%
  \else%
    \setlength{\unitlength}{\svgwidth}%
  \fi%
  \global\let\svgwidth\undefined%
  \global\let\svgscale\undefined%
  \makeatother%
  \begin{picture}(1,0.6)%
    \put(0,0){\includegraphics[width= \unitlength]{./Plots/n50l40.pdf}}%

\put(0.97 ,-0.02){\color[rgb]{0,0,0}{\makebox(0,0)[lb]{\smash{$\ell$}}}}

\put(0.87 ,-0.02){\color[rgb]{0,0,0}{\makebox(0,0)[lb]{\smash{$90$}}}}
\put(0.39,-0.02){\color[rgb]{0,0,0}{\makebox(0,0)[lb]{\smash{$40$}}}}
\put(0.01,-0.02){\color[rgb]{0,0,0}{\makebox(0,0)[lb]{\smash{$0$}}}}

\put(-0.01 ,0.6){\color[rgb]{0,0,0}{\makebox(0,0)[cc]{\smash{$m$}}}}
\put(-0.01 ,0.49){\color[rgb]{0,0,0}{\makebox(0,0)[cc]{\smash{$50$}}}}

  \end{picture}%
\endgroup%